\documentclass[prl,amsmath,amssymb,twocolumn, showpacs, superscriptaddress,10pt]{revtex4-1}

\usepackage{amsmath}
\usepackage{hyperref}
\usepackage{graphicx}
\usepackage{amsfonts}
\usepackage{amsthm}
\usepackage{cases}
\usepackage{mathtools,amssymb}
\usepackage{soul}
\usepackage{upgreek} 
\usepackage{dsfont}

\usepackage[dvipsnames]{xcolor}

\newcommand{\be}{\begin{equation}}
\newcommand{\ee}{\end{equation}}
\newcommand{\bea}{\begin{eqnarray}}
\newcommand{\eea}{\end{eqnarray}}

\newcommand{\I}{\ensuremath{\mathbf{i}}}
\newcommand{\temp}{\theta}
\newcommand{\Enorm}{\mathcal{E}}

\usepackage{tikz}
\usetikzlibrary{snakes}

\begin{document}

\title{Stochastic growth in time dependent environments}

\author{Guillaume Barraquand}
\affiliation{Laboratoire  de  Physique  de  l'\'Ecole  Normale  Sup\'erieure,  ENS, CNRS,  Universit\'e  PSL,   Sorbonne  Universit\'e,  Universit\'e de  Paris, 24 rue Lhomond, 75231 Paris, France}

\author{Pierre Le Doussal}
\affiliation{Laboratoire  de  Physique  de  l'\'Ecole  Normale  Sup\'erieure,  ENS, CNRS,  Universit\'e  PSL,  Sorbonne  Universit\'e,  Universit\'e de  Paris, 24 rue Lhomond, 75231 Paris, France}

\author{Alberto Rosso}
\affiliation{LPTMS, CNRS, Univ. Paris-Sud, Universit\'e Paris-Saclay, 91405 Orsay, France}

\date{\today}

\begin{abstract}
We study the Kardar-Parisi-Zhang (KPZ) growth equation in one dimension with a noise variance $c(t)$ depending on time. We find that for $c(t)\propto t^{-\alpha}$ there is a transition at $\alpha=1/2$. When $\alpha>1/2$, the solution saturates at large times towards a non-universal limiting distribution. When $\alpha<1/2$ the fluctuation field is governed by scaling exponents depending on $\alpha$ and the limiting statistics are similar to the case when  $c(t)$ is constant. We investigate this problem using different  methods: (1) Elementary changes of variables mapping the time dependent case to variants of the KPZ equation with constant variance of the noise but in a deformed potential (2) An exactly solvable discretization, the log-gamma polymer model (3) Numerical simulations.    
\end{abstract}



\maketitle

{\bf Introduction.} Many growth models in one spatial dimension share universal scaling properties. In the Kardar-Parisi-Zhang universality class, this phenomenon is particularly manifest since not only scaling exponents are universal, but  depending on initial data, the limiting distribution of fluctuations is universal as well. 
A much studied model in this class is the KPZ equation \cite{KPZ} in one dimension, where the height field $h(x,t)$ satisfies 
\be \label{kpz}
\partial_t h(x,t) = \partial_x^2 h(x,t) + (\partial_x h(x,t))^2 + \sqrt{2 c} ~ \xi(x,t) 
\ee
with $\overline{ \xi(x,t) \xi(x',t') } = \delta(x-x') \delta(t-t')$.
Its solution is related, via $h(x,t)= \log Z(x,t)$, to (minus) the free energy of a continuum directed polymer (DP) 
in the random potential $\xi(x,t)$ 
at finite temperature. The partition function of the DP with endpoint at $x,t$, $Z(x,t)$, obeys the stochastic heat equation (SHE). It is known \cite{amir2011probability, calabrese2010free, dotsenko2010replica, sasamoto2010exact} 
that
the local height fluctuations grow at large time $t$ as $\delta h \simeq (c^2 t)^{1/3} \chi  $  \cite{footnote2} where the 
probability distribution function (PDF) of the random variable  $\chi$ 
is related to random matrix theory and depends on some features of the initial condition (IC),
which fall into IC classes. For the class containing the droplet IC, $\chi$ follows the the GUE Tracy-Widom (TW) distribution. The integrability of the KPZ equation is related to the integrability of quantum models, e.g. of the
one dimensional (attractive) delta Bose gas with interaction parameter $-c<0$ \cite{footnote1}. 
One also defines a
spatial correlation scale for the height fluctuations, which grows as $x \propto t^{2/3}$.
The transverse wandering of the DP also grows with this length scale.

In this paper we study the case where the amplitude of the noise
depends itself on time, i.e.  $c$ becomes a function  $c(t)$ in \eqref{kpz}. One expects that if $c(t)$ decays very fast, 
some saturation of the fluctuations may occur, if $c(t)$ decays slowly, perhaps the fluctuations are similar as the homogeneous case. Considering a time dependent interaction parameter $c(t)$ is also of great 
importance in the related problem of quantum quenches \cite{Gritsev2010,KormosBosons,deNardisCaux,PCPLDQuench,ErmakovBethe, colcelli2019integrable}. 

There is to our knowledge only one exact result in the growth context. Before describing it, let us go back to the homogeneous case. In a seminal paper, Johansson \cite{johansson2000shape}
obtained an exact and concise formula for the optimal energy for a directed polymer at zero temperature
on the square lattice (of coordinate $i,j$) with random exponential on site energies. He then extended in \cite{johansson2008some} 
the solution to an inhomogeneous disorder, where the amplitude of the disorder
was chosen $1/(i^a+j^a)$, with $a \geqslant 0$. He discovered a sharp transition at $a=1/3$:  $\delta h\propto t^{1/3-a}$ with universal TW fluctuations if $a>1/3$ and $\delta h<\infty$ with non universal fluctuations if $a<1/3$. 

A natural question is whether, for the KPZ equation itself, or more generally for 
other finite temperature models, there is a similar transition, and 
how does it depend on the profile of $c(t)$? We probe this question here 
by 
considering three models, using 
complementary methods. 

(i) The first one is the KPZ equation, for which we use change of variables. Despite the simplicity of the method, 
it leads to interesting results. We obtain exact solutions for
the inhomogeneous KPZ equation
\be \label{kpz2}
\partial_t h = \partial_x^2 h + (\partial_x h)^2 + V(x,t) + \sqrt{2 c(t)} ~ \xi(x,t) 
\ee
with both time dependent noise amplitude and an external potential $V(x,t)= a(t)x^2 +b(t)x$,
for some specific relation between $a(t)$ and $c(t)$. One finds a 
transition from TW to non-universal fluctuations at large time
depending on whether $\int_0^{+\infty} dt \, c(t)^2$ is, respectively,
divergent or
convergent. In the case $c(t) \propto  t^{-\alpha}$, the transition is at $\alpha=1/2$.
We obtain the exponents for the height fluctuations $\delta h \propto t^{\beta(\alpha)}$
and for the correlation scale $x \propto t^{\zeta(\alpha)}$ and the PDF of
the height for various cases. 

(ii) Next, we study a discretization of the KPZ equation, a directed polymer on the square latttice
at finite temperature, called the log-gamma polymer. This model is known to be integrable in presence of
inhomogeneity parameters $\gamma_{i,j} = \temp (i^a+j^a)$ which control the strength of the disorder
at location $(i,j)$. 
As $\temp $ goes to zero, after rescaling, one recovers Johansson's model.
For the positive temperature model we find a transition at $a=1/2$:  {$\delta h \propto  t^{\frac{1-2a}{3}}$} with universal TW fluctuations if $a>1/2$ and $\delta h<\infty$ with non universal fluctuations if $a<1/2$. In particular, when the exponent $a\in(1/3,1/2)$, the free energy fluctuations are universal at positive temperature ($\temp>0$) and non-universal at zero-temperature ($\temp \to 0$).

\begin{figure}[h]
	\centering
	\includegraphics[width=8.5cm]{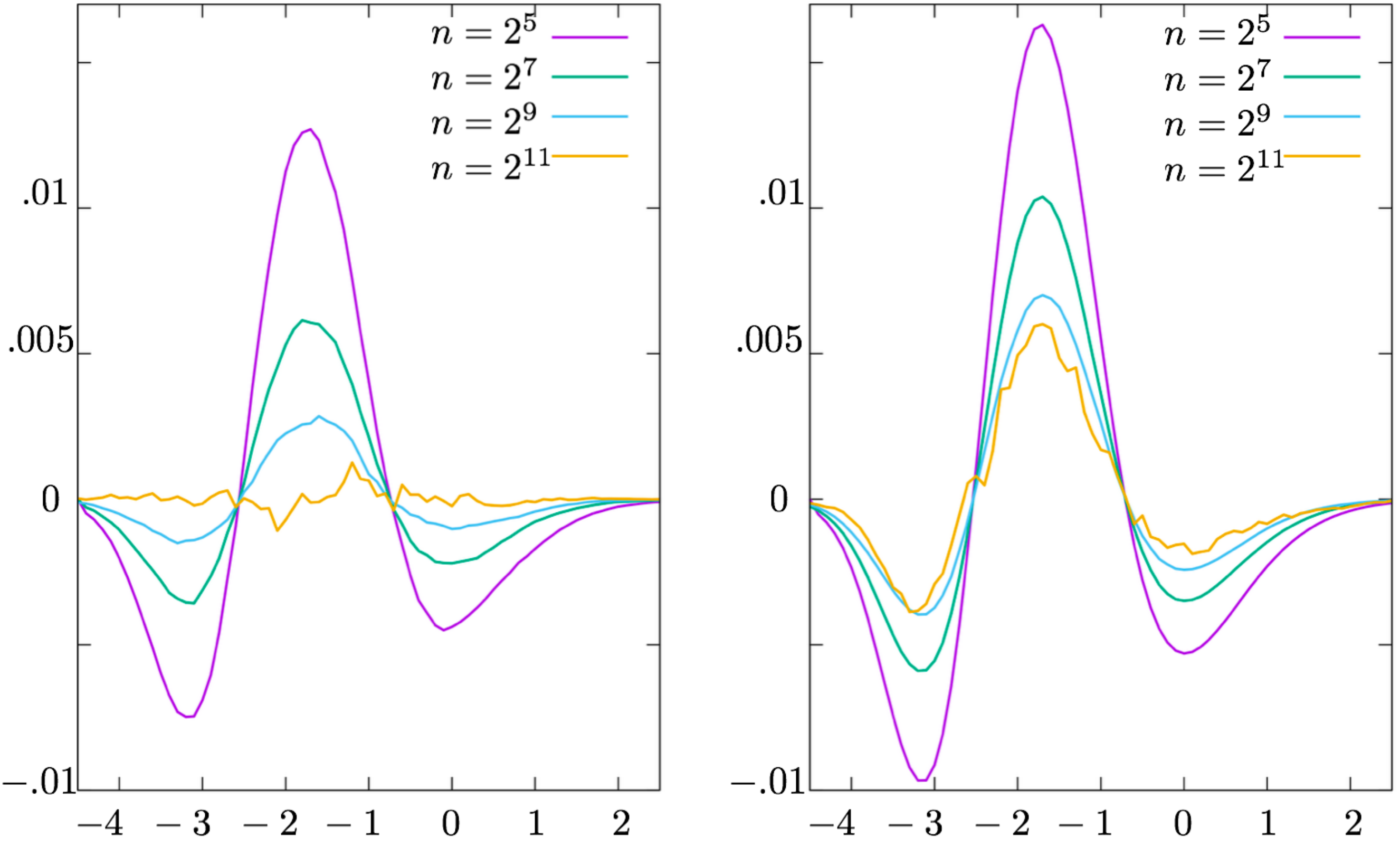}
 (a) \hspace{4cm} (b) 
	\caption{Model (iii) at zero temperature: difference between the empirical CDF of the ground state energy and the CDF of the GUE TW distribution (centered and scaled to the same mean and
		variance). (a): for $a=0.3$ and various polymer lengths $n$. (b): for $a=0.4$ and the same polymer lengths. See also a comparison of the tails in \cite{SM}.}
	\label{fig:DPTW}
\end{figure}

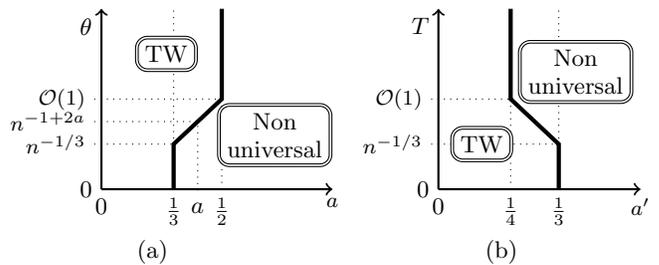
\begin{figure}[h]
	\begin{center}
		\begin{tikzpicture}[xscale=1.6, yscale=1 ,  every text node part/.style={align=center}]
		\begin{scope}[scale=0.6]
		\draw[thick, ->] (0,0) node[anchor=east]{$0$} -- (3.2,0) node[anchor=north]{{\footnotesize $a$}};
		\draw [thick, ->] (0,0) node[anchor=north]{$0$} -- (0,4) node[anchor=north east]{$\temp$};
		\draw [dotted] (1,0) node[anchor=north]{{\footnotesize $\frac 1 3 $}} -- (1,4); 
		\draw [dotted] (5/3,0) node[anchor=north]{{\footnotesize $\frac 1 2 $}} -- (5/3,2); 
		\draw[ultra thick] (1,0) -- (1,1) -- (5/3,2) -- (5/3,4);
		\draw[dotted] (-0.1,1) node[anchor=east]{{\footnotesize $n^{-1/3}$}} -- (1,1);
		\draw[dotted] (-0.1,2) node[anchor=east]{{\footnotesize $\mathcal O(1)$}} -- (5/3,2);
		\draw[dotted] (-0.1,1.5) node[anchor=east]{{\footnotesize $n^{-1+2a}$}} -- (4/3,1.5) -- (4/3,-0.1) node[anchor=north]{$a$};
		\draw (0.9,3) node[fill=white, draw,double,rounded corners]{TW};
		\draw (2.4,1.2) node[fill=white, draw,double,rounded corners]{Non \\ universal};
		\end{scope}
		\begin{scope}[xshift=2.8cm, scale=0.6] 
		\draw[thick, ->] (0,0) node[anchor=east]{$0$} -- (2.8,0) node[anchor=north]{{\footnotesize $a'$}};
		\draw [thick, ->] (0,0) node[anchor=north]{$0$} -- (0,4) node[anchor=north east]{$T$};
		\draw [dotted] (5/3,0) node[anchor=north]{{\footnotesize $\frac 1 3$}} -- (5/3,4); 
		\draw [dotted] (1,0) node[anchor=north]{{\footnotesize $\frac 1 4 $}} -- (1,2); 
		\draw[ultra thick] (5/3,0) -- (5/3,1) -- (1,2) -- (1,4);
		\draw[dotted] (-0.1,1) node[anchor=east]{{\footnotesize $n^{-1/3}$}} -- (5/3,1);
		\draw[dotted] (-0.1,2) node[anchor=east]{{\footnotesize $\mathcal O(1)$}} -- (1,2);
		\draw (0.6,1) node[fill=white, draw,double,rounded corners]{TW};
		\draw (1.9,2.6) node[fill=white, draw,double,rounded corners]{Non \\ universal};
		\end{scope}
		\end{tikzpicture}
	\end{center}
\vspace{-.4cm}
(a) \hspace{4cm} (b) 
	\caption{(a): The two phases depending on ``temperature'' $\temp$ and exponent $a$, for the log-gamma polymer (ii). The crossover arises when zooming at a point on the oblique thick line. (b): The two phases for the polymer (iii) with Boltzmann weights $e^{E_{i,j}/T}$ with exponentially distributed energies $E_{i,j}$ of parameter $\tilde \gamma_{i,j} = (i+j)^{a'}$.}
	\label{fig:crossover}
\end{figure}

(iii) Finally, we perform numerical simulations of a polymer model on the square lattice with exponentially distributed energies with rates $\tilde \gamma_{i,j}=(i+j)^{a'}$, both at zero and finite temperature. In this model the noise is thus purely time-dependent (without additional potential). The results at zero temperature are shown in Fig. \ref{fig:DPTW}. There is a strong evidence that the fluctuations are TW distributed for $a'=0.3 <1/3$ and converge to another limit when $a'=0.4>1/3$, consistent with the same critical value $a'=1/3$ as in the Johansson model. At positive temperature 
the numerics indicate that the transition occurs between $a'=0.2$ and $a'=0.3$. This is in agreement with a general criterium that we obtain for the occurrence of TW fluctuations in inhomogeneous models, which predicts
$a'<1/4$ for this model, while it correctly predicts $a<1/2$ for the log-gamma polymer.
The resulting phase diagram for models (ii) and (iii) is presented in Fig. \ref{fig:crossover}.

%

\begin{figure}[h]
\begin{center}
\begin{tikzpicture}[scale=.8,every text node part/.style={align=center}]
\begin{scope}[scale=0.9]
\begin{scope}[rotate=45]
\begin{scope}
\clip (0,0) rectangle (2.2,2.2);
\shade[shading=radial, inner color=black, outer color=white, fill opacity=1] (0,0) circle(2);
\draw[very thin, gray] (0,0) grid[step=0.1](2.2,2.2);
\end{scope}
\draw[thick, gray, ->] (0,0) -- (3,0) node[below]{$i$};
\draw[thick,gray,  ->] (0,0) -- (0,3) node[below]{$j$};
\end{scope}
\draw[thick,gray,  ->] (-2,0) -- (2,0) node[below]{$\varkappa$};
\draw[thick,gray,  ->] (0,0) -- (0,3.5) node[left]{$\uptau$};
\draw[thick] (0,0) -- ++(135:0.1) -- ++(45:0.3) -- ++(135:1) -- ++(45:1) -- ++(135:.4) -- ++(45:.3) -- ++(135:.3) -- ++(45:.6) node[right]{$(n,m)$};
\end{scope}

\draw[->, ultra thick] (2.5,1.5) -- (3.5,1.5);
\draw (3,1) node{continuum\\  limit};

\begin{scope}[xshift=6cm]
\shade[bottom color=black, top color=white, fill opacity=0.7] (-1.5,0) rectangle(2,1.5);
\draw[thick, gray, ->] (-1.5,0) -- (2.2,0) node[below]{$x$};
\draw[thick, gray, ->] (0,0) -- (0,3) node[left]{$t$};
\draw (0,0) node[below] {$0$};
\draw[thick] (0,0) .. controls  (.3,1) and (0,0)  ..  (-.2,1.5) .. controls (0,2.2) and (0,2)  ..  (.5,2.5) node[right] {$(x,t)$};
\draw[<->, thick, red] (0.2,1.5) -- (-.5,1.5) node[anchor=south] {$t^{\zeta(\alpha)}$};
\end{scope}
\end{tikzpicture}
\end{center}
\caption{Directed polymer paths in the discrete lattice (left panel) and in the continuous limit (right panel). We indicated the wandering exponent in red, that is the lateral extension of typical paths.}
\label{fig:discrete to continuous}
\end{figure}
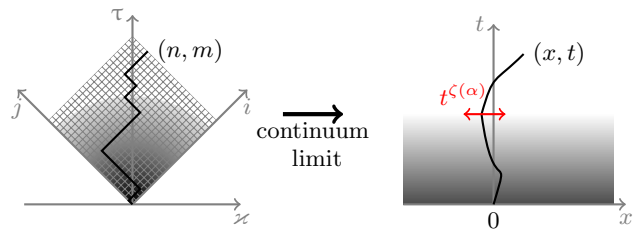

A property of the positive temperature  models (ii) and (iii), not shared by the Johansson model,
is that, in the discrete to continuous scaling limit at high temperature (see Fig. \ref{fig:discrete to continuous}), their partition functions converge to a solution of the inhomogeneous SHE 
\begin{equation}
\partial_{t} Z = \partial_x^2 Z + \big( V(x,t) + \sqrt{2c(t)}\xi(x, t) \big) Z
\label{eq:SHEtdependent}
\end{equation}
so that $h(x,t)=\log Z(x,t)$  solves the KPZ equation \eqref{kpz2}.
In that limit, their transition points $a=1/2, a'=1/4$ correspond to the inhomogeneous KPZ equation \eqref{kpz2} where $c(t)\propto t^{-a}, c(t) \propto  t^{-2a'}$ (i.e. $\alpha=a=2a'$).
Note that in the case of the integrable model (ii), the quadratic term $V(x,t)$, dictated by the form of inhomogeneity parameters $\gamma_{i,j}$ 
that preserve the integrability, turns out to be exactly the one that we found when performing changes of variables directly on the KPZ equation.

{\bf Inhomogeneous KPZ equation}. 
We start with studying the KPZ equation in presence of a time dependent noise
of amplitude $c(t)$, i.e. the equation \eqref{kpz2}.
It is convenient to also include a quadratic
external potential $V(x,t)=a(t) \frac{x^2}{2}$ with a time dependent curvature $a(t)$. The standard KPZ problem is recovered for $a(t)=0$, $c(t)=c(0)$. One can then ask what are the
space and time change of coordinates on the equation \eqref{kpz2} which retain its general form and lead to time independent noise. 
The answer is that the space transformation must be linear and one arrives at
\bea
&& h(x,t) = H(y,\tau(t)) +  \frac{c'(t)}{4 c(t)} x^2 + \frac{1}{2} \log \frac{c(t)}{c(0)}  \label{transf1} \\
&& y = c(t) x  \quad , \quad {\tau(t)= \int_0^t c(s)^2 ds} \label{transf2}
\eea
Under the transformation \eqref{transf1}-\eqref{transf2} the equation \eqref{kpz2} is mapped onto
the following equation for $H(y,\tau)$ 
\bea \label{eqH0} 
\partial_\tau H  = \partial_y^2 H + (\partial_y H)^2 - A(\tau) \frac{y^2}{2}  + \sqrt{2}~ \hat \xi(y,\tau) {, }
\eea
where $\hat \xi$ is again a standard white noise in the coordinates $y,\tau$, and 
\bea \label{A} 
&& {A(\tau(t)):=} \frac{a_c(t)-a(t)}{c(t)^4},  \\
&& a_c(t) := \frac{c(t) c''(t) - 2 c'(t)^2}{2 c(t)^2} = \frac{- c(t)}{2}\left(\frac{1}{c(t)}\right)'' .
\label{eq:defac}
\eea 
Note that similar transformations have been considered for 1D quantum systems 
\cite{Popov1969,Gritsev2010,UsFermionDynamics}
and for the Burgers equation \cite{MoreauVallee2005}. 
Here the mapping works because the white
noise is invariant by linear transformations. 
We assume $0< c(0) < +\infty$.
The correspondence
between the initial conditions at $t=\tau=0$ is then 
\bea \label{init} 
H(y,0) = h(y/c(0),0) - \frac{c'(0)}{4 c(0)^3} y^2 {. }
 \label{ini}
\eea 

Let us first consider the case where the functions $a(t)$ and $c(t)$ are
related by the condition $a(t)=a_c(t)$. In that case $A(\tau)=0$. Hence
the full solution of \eqref{kpz2} 
for $h(x,t)$ is given by \eqref{transf1}-\eqref{transf2}
where $H(y,\tau)$ is the solution of the standard KPZ equation
\eqref{kpz} with initial condition \eqref{ini}. Since a lot is
known {about} the statistics of the standard KPZ equation, 
a wealth of information can thus be obtained for
the case $a(t)=a_c(t)$. {Regarding the large time asymptotics,  there} is clearly a transition
depending on whether $\tau(t)$ diverges or remains
finite when $t \to +\infty$. In the first case large $t$
maps onto large $\tau$ and one can use the universal
results for the KPZ equation at large time (which are common
to the full KPZ class). Then the one 
point height fluctuations grow for large $t$ as $\delta h \propto \tau(t)^{1/3}$
with a $O(1)$ PDF depending on the initial condition, as discussed below. 
The correlation scale grows as 
$x \propto \tau(t)^{2/3}/c(t)$.
In the second case, $\tau(+\infty)<+\infty$, the growth saturates
and is described by finite time KPZ. Some results are available,
but they are not universal (unless $\tau(+\infty) \gg 1$). The mapping \eqref{transf1}-\eqref{transf2}
extends to several space-time points 
correlations.

To be specific consider now a noise amplitude decaying as $c(t) \propto t^{-\alpha}$, 
of the form 
\be \label{calpha} 
c(t) =  \Big(\frac{t_0}{t+t_0}\Big)^{\alpha}
\ee 
with $t_0>0$ a constant. Then the amplitude of the
quadratic potential decays as $t^{-2}$, i.e. 
$a(t) = a_c(t)=\frac{\alpha (1-\alpha)}{2} (t+t_0)^{-2}$.
There is thus one particular case, $\alpha=1$, 
where $a(t)=0$ 
and the present
solution is the full solution of the model \eqref{kpz2}
without external potential $V(x,t)=0$. For $\alpha>1/2$, the rescaled time is
\be \label{tau}
\tau(t) = \frac{t_0}{1-2 \alpha} \Big(\Big(1+ \frac{t}{t_0}\Big)^{1-2\alpha}-1\Big)
\ee  
with $\tau(t)=t_0 t/(t+t_0)$ for $\alpha=1$.
The transition thus occurs at $\alpha=1/2$. For $\alpha<1/2$ the
growth is unbounded, with $\delta h \propto t^{\beta(\alpha)}$ 
and the spatial scale grows as $x \propto t^{\zeta(\alpha)}$
with exponents
\be
\beta(\alpha)=\frac{1-2 \alpha}{3} \quad , \quad \zeta(\alpha) = \frac{2-\alpha}{3}
\ee
At the transition, for $\alpha=1/2$, $\tau(t)=t_0 \log(1+ \frac{t}{t_0})$,
hence $\delta h \propto (\log t)^{1/3}$, and the spatial scale 
is $x \propto t^{1/2} (\log t)^{2/3}$, barely superdiffusive.
For $\alpha>1/2$ we see from \eqref{tau} that $\tau(+\infty)=t_0/(2 \alpha-1)$.
The heuristics is that the KPZ noise/DP disorder acts only for
some finite time $\propto t_0$ and is absent beyond. One finds
that the transverse wandering of 
the polymer is diffusive $x \propto t^{1/2}$. However there is
a distinct and fast growing scale $x \propto 1/c(t) \propto t^{\alpha}$ 
which measures the spatial extent of regions which
are correlated by what happened at the earlier times 
$t<t_0$ (see Fig. \ref{fig:decorrelation} in \cite{SM}).

As is well-known for the standard KPZ equation, the precise distribution of the large time distribution can be classified according to initial condition. We now address this problem in the time dependent case and stress that the mapping  \eqref{transf1}-\eqref{transf2}  may map initial data for $h$ and $H$ from different IC classes. 
In terms of the DP partition functions  
$Z(x,t)=e^{h(x,t)}$, $\hat Z(y,\tau)=e^{\hat H(y,\tau)}$,
the mapping between initial conditions \eqref{init},
with the choice \eqref{calpha}, reads
\be \label{Zinit} 
\hat Z(y,0) = Z(y,0) e^{\frac{\alpha}{4 t_0} y^2}
\ee
The positive sign in the exponential makes the mapping of the IC classes a bit delicate. The
droplet IC for $h$, $Z(x,0)=\delta(x)$, clearly maps to $\hat Z(y,0)=\delta(y)$,
i.e. to the droplet IC for $H$, from \eqref{Zinit}, leading to the GUE-TW 
distribution for the scaled fluctuations of $h(0,t)$. This remains true for initial conditions 
$Z(x,t) = e^{- B z^2}$ with $B >  \frac{\alpha}{4 t_0}$.
Indeed, from \eqref{Zinit}, $\hat Z(y,0)$ decays fast
enough so that $H$ still belongs to the droplet IC class. 
However, for $B= \frac{\alpha}{4 t_0}$, $H$ 
now belongs to the flat IC class, since $\hat Z(y,0)=1$. 
It leads now to the GOE-TW distribution. 
Hence we see that many IC which belong to the droplet class when $c(t)=c(0)$,
such as the wedge $h(x,0)=-w |x|$,
are actually not in that class in the time dependent problem.

When $B< \frac{\alpha}{4 t_0}$ a blow up of the solution can occur.
Let us study here and below the flat IC $h(x,0)=0$, i.e. $B=0$. One finds that
for $0<\alpha<1$ the solution
blows up at finite time $t^*$. It can be estimated as $\tau(t^*) \simeq t_0/\alpha$,
i.e. $t^*=((\frac{1}{\alpha} -1)^{\frac{1}{1-2 \alpha}} -1) t_0$.
For $\alpha \geq 1$ there is no blow up, since $\tau(+\infty) < t_0/\alpha$. 
This change of behavior appears to be related to the sign of 
the quadratic term, $a_c(t)>0$ for $\alpha<1$ and 
$a_c(t)<0$ for $\alpha>1$. Let us focus on the case $\alpha=1$, i.e.
$c(t) \propto 1/t$, where there is no external potential,
$a_c(t)=0$, and the blow-up occurs at infinite time $t^*=+\infty$. 
Since in this case the growth saturates at time scales {$t \propto t_0$},
with $\tau(+\infty)=t_0$, we now study the universal limit in which both $t_0$
and $t$ are large with a fixed ratio $t/t_0$. One finds for the flat IC
\be
h(0,t) \simeq \Big(\frac{t_0 t}{t+t_0}\Big)^{1/3}  \max_{\hat z \in \mathbb{R}} \big\lbrace {\cal A}_2(\hat z)- \tfrac{t_0}{t+t_0} \hat z^2  \big\rbrace. 
\ee
where ${\cal A}_2(\hat z)$ is the so-called Airy$_2$ process
(see e.g. \cite{prolhac2011one, quastel2014airy}).
We can now use the results of 
\cite[Example 1.25]{quastel2019flat} 
and conclude that the 
CDF of the fluctuating part $h(0,t)  \propto (\frac{t_0 t}{t+t_0})^{1/3}  s$, 
is given by the universal ``parabolic IC'' function 
$F^{\beta,\beta}_{\rm parbl}(s)$, with $\beta=-\frac{t}{t+t_0}$.
It interpolates between the GOE-TW ($\frac{t}{t_0} \ll 1$) and Gumbel ($\frac{t}{t_0} \gg 1$) distributions. 
It can also be related to ingrowing circular interfaces 
\cite{KazPC} the blow-up time $\tau=\tau(+\infty)=t_0$ being
the time at which the circular droplet collapses.\\

An important question is how do the models 
with $a(t)=0$ and $a(t)=a_c(t)$ compare, i.e.
how does the presence of the (time dependent)
quadratic potential $\sim x^2/t^2$ changes the results.
We can safely surmise that it does not change the
scaling exponents. However it is probable that it
changes the PDF, as it also has some effect on
the classification of the IC. 

We now study the case \eqref{calpha}, i.e. $c(t) \propto t^{-\alpha}$,
with no external potential $a(t)=0$. From \eqref{A} it maps onto the usual KPZ equation
plus a quadratic potential of curvature $A(\tau)
=  \frac{\alpha (1-\alpha) }{2 (t_0 + (1-2 \alpha) \tau)^2 }$.
For $\alpha<1/2$ is it again a potential of the form 
$y^2/\tau^2$. For $\alpha>1/2$, $A(\tau)$ diverges at $\tau=t_0/(2 \alpha-1)$ 
which corresponds to $t=\infty$. 

Let us study the marginal case $\alpha=1/2$
\bea
c(t) = \frac{1}{\sqrt{1 + \frac{t}{t_0}}}  \quad , \quad \tau(t) = t_0 \log\Big(1 + \frac{t}{t_0}\Big) 
\eea 
Then one has $A(\tau)=A=\frac{1}{8 t_0^2}$, hence the
initial problem maps to a DP in a static $- \frac{1}{2} A y^2$ confining potential. 
Although no exact result is known, heuristics is easy. 

Let us consider the fixed endpoint DP (i.e. droplet IC) and
$t_0 \gg 1$ where universal results can be obtained. 
For $1 < t \ll t_0$, $\tau \simeq t$, the quadratic potential can be neglected and
the fluctuations are the standard TW ones for KPZ. For $t \gg t_0$ 
the quadratic well confines the DP, i.e. the variance of the
endpoint distribution saturates as $\langle y^2 \rangle \propto t_0^{4/3}$,
and segments of length $\tau \propto t_0$ become uncorrelated. 
For the initial model it implies that the variance of the
endpoint distribution behaves as
\be
\langle x^2 \rangle \propto c(t)^{-2} \langle y^2 \rangle
\propto t_0^{4/3} \Big(1 + \frac{t}{t_0}\Big) \simeq  t_0^{1/3} t
\ee 
i.e. diffusion. The free energy fluctuations scale as
\bea {
\delta h \propto t_0^{1/3} \sqrt{ \log\Big(1 + \frac{t}{t_0}\Big) } \propto t_0^{1/3} \sqrt{ \log\Big(\frac{t}{t_0}\Big) }}
\eea 
but now they have a Gaussian distribution. There is thus, for large $t_0$,
a crossover from TW to Gaussian. For $t_0=O(1)$ the above scaling
still holds but the intermediate time
distribution and prefactors are non-universal.

\bigskip 

{\bf Inhomogeneous discrete model}.
We consider now an integrable discretization of the KPZ equation, 
the so-called log-gamma directed polymer
on the square lattice $\mathbb Z_{>0}^2$.
The (point-to-point) partition function of the model is defined by 
\begin{equation}
\mathcal Z(n,m) = \sum_{\pi} \prod_{(i,j)\in \pi} w_{i,j},
\label{eq:defZ}
\end{equation}
where the sum is over all up-right directed paths from $(1,1)$ to $(n,m)$ in the square lattice. The model is integrable \cite{seppalainen2012scaling} when the random weights $w_{i,j}$ are independent and distributed according to the inverse of a gamma random variable with parameter $\gamma$, i.e. its PDF $P(w)$ is $P(w) = \frac{1}{\Gamma(\gamma)} w^{-\gamma-1}\exp(-1/w) $. 
It was noted in \cite{corwin2014tropical} that the model remains exactly solvable when the parameter $\gamma$ depends on the position $i,j$ as $\gamma_{i,j} = \alpha_i+\beta_j$, where $\alpha_i$ and $\beta_j$ are any sequences of real numbers such that the $\gamma_{i,j}$ are positive. In order to emulate the case of a disorder whose amplitude decays with time as a power law, we will consider the case where $\gamma_{i,j} = \temp (i^a+j^a)$ for some parameter $\temp >0$. 

 The Laplace transform of the partition function of a polymer of length $n$ can be written as a Fredholm determinant \cite{borodin2013log, corwin2014tropical}
\begin{equation}
\mathbb E[e^{-u\mathcal Z(n,n)}] = \det(I+K)_{\mathbb{L}^2(\mathcal C)}, 
\end{equation}
where the operator $K$ 
is defined by its integral kernel as 
\begin{equation}
K(v,v') = \int_{-\I\infty}^{\I\infty} \frac{\mathrm{d}z}{2\I\pi}\frac{\pi}{\sin(\pi(v-z))}\frac{ 1}{z-v'}\frac{e^{G(z)}}{e^{G(v)}},
\label{eq:kernelavecfonctionG}
\end{equation}
with 
\begin{equation}
G(z) = z\log(u) + \sum_{i=1}^n \log \frac{\Gamma(\temp  i^a-z)}{ \Gamma(\temp i^a+z)}.
\label{eq:defG}
\end{equation}
The kernel is acting on a contour $\mathcal C$ of the complex plane enclosing all singularities of the kernel at the points $-\temp i^a$ for all $i\geqslant 1$. We now analyze the large $n$ asymptotics of
\eqref{eq:kernelavecfonctionG} using a saddle point method.
It is easy to notice that $G''(0)=0$, so that by Taylor expansion,
\begin{equation} \label{19} 
G(z) = z\log(u) + zf_n/\temp + \sigma_n^3 z^3/(3\temp^3) + O(z^5)
\end{equation}
where 
 $f_n= -2 \temp \sum_{i=1}^n \psi(\temp  i^a)$,
$\psi(x)=\tfrac{d}{dx}\log(\Gamma(x))$, and 
\begin{equation}
\sigma_n^3 = \sum_{i=1}^n -\temp^3 \psi''(\temp i^a). 
\label{eq:sigman}
\end{equation}
The quantity $f_n$ is the leading order of (minus) the free energy $\mathcal F_n = \temp\log \mathcal Z(n,n)$, and  $\sigma_n$ should be understood as the amplitude of free energy fluctuations. 
The asymptotic behaviour of $\mathcal F_n$ will depend on whether 
$\sigma_n$ 
stays bounded or diverges as $n\to\infty$. In the zero temperature limit $\temp\to 0$, the threshold found in \cite{johansson2008some} was for $a=1/3$. Interestingly, the result is different for positive $\temp$.  For $\temp >0$, a careful analysis of \eqref{eq:sigman} shows that it diverges for $a\leqslant 1/2$ and converges for $a>1/2$ (this is due to the fact that $ \psi''(x)\simeq -1/x^2$ as $x\to+\infty$).

When $a>1/2$, 
the weights decay rapidly away from the origin. The parameter $\sigma_n$ converges to a constant, i.e.
the 
weights which contribute 
significantly to free energy fluctuations is a finite set 
near the origin. The one point distribution of those fluctuations can be computed explicitly (see \cite{SM} Eq. (136)).

When $a\leqslant 1/2$, the magnitude of free energy fluctuations diverge as 
\begin{equation}
\sigma_n \simeq \begin{cases} \left(\temp  n^{1-2a}/(1-2a)\right)^{1/3} &\text{ for }a<1/2,\\   \left(\temp \log n\right)^{1/3} &\text{ for }a=1/2.\end{cases} 
\end{equation}
To analyze the limit distribution, let us choose $u  =  e^{-f_n- r \sigma_n} $. Since $\sigma_n \to +\infty$ at large $n$ one has
\begin{equation}
\mathbb E\left[e^{-u \mathcal Z(n,n)}\right] \simeq \mathbb P\left( \frac{\mathcal F_n -f_n }{\sigma_n} \leqslant r \right).
\end{equation} 
The Fredholm determinant $\det(I+K)_{\mathbb L^2(\mathcal C)}$ converges to $\det(I+K^{GUE})_{\mathbb L^2(-1+\I\mathbb R)}$
where
\begin{equation}
 K^{\rm GUE}(v,v')  =  \int_{1+\I\mathbb R} \frac{\mathrm{d}z}{2\I\pi}
\frac{1}{v-z} \frac{1}{z-v'}\frac{e^{\frac{z^3}{3} - r z}}{e^{\frac{v^3}{3} - r v}},
\label{eq:kernelGUEletter}
\end{equation} 
This comes from \eqref{19} upon rescaling $z$ by $\temp/\sigma_n$,
implying
\begin{equation}
 \lim_{n\to \infty} \mathbb P\left( \frac{ \mathcal F_n -f_n }{\sigma_n} \leqslant r \right) = F_{\rm GUE}(r),
 \label{eq:cvtoTW}
\end{equation}
where $F_{\rm GUE}$ is the CDF of the TW GUE distribution. 

When the parameter $\temp $ goes to $0$, one recovers the zero-temperature model studied by Johannsson \cite{johansson2008some} (in the sense that $\mathcal F_n$ goes to the zero-temperature free energy defined in \cite{johansson2008some}). Using the asymptotics 
$\psi''(x) \underset{x\to 0^+}{\simeq} \frac{-1}{x^3}$,
one readily sees, taking $\temp \to 0$ in \eqref{eq:sigman}, that 
\begin{equation}
 \sigma_n^3 \simeq \sum_{i=1}^n \frac{1}{i^{3a}},\quad n\to\infty, \quad \temp \ll n^{-a},
\end{equation}
hence the transition at $a=1/3$ at zero-temperature. Considering the 
limit $\temp \to 0$ of \eqref{eq:kernelavecfonctionG} 
an asymptotic analysis recovers 
the one-point distribution 
results from \cite{johansson2008some}.

We may also let $\temp $ go to zero simultaneously as $n$ goes to infinity and study the crossover between zero and finite temperature. Let $a\in (1/3,1/2)$.  The relevant scale to see a crossover is $\temp =A n^{2a-1}$, where
$A$ is a free parameter. For higher or lower values of $\temp$, the free energy fluctuations will be either of TW type or non-universal according to the two phases depicted in Fig. \ref{fig:crossover}. 

At the crossover scale, we obtain
\bea
&& \lim_{n\to \infty} \mathbb P\left( \mathcal F_n - f_n \leqslant  s \right)  = \det(I+K^{\rm cross}) \\
&& K^{\rm cross}(v,v')  = \int_{-\I\infty}^{\I\infty} \frac{\mathrm{d}z}{2\I\pi} \frac{1}{v-z}\frac{1}{z-v'} \frac{e^{F^{\rm cross}(z)-sz}}{e^{F^{\rm cross}(v)-sv}}, \nonumber
\eea
The function $F_{\rm cross}$ interpolates between the zero temperature case  and  a cubic behaviour as in the Airy kernel. It depends on $A$ as
\begin{align}
F_{\rm cross}(z)  &= \frac{A z^3}{3(1-2a)} + F_{\temp\to 0}(z),\\
F_{\temp\to 0}(z) &=  \sum_{k=1}^{\infty} \log \left(1+\frac{z}{k^a} \right) - \log \left(1-\frac{z}{k^a} \right) -\frac{2z}{k^a} 
\end{align} 
where $ F_{\temp\to 0}$ is the function arising in the zero-temperature kernel as in  \cite{johansson2008some}. 
\bigskip 

{\bf From discrete to continuous}. It was shown in 
\cite{alberts2014intermediate} that the free energy of the (homogeneous) log-gamma polymer model converges to the solution to the KPZ equation.  We use the convenient new coordinates $\uptau=n+m$, $\varkappa=n-m$, and denote  $\mathcal Z(n,m) = Z_d(\varkappa,\uptau)$ \cite{footnote3}. The subscript $d$ means that $Z_d$ satisfies a discrete version of the stochastic heat equation
\begin{equation}
Z_d(\varkappa,\uptau) = w_{\varkappa,\uptau} (Z_d(\varkappa-1,\uptau-1)+Z_d(\varkappa+1,\uptau-1)),
\label{eq:discreterecurrence}
\end{equation}
where $w_{\varkappa,\uptau}$ is an inverse gamma random variable with parameter $\gamma_{\varkappa,\uptau}$ (independent for each $\varkappa,\uptau$).  
Let us rescale $Z_d$ and denote 
$Z_r(\varkappa, \uptau) = Z_d(\varkappa, \uptau) \left(\prod_{s=1}^{\uptau}C_{s}\right)^{-1}$.
A natural choice for the function $C_{\uptau}$ would be $\mathbb E [w_{\varkappa, \uptau}]$ (when it does not depend on $\varkappa$). We may rewrite \eqref{eq:discreterecurrence} as 
\begin{equation} 
\nabla_{\uptau} Z_r(\varkappa, \uptau) = \tfrac{1+\eta_{\varkappa, \uptau}}{2} \Delta_\varkappa Z_r(\varkappa, \uptau-1) +\eta_{\varkappa, \uptau} Z_r(\varkappa, \uptau-1),
\label{eq:discreteSHE}
\end{equation}
where $\eta_{\varkappa, \uptau}= \frac{2 w_{\varkappa,\uptau}}{C_{\uptau}}-1 $, $\nabla_{\uptau}$ is
the discrete time derivative and $ \Delta_\varkappa$ is the discrete Laplacian. Let us use the scalings 
\begin{equation}
\uptau=2 n t, \quad \varkappa=\sqrt{n}x.
\label{eq:scalingsloggamma}
\end{equation}
In order to obtain a time-inhomogeneous variance of the noise, let us  scale the parameter $\gamma$ as $\gamma_{\varkappa, \uptau}=\sqrt{n}/ c(t)$.  In this case, one takes $C_{\uptau} = 2 \mathbb E\, [w_{\varkappa, \uptau}]$ and the family of random variables $w_{\varkappa,\uptau}$ rescales to a white noise in the sense that $n\, \eta_{\varkappa, \uptau} = \sqrt{c(t)/2} \xi(x,t)$. Multiplying Eq. \eqref{eq:discreteSHE} by $n$ and taking the continuum limit, we obtain the
SHE \eqref{eq:SHEtdependent} with $V(x,t)=0$. 

However, in the inhomogeneous log-gamma polymer, one cannot exactly take inhomogeneity parameters depending only on $\uptau$. Recall that $\gamma_{\varkappa,\uptau}=\alpha_i+\beta_j$ where $\uptau=i+j$ and $\varkappa=i-j$. Let us consider the case where 
\begin{equation}
\gamma_{\varkappa, \uptau} = \frac{\sqrt{n}}{2 c(i/n)} + \frac{\sqrt{n}}{2 c(j/n)}. 
\end{equation}
We set now $C_{\uptau}=\frac{2}{\sqrt{n}/c(t)-1}\neq 2 \mathbb E w_{\varkappa, \uptau}$, and the noise converges to a white noise with an extra potential $a_c(t)x^2/2$. In the continuum limit, we obtain 
the
SHE \eqref{eq:SHEtdependent} with $V(x,t)=a_c(t) \frac{x^2}{2}$.
In particular, choosing $\gamma$ as
\begin{eqnarray}
\text{Model I: }&\gamma_{i,j} =& n^{\frac{1}{2}-a} \left(\frac{i+j}{2}+t_0n \right)^a  \label{eq:model1}, \\
\text{Model II: } &\gamma_{i,j} =& \frac{1}{2} n^{\frac{1}{2}-a} \big((i+t_0n)^a + (j+t_0n)^a \big),\nonumber 
\end{eqnarray}
one obtains for large $n$ the continuous SHE \eqref{eq:SHEtdependent} 
for $c(t)=(t+t_0)^{-\alpha}$, $\alpha=a$. In Model I $a(t)=0$, in Model II  $a(t)=a_c(t)$. The latter, when $a=1/2$ and $t_0=0$ corresponds to the discrete model 
analyzed above.

Further, the criterium $\int_0^{+\infty} dt ~ c(t)^2 = \infty$ that we found for TW fluctuations in the continuous model \eqref{kpz2} becomes  equivalent to the criterium $\lim_{n \to + \infty} \sigma_n = \infty$ that we have used in the study of the discrete model. In addition, the two critical models $\alpha=\frac{1}{2}$ and
$a=\frac{1}{2}$ match in the double limit $n\to\infty$, $t_0\to 0$.

We now discuss when a time-inhomogeneous discrete model at finite temperature leads to TW fluctuations. We find that these arise if and only if the sum along the polymer of $(\mathrm{Var} \log w)^2$ (i.e. the discrete analogue of $\int_{0}^t c^2(s)ds$) diverges as the length of the polymer goes to infinity \cite{SM}.
This criterium predicts a transition at $a'=1/4$ for the model
with exponentially distributed energies $E_{ij}$ and rates $\tilde \gamma_{i,j}=(i+j)^{a'}$ (with Boltzmann weights $w_{i,j}=e^{E_{i,j}/T}$), also supported by our numerics \cite{SM}. It 
predicts a transition at $a=1/2$ for the log-gamma polymer model with $\gamma_{ij}=(i+j)^a$. 
While both models are identical at zero temperature with $a=a'$, their critical values at finite temperature are distinct. Indeed, the log-gamma distribution of Boltzmann weights induces a 
 temperature-dependence on the distribution of energies.

%

\bigskip 
{\bf Linear potential} Finally, the KPZ equation \eqref{kpz2} with $c(t)=1$ and a linear potential 
$V(x,t)=b x$ is solved as
\be 
 h(x,t) = H(x + b t^2,t) + b x t  + \frac{1}{3} b^2 t^3 
\ee 
where $H(y,t)$ is the solution of the standard KPZ equation with the same IC $H(x,0)=h(x,0)$. For the droplet IC, the one point
height distribution 
is $h(x,t) \equiv - \frac{(x-b t^2)^2}{4 t} + \frac{1}{3} b^2 t^3 + H_{\rm droplet}$
where $H_{\rm droplet}$ is the one-point droplet KPZ height, and the
profile has a maximum at $x \simeq b t^2$. This holds for sufficiently
localized IC.

\bigskip 
\textbf{Outlook:} Using three complementary methods, we have obtained results for growth in presence of time dependent noise and investigated how KPZ universality extends to this setting. 
This may be of interest for experiments where the variance of the noise can be controlled, see e.g.  \cite{takeuchi2012evidence}.

\bigskip

\textbf{Acknowledgements:} We thank Thimoth\'ee Thiery for initiating this project and sharing his ideas with us and M.-T. Commault for the help in the data analysis. 
We acknowledge support from ANR grant ANR-17-CE30-0027-01 RaMaTraF.

\let\oldaddcontentsline\addcontentsline
\renewcommand{\addcontentsline}[3]{}

{}

\let\addcontentsline\oldaddcontentsline

\begin{widetext} 
\newpage 
\bigskip

\bigskip

\begin{large}
\begin{center}

SUPPLEMENTARY MATERIAL \\
to ``Stochastic growth in time dependent environments''
\setcounter{page}{1}

\end{center}
\end{large}

\bigskip

We give here the details of the calculations and their applications, as
described in the main text of the Letter.

\bigskip

\tableofcontents

\section{I Spatially linear change of variables on the KPZ equation}

\subsection{1) General time-inhomogeneous KPZ equation}

Let us consider the general time inhomogeneous KPZ equation
\be \label{kpz3}
\partial_t h = \nu(t) \partial_x^2 h + \frac{\lambda(t)}{2} (\partial_x h)^2 + V(x,t) + \sqrt{2 c(t)} ~ \xi(x,t) 
\ee
in presence of an external potential $V(x,t)$. The change of variable method considered here
works only for the case $\lambda(t) \propto \nu(t)$, which we will assume from now on (see \cite{footnote9} for a study of the general $\lambda(t)$). 
Since an additional rescaling by space-time independent coefficients $x,t,h \to x/x^*,t/t^*,h/h^*$ 
is always possible, akin to a choice of units, we assume from now 
$\lambda(t) = 2 \nu(t)$. If this condition is violated, a new term $\propto h$ appears
in the transformed equation, not studied here.

To treat also cases including linear potentials, we consider the following change of variable
from $h(x,t)$ to $H(y,\tau)$
\be \label{rel1} 
h(x,t) = H(c(t) x + y_0(t),\tau(t)) + \frac{c'(t)}{4 c(t) \nu(t)}  x^2 + \frac{y_0'(t)}{2 c(t) \nu(t)} x +
\frac{1}{2} \log \frac{c(t)}{c(0)} + J(t)  \quad , \quad \tau(t)= \int_0^t ds \, c(s)^2 \nu(s)  
\ee 
Then we find that if $h(x,t)$ satisfies the time inhomogeneous equation \eqref{kpz3} with white noise,
then $H(y,\tau)$ satisfies the time-homogeneous equation
\be \label{EqH} 
\partial_\tau H = \partial_y^2 H + (\partial_y H)^2 + W(y,\tau) + \sqrt{2} ~ \tilde  \xi(y,\tau) 
\ee
in the external potential
\bea
&&  W(y,\tau) = \frac{V(x=\frac{y-y_0(t)}{c(t)},t)}{c(t)^2 \nu(t)} 
- A(\tau) \frac{y^2}{2} - B(\tau) y  + W_0(\tau) \\
&& A(\tau(t)) = \frac{1}{2 c(t)^6 \nu(t)^2} 
\left(c(t) c''(t) - 2 c'(t)^2 -  c(t) c'(t) 
\frac{\nu'(t)}{\nu(t)}\right) = \frac{- 1}{2 c(t)^3 \nu(t)} \frac{d}{dt} \left(\frac{1}{\nu(t)} \frac{d}{dt} \frac{1}{c(t)}\right)
\\
&& B(\tau(t)) = \frac{1}{2 c(t)^3 \nu(t)}  \frac{d}{dt} \left(\frac{1}{\nu(t)} \frac{d}{dt} \frac{y_0(t)}{c(t)} \right)
\eea 
The coefficients in \eqref{rel1} have been determined so that no term 
linear in $\partial_y H$ appears in the equation \eqref{EqH}. In \eqref{rel1} the last term
reads
\bea
J(t) = \int_0^t ds \bigg[ \frac{1}{4 c(s)} \frac{d}{ds} \left(\frac{1}{\nu(s)} \frac{d}{ds} \frac{y_0(s)^2}{c(s)} \right) 
- \frac{y_0'(s)^2}{4 \nu(s) c(s)^2} - c(s)^2 \nu(s) W_0(\tau(s)) \bigg]
\eea 
Note that the function $W_0$ can be chosen arbitrarily, for convenience. 

A case of particular interest is when the initial equation \eqref{kpz3} 
contains no external potential, i.e. $V(x,t)=0$. Then, in the subcase such that 
\be
c(t) = \frac{c(0)}{1 + \int_0^t \nu(t') dt'} 
\ee 
the transform \eqref{rel1} with the choice $y_0(t)=0$, $W_0(\tau)=0$,
maps the problem to the standard KPZ equation with $W(y,\tau)=0$. 

In the case $\nu(t)=1$ and $V(x,t) = a(t) \frac{x^2}{2}$, choosing $y_0(t)=0$ and $W_0(\tau)=0$, one recovers the formula 
\eqref{transf1}, \eqref{transf2} and \eqref{eq:defac} given in the text. \\

{\bf KPZ equation in presence of a linear potential}. Consider now the usual KPZ equation in presence of a linear time-dependent potential
$V(x,t)= b(t) x$ 
\bea
\partial_t h(x,t) = \partial_x^2 h(x,t) + (\partial_x h(x,t))^2 + b(t) x + \sqrt{2} \xi(x,t) 
\label{eq:KPZlinearpotential}
\eea
With $\nu(t)=1$, $c(t)=1$, choosing 
\be
y_0(t) = 2 \int_0^t ds \int_0^s du \, b(u) 
\ee 
and $W_0(\tau(t))=b(t) y_0(t)$, it is mapped under the shift
\be \label{rel1badname} 
h(x,t) = H(x + y_0(t),t) + \frac{y_0'(t)}{2} x + \frac{1}{4} \int_0^t ds \, y_0'(s)^2  
\ee 
to the standard KPZ equation for $H(y,\tau)$ without external potential $W(y,\tau)=0$
and the same initial condition $H(y,\tau=0)=h(y,t=0)$ (since $y_0(0)=y_0'(0)=0$).

\subsection{2) Mapping and solution in the absence of noise: time-dependent harmonic oscillator and blow up}

In the absence of noise, the equation $\partial_t h = \partial_x^2 h + (\partial_x^2 h)^2 +
a(t) \frac{x^2}{2}$ (related to the quantum time-dependent harmonic oscillator -- in imaginary time) 
can be solved by the rescaling method \cite{Popov1969} as $h(x,t)=H(\frac{x}{L(t)},\tau(t)) - \frac{L'(t)}{4 L(t)} x^2 - \frac{1}{2} \log (L(t)/L(0))$, where
$\partial_\tau H = \partial_y^2 H + (\partial_y^2 H)^2 - \frac{A}{2} y^2$, and $L(t)$ satisfies the
Ermakov equation
\cite{Ermakov,PainleveErmakov,ErmakovReview,ErmakovCosmology}, i.e. $L''(t) + 2 a(t) L(t) + \frac{2 A}{L(t)^3}=0$ and $\tau'(t)=1/L(t)^2$. Here $A$
is an arbitrary constant. Since it is a second order differential equation, for any given $a(t)$ there is in addition a two-parameter family of solutions $L(t)$ indexed by $(L(0),L'(0))$. The solution for $h(x,t)$ should be invariant under
the possible choices of these three parameters, provided the initial condition is modified correspondingly.

For instance, choosing $A=0$ and using the solution of the standard heat equation, one obtains $h(x,t)$ as
\be \label{solua} 
e^{h(x,t)} = \sqrt{\frac{L(0)}{L(t)}} 
\frac{e^{- \frac{L'(t)}{4 L(t)} x^2}}{\sqrt{ 4 \pi \int_0^t \frac{du}{L(u)^2}}} 
 \int_{- \infty}^{+\infty} dy \, 
e^{ - \frac{\left(\frac{x}{L(t)} - y\right)^2}{4 \int_0^t \frac{du}{L(u)^2}} + h(L(0) y,0) + \frac{L'(0) L(0)}{4} y^2 }
\ee 
where $L(t)$ satisfies $L''(t)=-2 a(t) L(t)$. Let us check that \eqref{solua} is indeed independent of the choice $(L(0),L'(0))$, which is not immediately obvious. First note that the r.h.s. of \eqref{solua}
is invariant by the rescaling $L(t) \to \lambda L(t)$, $y \to y/\lambda$, hence one can always choose $L(0)=1$, 
which we do from now on (we will not consider the case $L(0)=0$).
To see that \eqref{solua} does not depend on the choice of $L'(0)$, let us consider the Wronskian of two 
solutions $L_1(t)$, $L_2(t)$ (with $L_1(0)=L_2(0)=1$) for the same $a(t)$. One has $L'_1(t) L_2(t)- L'_2(t) L_1(t)=L_1'(0) - L_2'(0)$. One can solve this equation for $L_2(t)$ as a function of $L_1(t)$ and one obtains
\be \label{L2} 
L_2(t) = L_1(t) (1 + (L_2'(0) - L_1'(0)) \int_0^t \frac{du}{L_1(u)^2} )
\ee 
Solving instead for $L_1(t)$ as a function of $L_2(t)$ leads to the same equation with $L_1$ and $L_2$ exchanged.
Combining both equations we obtain
\be \label{eqnew} 
L_1(t) \int_0^t \frac{du}{L_1(u)^2} = L_2(t) \int_0^t \frac{du}{L_2(u)^2} = \frac{L_2(t)-L_1(t)}{L_2'(0) - L_1'(0)} 
\ee 
Hence the pre-exponential factor in \eqref{solua}, as well as the term proportional to $x y$ in the
exponential, take the same value for both solutions. Now, \eqref{eqnew} implies
\be
L_2'(0) - L_1'(0) = \frac{L_2(t)-L_1(t)}{L_1(t) \int_0^t \frac{du}{L_1(u)^2}} 
= \frac{L_2(t)}{L_1(t) \int_0^t \frac{du}{L_1(u)^2} } - \frac{1}{\int_0^t \frac{du}{L_1(u)^2} }
= \frac{1}{\int_0^t \frac{du}{L_2(u)^2} }- \frac{1}{\int_0^t \frac{du}{L_1(u)^2} }
\ee 
hence
\be \label{indep} 
L_2'(0) - \frac{1}{\int_0^t \frac{du}{L_2(u)^2} } = L_1'(0) - \frac{1}{\int_0^t \frac{du}{L_1(u)^2}}
\ee 
which is precisely the coefficient of $y^2/4$ in the exponential in \eqref{solua}. Finally, dividing the
Wronskian by $L_1(t) L_2(t)$ one obtains, using \eqref{eqnew}
\be
\frac{L'_1(t)}{L_1(t)} - \frac{L'_2(t)}{L_2(t)} = \frac{L'_1(0)-L'_2(0)}{L_1(t) L_2(t)} = \frac{L_1(t)-L_2(t)}{L_1(t) L_2(t)
\int_0^t \frac{du}{L_1(u)^2}} = \frac{1}{L_2(t) L_1(t) \int_0^t \frac{du}{L_1(u)^2}} - \frac{1}{L_1(t)^2 \int_0^t \frac{du}{L_1(u)^2}}
\ee 
Using \eqref{eqnew} once more we finally obtain 
\be
\frac{L'_1(t)}{L_1(t) } + \frac{1}{L_1(t)^2 \int_0^t \frac{du}{L_1(u)^2}} =
\frac{L'_2(t)}{L_2(t) } + \frac{1}{L_2(t)^2 \int_0^t \frac{du}{L_2(u)^2}} 
\ee  
which is the coefficient of $- x^2/4$ in the exponential in \eqref{solua}. Hence all combinations of $L(t)$ appearing in \eqref{solua} are indeed independent of the choice of $L'(0)$, via some elementary identities.
 \bigskip 
 
{\bf Blow-up}. For each choice of $a(t)$ there is a class of initial conditions which lead to blow-up.
Blow-up occurs when the integral over $y$ in \eqref{solua} diverges. For simplicity let us consider IC of the type $h(x,t=0)=- B x^2$, $B=0$ being the flat IC. For each $a(t)$ there is a $B_c$ such that blow-up occurs for
$B<B_c$ and no blow-up for $B \geq B_c$. For the standard noiseless KPZ equation $a(t)=0$, a blow-up occurs when $B<0$ and no blow up for $B \geq 0$ (as seen choosing $L(t)=1$ in \eqref{solua}). Hence $B_c=0$ in that case.
More generally the condition for absence of blow up is that the coefficient of $y^2/4$ in \eqref{solua} remains negative, i.e. $L'(0) - \frac{1}{\int_0^t \frac{du}{L(u)^2}} - 4 B <0$ for all $t$. It this quantity changes sign for the first time at some $t^*$, there is a blow-up at $t=t^*$ which satisfies
\be \label{tstar} 
\tau(t^*) =  \int_0^{t^*} \frac{du}{L(u)^2} = \frac{1}{L'(0) - B}
\ee
leading to a blow-up of the solution $h(x,t)$ towards $+\infty$ at $t=t^*$. From \eqref{indep} we see that $t^*$ is
independent of the choice of $L'(0)$, as expected, since the blow-up is an intrinsic property which depends only on the choice of $a(t)$ and $B$. The critical value is thus given $4 B_c = \max_t [ L'(0) - \frac{1}{\int_0^t \frac{du}{L(u)^2}} ]$.

Let us give an example. Consider $L_1(t)=\sqrt{1+t}$ and flat IC $B=0$. It corresponds 
to $a(t)=- \frac{L_1''(t)}{2 L_1(t)} 
= \frac{1}{8(1+t)^2}$. The coefficient of $y^2/4$ in the exponential in \eqref{solua} is given by \eqref{indep} and
equal to $\frac{1}{2} - (1/\log(1+t))$. There is thus a blow-up of $h(x,t)$ at time $t^*=e^2-1=6.389..$ where
this coefficient vanishes. One could naively argue that the blow-up arises from choosing a function
$L(t)$ such that $L'(0)>0$, which gives a positive contribution to the coefficient of $y^2/4$. Choosing
$L'(0)=0$ would naively seem as a way to push the root for $t^*$ in \eqref{tstar} to infinity. This is not the case however.
Indeed let us make the equivalent choice $L_2(t)$ such that $L_2'(0)=0$. One finds
from \eqref{L2}, that it is given by $L_2(t)=\sqrt{1+t} (1- \frac{1}{2} \log(1+t))$. The associated 
$a(t)= \frac{1}{8(1+t)^2}$ is the same as for $L_1(t)$, and the blowup still occurs at the same
$t^*$ but it is because $L_2(t)$ changes sign at $t=t^*$, and the integral on the l.h.s. of \eqref{tstar}
diverges at $t=t^*$. 

It is interesting to ask which choices of $a(t)$ and $B$ lead to blow-up and which do not. The equation
$L''(t) + 2 a(t) L(t)=0$ can be interpreted as a Schrodinger eigenvalue equation for
a wave-function $\psi(t) \equiv L(t)$ in one space dimension, where the space variable is $t$,
i.e. $- \psi''(t) + V(t) \psi(t)=E \psi(t)$. One can choose the potential $V(t)=-2 a(t)$ and $L(t)$ corresponds to 
any zero energy solution at $E=0$. Let us consider the flat IC, $B=0$, in which case it is
natural to choose $L'(0)=0$ (i.e. $\psi'(0)=0$) as the boundary condition for the Schrodinger operator on $t \in [0,+\infty[$. Generally, let us denote $E_0$ the lowest energy of the
spectrum of the quantum potential $V(t)$ with this boundary condition. If $E_0 <0$ one expects that the solutions for $L(t)$ at zero energy $E=0$ will oscillate and change sign, leading to a blow-up (see the example of the previous paragraph). At the contrary, if $E_0>0$ one expects exponentially decaying or exploding solutions for $L(t)$, with no sign change, hence no blow-up. 

Conversely one can pick any explicit function $L(t)$ and obtain a function $a(t)$ for which the solution $h(x,t)$ can be written explicitely. A case of interest in this paper is $L(t)=(1+t)^{\alpha}$ in which case 
$a(t)=\frac{\alpha(1-\alpha)}{2 (1+t)^2}$. For flat IC, $B=0$, there is a blowup for $\alpha<1$ at time $t^*$ given by
$t^* = (\frac{1-\alpha}{\alpha})^{\frac{1}{1-2\alpha}} -1$. For $\alpha \geq 1$ there is no blow up.
This can be understood since $a(t)>0$ for $\alpha<1$, while $a(t) <0$ for $\alpha>1$, hence the
potential $V(t)=-2 a(t)<0$ for $\alpha<1$ with $E_0<0$ and positive for $\alpha>1$ with $E_0>0$.

It is interesting to consider the case $L(t)=(1+ t^2)^{\alpha/2}$, which has the same large time behavior as the previous example, but has $L'(0)=0$. It corresponds to 
$a(t)= -\frac{\alpha \left(1- (1-\alpha) t^2\right)}{2 \left(t^2+1\right)^2}$. It is easy to see from the condition
\eqref{tstar} that for flat IC $B=0$ there is never a blow-up for any $\alpha >0$. The function $a(t)$ starts negative at small times with  
$a(0)=-\alpha/2$, hence $V(0)=\alpha>0$, 
which seems to be sufficient to avoid the blow up (and leads to $E_0>0$).

\bigskip 

{\bf Connection to the noisy KPZ equation}. The main point of the present paper is to note that the above mapping
for the noiseless KPZ equation, which we can denote
$(a(t),0) \to (-A,0)$, extends in presence of white noise, to the mapping
$(a(t),c(t)) \to (-A,1)$, if one chooses $c(t)=1/L(t)$ (where $L(t)$ is associated to $a(t)$ as described above). One can check that indeed Eqs. \eqref{eq:defac} and
\eqref{A} are equivalent in that sense to Ermakov's equation. Although the noise generates fluctuations in $h(x,t)$, the question of the blowup for a given $(a(t),c(t))$ model can be discussed already in the absence of noise, as we have shown (see also below).

It is interesting to note that the solution of the noiseless KPZ problem, i.e. $(a(t),0)$, with the droplet initial condition $e^{h(y,0)} \to \delta(y)$ is a simple gaussian
\be \label{noiseless} 
e^{h(x,t)} = \frac{\sqrt{c(t)}}{\sqrt{ 4 \pi c(0) \tau(t)} }
e^{- \frac{x^2}{4} [ \frac{c(t)^2}{\tau(t)}  - \frac{c'(t)}{c(t)}  ] } \Big\vert_{c(t)=1/L(t)}
\ee 
This solution will provide the ``mean profile'' for the droplet IC in presence of noise. 
Indeed consider Eq. (4) of the text for the problem $(a(t)=a_c(t),c(t))$. Since $A(\tau)=0$ 
in (6), we know that one has equivalence in one-point PDF law
$H(y,\tau) \equiv H(0,\tau) - \frac{y^2}{4 \tau}$. This implies the equality in law
\be \label{sts2} 
h(x,t) \equiv h(0,t) - \frac{x^2}{4} \left[ \frac{c(t)^2}{\tau(t)}  - \frac{c'(t)}{c(t)}  \right] 
\ee
This can be obtained, more generally, by studying the symmetries of the inhomogeneous KPZ equation as we now discuss

\subsection{3) Statistical tilt symmetry (STS), Galilean invariance}

Consider the case $\nu(t)=1$, i.e. the model $(a(t),b(t),c(t))$
\bea
\partial_t h(x,t) = \partial_x^2 h(x,t) + (\partial_x h(x,t))^2 + \frac{1}{2} a(t) x^2 + b(t) x + \sqrt{2 c(t)} \xi(x,t) 
\label{eq:KPZ44}
\eea
Let us define $\tilde h(y,t)$ via the relation $h(x,t)= \tilde h(x + f(t),t)  + \frac{f'(t)}{2} x + F(t)$. If we choose
$f(t)$ and $F'(t)$ such that
\be \label{cond1} 
f''(t) + 2 a(t) f(t) = 0 \quad , \quad F'(t) = \frac{1}{4} f'(t)^2 - \frac{1}{2} a(t) f(t)^2 - b(t) f(t) 
\ee
then $\tilde h(y,t)$ satisfies exactly the same equation \eqref{eq:KPZ44} with $x \to y$, and 
$\sqrt{2 c(t)} \xi(x,t) \to \sqrt{2 c(t)} \tilde \xi(y,t)$, with $\tilde \xi$ an equivalent white noise in $y$.
If we further choose $f(0)=F(0)=0$, and the droplet IC for $h$, i.e. $e^{h(x,0)}=\delta(x)$, this also corresponds
also to droplet IC for $\tilde h$, $e^{\tilde h(y,0)}=\delta(y)$. 
Hence the two fields $h(x,t)$ and $\tilde h(x,t)$ 
have the same statistics. Note that this is independent of the choice of $c(t)$.

In the case $a(t)=b(t)=0$, $f(t)= v t$, $F(t)= \frac{v^2}{4} t$, and this is the usual Galilean
invariance of the KPZ equation (and its derivative, Burger's equation). It states that for droplet IC
\be
h(x,t) \equiv h(x + v t ,t)  + \frac{v}{2} x + \frac{v^2}{4} 
\ee 
where $\equiv$ means the same statistics. For a fixed $x,t$, choosing $v=-x/t$ we obtain the celebrated identity of the one point PDF's, which we note here extends to any choice of $c(t)$
\be \label{sts0} 
h(x,t) \equiv h(0,t) - \frac{x^2}{4 t} 
\ee 

Let us take $b(t)=0$. Then from \eqref{cond1} one finds $F(t) =\frac{1}{4} f(t) f'(t)$. We can write the solution
$f(t)=v f_0(t)$ where $f_0'(0)=1$. Then we can choose $v=-x/f_0(t)$ and we obtain the general STS relation (equivalence in law) 
\be \label{equiv2} 
h(x,t) \equiv  h(x + v f_0(t),t)  + v \frac{f_0'(t)}{2} x + \frac{v^2}{4} f_0(t) f_0'(t) = 
h(0 ,t)   - \frac{x^2}{4} \frac{f_0'(t)}{f_0(t)}
\ee 
where we recall that $f_0(t)$ is the unique solution of $f_0''(t) + 2 a(t) f_0(t) = 0$ 
with $f_0(0)=0$ and $f_0'(0)=1$. Again we stress that this is valid for any $(a(t),b(t)=0,c(t))$. 

Suppose now that we choose $a(t)=a_c(t) := - \frac{1}{2} c(t) (\frac{1}{c(t)})''$. One can write the Wronskian
of the two solutions $f_0(t)$ and $1/c(t)$ (using $c(0)=1$), as $\frac{f_0'(t)}{c(t)} - f_0(t) (\frac{1}{c(t)})' = 1$.
This leads to $f_0(t) =\tau(t)/c(t)$ with $\tau(t)=\int_0^t c(t)^2$. Then \eqref{equiv2} leads to
\be
h(x,t) \equiv  h(0 ,t)   - \frac{x^2}{4} \left[ \frac{c(t)^2}{\tau(t)}  - \frac{c'(t)}{c(t)}  \right]
\ee 
as anticipated in \eqref{sts2}. 

One can consider other choices, e.g. $a(t)=a$. One has $f_0(t)= \frac{\sin \sqrt{2 a} t}{\sqrt{2 a}}$ for $a>0$
and
$f_0(t)= \frac{\sinh \sqrt{2 |a|} t}{\sqrt{2 |a|}}$ for $a<0$, hence
\bea \label{sts3} 
h(x,t) &\equiv&  h(0 ,t)   - \frac{x^2}{4} \sqrt{2 a} \cot(\sqrt{2 a} t) \quad , \quad a>0 \\
&\equiv&  h(0 ,t)   - \frac{x^2}{4} \sqrt{2 |a|} \coth(\sqrt{2 |a|} t) \quad , \quad a<0
\eea 
which will be used below.

\subsection{4) IC classes: known results for the standard KPZ equation} 
\label{SM:sub2} 

For later use below, we recall here some known results for the standard KPZ equation, satisfied by
$H(y,\tau)$ i.e. \eqref{EqH} with $W(y,\tau)=0$. In particular about the one-time statistics 
of the KPZ field in the limit of large $\tau$. From the Cole-Hopf
mapping one has
\be
\hat Z(y,\tau)= e^{H(y,\tau)} = \int dz \hat Z(y,\tau|z,0) e^{H(z,0)}
\ee 
where $\hat Z(y,\tau|z,0)$ is the partition function of the continuous directed polymer from space-time point $z,0$ to $y, \tau$.
For large $\tau \gg 1$ the integral on the r.h.s. is dominated 
by its maximum. The scaled droplet solution $\tau^{-1/3} \log \hat Z(y,\tau|0,0)$
is conjectured to converge, in rescaled coordinates $\hat y =\frac{y}{2 \tau^{2/3}}$
to the so-called Airy$_2$ process minus a parabola,
${\cal A}_2(\hat y)- \hat y^2$ \cite{prolhac2011one, quastel2014airy}.
The height field is then determined
\cite{footnote4}, as a process in $\hat y$,
by a variational problem
\be \label{Hlarge} 
H(y,\tau) + \frac{\tau}{12}  \simeq \tau^{1/3} \max_{\hat z} \big\lbrace {\cal A}_2(\hat z-\hat y)- (\hat z-\hat y)^2 + {\sf H}_0(\hat z) \big\rbrace. 
\ee
Here 
\be \label{defH0} 
{\sf H}_0(\hat z)= \tau^{-1/3} H(2 \hat z \tau^{2/3},0)
\ee
is the so-called rescaled IC (in \eqref{Hlarge} and \eqref{defH0} the limit of large $\tau \gg 1$ is understood).
All IC which share the same ${\sf H}_0$ lead
to the same universal height PDF at large $\tau$. The droplet IC class
corresponds to ${\sf H}_0(\hat z)=-\infty$ for $\hat z \neq 0$
and ${\sf H}_0(0)=1$, shared e.g. by any wedge, $H(z,0)=-w|z|$,
of a large class of IC where $e^{H(x,0)}$ is localized in space. 
In that case the maximum in \eqref{Hlarge} is attained at $\hat z=0$,
and the one point PDF of $H$ is related to the one of ${\cal A}_2(0)$ which is the GUE-TW
distribution. The flat IC corresponds to ${\sf H}_0(\hat z)=0$ and includes a class of IC extended over the whole axis. It leads to the GOE TW one point distribution. 
Eq. \eqref{Hlarge} expresses the solution
for arbitrary IC, and its one point distribution can be expressed 
in terms of a Fredholm determinant in terms of a kernel depending on ${\sf H}_0$, in general quite complicated
\cite{KPZFixedPoint,quastel2019flat}.

Let us recall also that for small $\tau \ll 1$, the KPZ field has Gaussian statistics: this is the so-called
Edwards-Wilkinson (EW) regime, with fluctuations growing as $\delta H \sim \tau^{1/4}$. The spatial correlation
scale of the standard KPZ field, denoted here $y(\tau)$, changes from $y(\tau) \propto \tau^{1/2}$ for
$\tau \ll 1$ to $y(\tau)= 2 \tau^{2/3}$ at large $\tau \gg 1$.

\subsection{5) Large time asymptotics for the time-inhomogeneous KPZ equation}

In this section we provide more details to the study in the main text of the statistics of the height field \eqref{kpz2} 
for various initial conditions. 
The questions are (i) for a given IC for $h$, what is the effective initial condition to use which determine $\chi$ (ii) what are the statistics of
the full height profile $h(x,t)$ at large time. 

We will center the discussion on the model $(a(t)=a_c(t),c(t))$ with noise variance $c(t) =  \Big(\frac{t_0}{t+t_0}\Big)^{\alpha}$ and external potential $V(x,t) = a(t)\frac{x^2}{2}$ and $a(t) = a_c(t) = \frac{\alpha (1-\alpha)}{2} (t+t_0)^{-2}$, although we will consider a few other cases below (we always assume $c(0)=1$). The variations of $c(t)$ thus occur on a time scale $t_0$. There are two distinct cases to be studied.
One is $t_0=O(1)$ fixed and large $t \gg 1$, which leads to non-universal results for $\alpha>1/2$. 
The other is $t_0 \gg 1$ large. In the latter case one can study the regime where both $t$ and $t_0$ large with a fixed ratio, and the results are always universal and can be quantified more precisely. Under the transformation
\eqref{transf1}, \eqref{transf2}, 
\be \label{transf5} 
h(x,t)=H(y,\tau(t)) - \frac{\alpha x^2}{4(t+t_0)} + \frac{\alpha}{2} \log \frac{t_0}{t+t_0}, \quad y = \Big(\frac{t_0}{t+t_0}\Big)^{\alpha} x , \quad \tau(t) = \frac{t_0}{1-2 \alpha} \Big(\Big(1+ \frac{t}{t_0}\Big)^{1-2\alpha}-1\Big), 
\ee
the inhomogeneous KPZ equation \eqref{kpz2} is mapped to the standard KPZ equation for $H(y,\tau)$, i.e.
\eqref{EqH} with $A(\tau)=0$, with the new time 
$\tau(t)$, for which we can use the results of Section I 4). In particular, the spatial correlation
scale of the growth can be defined as $x(t)=\frac{y(\tau(t))}{c(t)}$, where $y(\tau)$ is the 
spatial correlation scale associated to the standard KPZ equation \eqref{EqH}. We also recall the relation between the 
initial conditions, i.e. \eqref{transf5} at $t=0$
\be \label{init0} 
H(y,0) = h(y,0) - \frac{c'(0)}{4} y^2  = h(y,0) + \frac{\alpha}{4 t_0} y^2
\ee 

\bigskip 

\textbf{Asymptotics for $t_0=O(1)$. } Let us first consider the case where $t_0=O(1)$ is fixed. The new time $\tau(t)$ has a different behavior depending on whether $\alpha < 1/2$ or $\alpha > 1/2$.
In the limit $t/t_0 \gg 1$, it diverges for $\alpha<1/2$ as
$\tau(t) \simeq \frac{t_0}{1-2 \alpha} (\frac{t}{t_0})^{1-2\alpha}$, while it
saturates to a finite value for $\alpha>1/2$, as $\tau(t) \to \frac{t_0}{2 \alpha-1}$.\\ 

\begin{itemize} 

\item
For $\alpha <1/2$ we thus predict that the one point PDF of $h(x,t)$, e.g. at $x=0$, behaves as
$t \to +\infty$ as
\be
h(0,t) \simeq - \frac{1}{12} c_1 t_0 \left(\frac{t}{t_0}\right)^{1- 2 \alpha}  + (c_1 t_0)^{1/3} \left(\frac{t}{t_0}\right)^{\beta(\alpha)} \chi, \quad 
\beta(\alpha) = \frac{1-2 \alpha}{3} 
\label{eq:asymptotics}
\ee
with $c_1 = \frac{1}{1-2 \alpha}$,
and where the random variable $\chi$ is the TW type distribution associated to the KPZ fixed point
with the initial condition given by \eqref{init0}. Because of the positive quadratic part in \eqref{init0}
there is an important restriction on the class of IC for $h$ which lead to \eqref{eq:asymptotics}. Consider
the subset of IC such that $h(x,t=0) \simeq - B x^2$ for $x \gg 1$. Let us use the results of Section I 4). It is easy to see from \eqref{Hlarge}, \eqref{init0} and \eqref{defH0} that if $B>B_c=\frac{c'(0)}{4}= \frac{\alpha}{4 t_0}$ then ${\sf H}_0$ belongs to the droplet IC
and $\chi$ is GUE-TW distributed. If $B=B_c$ and if at large $x$, $| h(x,t=0)+B_c x^2 | < |x|^r$ with $r<1/2$,
then \eqref{eq:asymptotics} holds, with ${\sf H}_0(\hat z)=0$ implying that $\chi$ is GOE-TW distributed. The case $B < B_c$ maps to the standard KPZ equation for $H$ with a convex parabolic initial condition. It is well
known that this leads to a finite-time blow-up, i.e. the solution for $H(y,\tau)$, hence also for $h(x,t)$, blows up to $+\infty$ at a finite time $t^*$, hence \eqref{eq:asymptotics} does not hold.
The existence of a blow up for IC with $B<B_c$ is in fact already a property of the equation without noise, i.e. of the model $(a(t),0)$ studied in Section I 2).
It is related to the special form of $a(t)=a_c(t)$, dictated by the choice of $c(t)$, here
$c(t)=(1+\frac{t}{t_0})^{-\alpha}$. For $\alpha<1$, $a(t)<0$ which, for a flat IC, $B=0$, leads to a blow up,
while for $\alpha \geq 1$, $a(t)>0$ and there is no blow up.
This is discussed in detail in Section I 2). 
In particular, choosing e.g. $c(t)=(1+ t^2)^{-\alpha/2}$, 
does not change the large time behavior, but leads to $a(t)$ which is negative at small time
and avoids the blow up (in that case $B_c=0$). 

In the cases where \eqref{eq:asymptotics} holds ($B \geq B_c$, no blow up) the spatial correlation scale behaves as $t \to +\infty$ as
\be \label{xt1} 
x(t) = \frac{2 \tau(t)^{2/3}}{c(t)} \simeq \frac{2 t_0^{\frac{\alpha}{3}} }{(1-2 \alpha)^{2/3}} 
\, t^{\zeta(\alpha)}, \quad  \zeta(\alpha)=\frac{2-\alpha}{3} \quad , \quad \alpha < \frac{1}{2}
\ee 
For $B>B_c$, one obtains from \eqref{Hlarge} the one time statistics of the field $h(x,t)$ in the large $t$ limit as
\cite{footnote5}  
\bea 
 h(x,t) + \frac{\tau(t)}{12} &\simeq& \tau(t)^{1/3} 
\left[ {\cal A}_2(\tilde x) - \tilde x^2 \right ] - \frac{ \alpha x^2}{4 t} 
= \tau(t)^{1/3}  {\cal A}_2(\tilde x) - \frac{(1-\alpha) x^2}{4 t} \\
&=&
 \frac{t_0^{\frac{2 \alpha}{3}}}{(1-2 \alpha)^{1/3}}  t^{\frac{1-2 \alpha}{3}} 
\left[ {\cal A}_2(\tilde x) - \frac{1-\alpha}{1-2 \alpha} \tilde x^2 \right] 
 \quad , \quad \tilde x =  \frac{x}{x(t)} \label{drop00} 
\eea
i.e. GUE-TW for the one point PDF. Note however that the deterministic quadratic dependence in $\tilde x$ 
is different from the one for the standard KPZ equation (recovered for $\alpha=0$). It will be of importance
when analyzing the endpoint PDF of the directed polymer, see below. Finally, for $B=B_c$ one finds from \eqref{Hlarge} for $t \to +\infty$
\be
 h(x,t) + \frac{\tau(t)}{12} \simeq \tau(t)^{1/3} 2^{-2/3} \chi_1 - \frac{ \alpha x^2}{4 t} 
\ee
where $\chi_1=2^{2/3}  \max_z [ {\cal A}_2(\hat z) - \hat z^2 ]$ is distributed according to GOE-TW.
The spatial dependence of the field is trivial in that case. 

\item For $\alpha=1/2$ one has $\tau(t)= t_0 \log(1 + \frac{t}{t_0})$ and $x(t) = \frac{2 \tau(t)^{2/3}}{c(t)} \simeq 2 t_0^{1/6} t^{1/2} [\log(t/t_0)]^{2/3}$ at
large $t$. For $B>B_c=\frac{1}{8 t_0}$ the first equation in \eqref{drop00} is still valid, but not the second,
because the quadratic terms are now, at large $t$, equivalent to $ - \frac{x^2}{4 t \log(t/t_0)} - \frac{x^2}{8 t}$. 
The first one is now negligible compared to the second. Hence we obtain
\be
 h(x,t) + \frac{\tau(t)}{12} \simeq t_0^{1/3} [\log(t/t_0)]^{1/3} {\cal A}_2(\tilde x) -  \frac{x^2}{8 t}
 \quad , \quad \tilde x =  \frac{x}{x(t)} \label{drop01} 
\ee

\item For $\alpha>1/2$ the fluctuations of the field $h(x,t)$ saturate at $t \to +\infty$, and the limit 
statistics, including the one-point PDF, is related via \eqref{transf5} to the one of the standard KPZ equation at finite time $\tau(+\infty)=\frac{t_0}{2 \alpha-1}$, with the modified initial condition \eqref{init0}. These asymptotic distributions are specific to the standard KPZ equation, and are not universal across the KPZ class. They are not known analytically except for the one-point PDF and only for a few special IC. For $h(x,t)$ these are (i) the droplet IC, which maps to the droplet IC for $H$, for which one can use
the finite time results of \cite{amir2011probability,calabrese2010free,dotsenko2010replica,sasamoto2010exact}
(ii) $h(x,t=0)= e^{- \frac{\alpha}{4 t_0} y^2}$, which corresponds to the flat IC for $H$, for which
one can use the finite time results of \cite{PCPLDFlat}. More generally, one can define a correlation
scale, as above, which grows at large time $t$ as
\be
x(t)=\frac{y(\tau(+\infty))}{c(t)} \propto \, t^\alpha, \quad  \alpha>1/2.
\ee
The interpretation of this scale is discussed in Fig. \eqref{fig:decorrelation}. Again these results at large time
hold only when there is no blow-up. From Section I 2), we can surmise that this is the case for 
$B \geq B_c$ with $4 B_c = - c'(0) -  \frac{1}{\tau(+\infty)} = \frac{1-\alpha}{t_0}$. 

\end{itemize}

Thus for $\alpha<1/2$ and $t_0=O(1)$, in particular for $\alpha=1$ a case of special interest, the limiting height distribution at $t=+\infty$ is non-universal. There are
two cases however where it can be characterized more precisely. Small $t_0 \ll 1$, in which case it becomes
Gaussian and described by the EW fixed point (we will not study that case). And large $t_0 \gg 1$, in which case
it becomes again universal and described by the KPZ fixed point, as we now discuss. 

\bigskip 

\textbf{Asymptotics for large $t_0 \gg 1$. } We now study the case where $t_0$, the time scale over which
$c(t)$ varies, is chosen large, $t_0 \gg 1$. Note that this situation is natural in disordered systems undergoing aging or coarsening dynamics (in that case $t_0$ is the waiting time, which maybe large). In that case it is natural to study the regime where both times are large $t,t_0 \gg 1$,
with $t/t_0$ fixed. As a result $\tau(t)$ is also large, i.e. $\tau(t) \gg 1$ and 
$\tau(t)/t_0$ is a function of $t/t_0$. All asymptotics below are thus controled by $t_0 \gg 1$, at fixed ratio $t/t_0$.
The most general such model is defined by a shape function $\hat c(s)$ 
\be \label{models}
c(t)=\hat c(s) \quad , \quad s=\frac{t}{t_0} \quad , \quad \tau(t)/t_0 = \hat \tau(s)=  \int_0^{s} \hat c(u)^2 du 
\ee
where $\hat \tau(s)$ is the shape function of the new time (we impose $\hat c(0)=1$ for
simplicity). We recall that the equation for $h(x,t)$ contains also a quadratic external
potential $V(x,t)=\hat a_c(s) \frac{x^2}{2 t_0}$ with $\hat a_c(s) = \frac{- \hat c(s)}{2}(\frac{1}{\hat c(s)})''$.
Our main example here is a shape function chosen as
$\hat c(u)=(1+u)^{-\alpha}$. Hence in this case, for any $\alpha$ we may use the known asymptotics of $H$ 
in \eqref{Hlarge} and the limit is universal. We obtain the one point statistics as
\begin{equation} 
h(0,t) \simeq - \frac{1}{12} \hat \tau(s) \, t_0 + (\hat \tau(s) \, t_0)^{1/3} \chi_s  \quad , \quad s = \frac{t}{t_0} = O(1) \quad , \quad s ~ \text{fixed}
\label{eq:exampleinterpolation}
\end{equation}
where $\hat \tau(s)$ is given in \eqref{models}. Here
$\chi_s$ is a TW type distribution depending on the initial data {\it and} on (i) the shape function $\hat c$,
(ii) the parameter $s$: the dependence of $\chi_s$ in $s$ is non trivial, and the IC classes of  $H$ and $h$ are not identical anymore (see the discussion below).

When $\alpha <1/2$ one has, from \eqref{transf5}, $\hat \tau(s) \simeq \frac{1}{1-2 \alpha} s^{1-2 \alpha}$
at large $s \gg 1$, and one recovers, for $t/t_0 \gg 1$, the same behavior as in Eq. \eqref{eq:asymptotics}
with $\chi=\chi_{+\infty}$.
The additional information in \eqref{eq:exampleinterpolation} is the complete dependence in 
the parameter $s$. When $\alpha<1/2$, the prefactor $\hat \tau(s)$ saturates at large
$s \gg 1$, $\hat \tau(s) \simeq \frac{1}{2 \alpha -1}$, and the limit $t \to +\infty$, that is $t \gg t_0$, leads to interesting new results (see below).

\bigskip
 
\textbf{Mapping initial conditions.} Let us address the one-time, full space statistics of the field $h(x,t)$, and identify the
IC classes. We use \eqref{transf5} together with the asymptotic large $\tau$ result for
$H(y,\tau)$, as given by \eqref{Hlarge}. From \eqref{init0} one now finds that 
${\sf H}_0(\hat z) = {\sf h}_0(\hat z) + \frac{c_2 \tau(t)}{t_0} \hat z^2$, 
with $c_2=- \hat c'(0)/\hat c(0)= \alpha$. Here, for fixed $t/t_0=s=O(1)$, 
$\frac{\tau(t)}{t_0} = \hat \tau(s)= \int_0^s \hat c(u)^2 du$ is a fixed number.
Hence the shift between IC of $h$ and $H$ (an additional parabola)
remains important in this regime. We then find
\be \label{var} 
h(x,t) + \frac{\hat \tau(s) t_0}{12} \simeq ( \hat \tau(s) t_0)^{1/3} 
\left[ \max_{\hat z} \left({\cal A}_2(\hat z  -  \tilde x
)- (\hat z-\tilde x)^2 + {\sf h}_0(\hat z) + c_2 \hat \tau(s) \hat z^2\right)
+ \frac{\hat c'(s) \hat \tau(s)}{\hat c(s)^3} \tilde x^2 \right], \quad \tilde x =  \frac{x}{x(t)}
\ee
where we recall that $x(t)=\frac{2 \hat \tau(s)^{2/3}}{\hat c(s)} t_0^{2/3}$ is the spatial correlation scale
defined above, and ${\sf h}_0(\hat z)  = (t_0 \hat \tau(s))^{1/3} h(2 (t_0 \hat \tau(s))^{2/3},0)$ (with $t_0\gg 1$).
The variable $\chi_s$ introduced in \eqref{eq:exampleinterpolation} is thus equal to
the square bracket in \eqref{var}. The variational equation \eqref{var} characterizes completely the scaled
height field as a process in the variable $\tilde x=x/x(t)$. One can see that is {\it does not depend
only} on the final value $\hat c(s)$ of the noise, but on integrated information on the full shape function, $\hat c(u)$,
e.g. via $\hat \tau(s)$, as well as $c_2=- \hat c'(0)/\hat c(0)$.

Consider initial conditions such that $h_0(\hat z) \simeq - \hat B \hat z^2$ at large $\hat z$. We 
see that the result in \eqref{var}, for a given value of $s$, is finite if and only if $\hat B \geq c_2 \hat \tau(s) - 1$. 
If not, there is a blow-up. This condition corresponds, upon rescaling, to the one given in Section I 2) and
in the discussion above. For $\alpha < 1/2$ since $\hat \tau(s)$ diverges at large $s$ there
is always a blow up for some $s=s^*$. Only the droplet IC, with $\hat B=+\infty$, has no blow up. 
For $\alpha>1/2$ the IC which have no blow up are such that $\hat B \geq \hat B_c = c_2 \hat \tau(+\infty)-1
= \frac{1-\alpha}{2 \alpha -1}$. The flat IC thus has no blow-up for $\alpha \geq 1$. 
Interestingly, the absence of a blow-up for the flat IC is also guaranteed if $c_2=0$, i.e. $\hat c'(0)=0$, 
that is if $\hat c(s)$ is sufficiently smooth around the origin. In that case $\hat B_c=-1$. This is the case
for instance for $\hat c(s)=(1+s^2)^{\alpha/2}$, with $\hat a(s)= - \frac{\alpha(1+ s^2 (\alpha-1))}{2 (1+ s^2)^2}$
(see discussion in Section I 2)). \\

Let us rewrite \eqref{var} in the special case $\alpha=1$, for which $a(t)=a_c(t)=0$, in the more explicit form
\be \label{a1} 
h(x,t) + \frac{t}{12(t+t_0)}\simeq \left(\frac{t_0 t}{t+t_0}\right)^{1/3} 
\left[ \max_{\hat z} \left({\cal A}_2(\hat z  -  \tilde x
)- (\hat z-\tilde x)^2 + {\sf h}_0(\hat z) + \frac{t}{t+t_0} \hat z^2\right)
- \frac{t}{t_0} \tilde x^2 \right], \, \tilde x = \frac{x}{2 t^{2/3} (1+ \frac{t}{t_0})^{1/3}}.
\ee

\bigskip 
\textbf{Droplet IC.}
The droplet IC formally corresponds to ${\sf h}_0(0)=0$ and ${\sf h}_0(\hat z \neq 0)=-\infty$. 
In that case the maximum in \eqref{var} is attained for $\hat z=0$ and \cite{footnote5}
\be \label{vardrop} 
h(x,t) + \frac{\hat \tau(s) t_0}{12} \simeq ( \hat \tau(s) t_0)^{1/3} 
\left[{\cal A}_2(\tilde x) - \omega(s)  \tilde x^2 \right], \quad 
\omega(s) = (1- \frac{\hat c'(s) \hat \tau(s)}{\hat c(s)^3} )
\ee
Hence the one-point statistics is GUE-TW and the height field statistics is the Airy$_2$ process plus, however, a 
parabola with amplitude depending continuously on $s=t/t_0$. In the units of the correlation scale
$x(t)$, the amplitude of the parabola saturates at large $s=t/t_0$ for $\alpha<1/2$ as
$\omega(+\infty) = \frac{1-\alpha}{1-2 \alpha}$. This limit is consistent with the result 
\eqref{drop00} obtained there for $t/t_0 \gg 1$ at fixed $t_0=O(1)$. 

On the contrary, for $\alpha=1/2$ one has $\hat \tau(s)= \log(1+s)$ and $\omega(s)=1 + \frac{1}{2} \log(1+s)$,
which diverges as $\omega(s) \simeq \frac{1}{2} \log(s)$ at large $s$. For $\alpha>1/2$, the large $s$ divergence is 
$\omega(s) \simeq \frac{\alpha}{2 \alpha-1} s^{2 \alpha -1}$ for $s=t/t_0 \gg 1$. In the case
$\alpha=1$ 
\be \label{var11} 
h(x,t) + \frac{t}{12(t+t_0)} \simeq \left(\frac{t_0 t}{t+t_0}\right)^{1/3} \left[ {\cal A}_2(\tilde x) - (1+ \tfrac{t}{t_0}) \tilde x^2\right] = \left(\frac{t_0 t}{t+t_0}\right)^{1/3} {\cal A}_2(\tilde x) - \frac{x^2}{4 t} 
\ee
where we recall that the Airy$_2$ process is statistically invariant by translation and reflection.

\bigskip 
\textbf{Flat IC.}
Let us consider now the flat initial condition, $h(x,0)=0$, i.e. ${\sf h}_0(\hat z)=0$, and focus on
$\alpha=1$ for simplicity. In Eq. \eqref{a1} we can redefine $\hat z \to \hat z+\tilde x$. 
Then we observe that the remaining deterministic terms form a perfect square. Hence
we obtain 
\be \label{flat1} 
h(x,t) + \frac{t}{12(t+t_0)} \simeq \left(\frac{t_0 t}{t+t_0}\right)^{1/3} 
\max_{\hat z} \left({\cal A}_2(\hat z)
- \frac{t_0}{t+t_0} \left(\hat z - \frac{t}{t_0} \tilde x\right)^2  \right), \, \tilde x = \frac{x}{2 t^{2/3} (1+ \frac{t}{t_0})^{1/3}}.
\ee

For $\tilde x=0$ we find the result mentionned in the main text, namely that the
CDF of the scaled fluctuating part $(\frac{t_0 t}{t+t_0})^{-1/3}\delta h(0,t) $
is given by 
$F^{\beta,\beta}_{\rm parbl}(s) = {\rm Prob} \left( \max_{\hat z} ({\cal A}_2(\hat z) - (1+\beta) \hat z^2) \leq s \right)$
, with $\beta=-\frac{\tau(t)}{t_0}= - \frac{t}{t+t_0}$, a distribution
for which a formula was obtained in \cite[Example 1.25]{quastel2019flat}.
It interpolates between the GOE TW for small $t/t_0$ (small negative $\beta$) and the Gumbel distribution for
$t/t_0 \to +\infty$ ($\beta \to -1$) \cite{QuastelPrivateComm}. The latter can be seen from the following heuristics.
As $\beta \to -1$, the parabola weakens and 
$\hat z$ explores a larger region $|\hat z| \propto (1+\beta)^{-1/2}$.
Since correlations of $\mathcal{A}_2(\hat z)$
decay fast enough (as $1/\hat z^2$) on scales $\hat z = O(1)$ 
the problem becomes similar to the maximum of $M \propto (1+\beta)^{-1/2}$ i.i.d. random variables. One 
obtains the estimate
\begin{equation}F_{\rm parbl}^{\beta,\beta}\left(A_\beta + \frac{s}{2 \sqrt{A_\beta}}\right)\to e^{-e^{-s}}\end{equation}
as $\beta$ goes to $-1$ with $A_\beta \simeq ( - \frac{3}{8} \log(1+\beta))^{2/3}$,
where we used that the CDF of $v=\mathcal{A}_2(0)$ decays as $\propto v^{-3/2} e^{- \frac{4}{3} v^{3/2}}$ for large positive $v$.

 Note that in \cite{KazPC} the above optimisation problem was simulated using the Dyson Brownian motion and compared to inward KPZ growth experiments. In particular the few lowest cumulants of the distribution $F_{\rm parbl}^{\beta,\beta}$ have been computed numerically.  
This thus provides another nice example where this ``parabolic'' KPZ fixed point distribution 
appears.

Note also  that in \eqref{flat1}, since the Airy$_2$ process is statistically translationally invariant, the one point PDF of $h(x,t)$ is independent of $x$, as is expected for a flat initial condition.

\bigskip 

{\bf Remark}. The fact that the case $c(t)=1/t$ has special properties can also 
be seen from the invariance of the Brownian
motion under the transformation $t \to 1/t$, i.e. $\hat B(1/t)= B(t)/t$, where $B$ and $\hat B$ are two
unit Brownians. For the point to point DP partition sum with $V(x,t)=0$ and noise $c(t)=1/t$, one can write the solution of the SHE in the time interval $[t_1,t_2]$ as an expectation over a Brownian
\bea
&& Z(x_2,t_2|x_1,t_1)= \mathbb{E}\left[ \exp\left(\int_{t_1}^{t_2} dt \sqrt{\frac{2}{t}} \xi(B(t),t) \right) | B(t_1)=x_1, B(t_2)=x_2\right]  \\
&& = \mathbb{E}\left[ \exp\left(\int_{t_1}^{t_2} \frac{d t}{t^2} \sqrt{2}  \hat \xi(\frac{B(t)}{t},\frac{1}{t}) \right) | B(t_1)=x_1, B(t_2)=x_2\right]  
\\
&&  = \mathbb{E}\left[ \exp\left(\int_{u_2}^{u_1} du \sqrt{2}  \hat \xi(\hat B(u),u) \right) | \hat B(u_1)=\frac{x_1}{t_1}, 
\hat B(u_2)= \frac{x_2}{t_2}\right]
= \hat Z(y_1, u_1=1/t_1|y_2, u_2=1/t_2) 
\eea
with $y_i=x_i/t_i$. In the second identity we only used the scale invariance of the space-time white noise ($\hat \xi$ being here another unit space-time white noise) and in the third we used the change of variable $t_i = 1/u_i$ and the  
above property of the Brownian motion. Note that here the time change $t \to 1/u$ has reversed time order, and
to connect to our result for $\alpha=1$ we can use the reversibility symmetry 
$\hat Z_{\hat \xi(y,u)}(y_1, u_1|y_2, u_2) = \hat Z_{\hat \xi(y,u_1+u_2-u)}(y_2, u_1|y_1, u_2)$.

\bigskip

\textbf{Case of a linear potential.} We now discuss 
the case of the KPZ equation with linear potential  \eqref{eq:KPZlinearpotential}. 
For droplet initial conditions we
know that for the one point PDF, $H(y,t) \equiv H(0,t) - \frac{y^2}{4 t}$, where $\equiv$ means equality in distribution.
Hence we have using \eqref{rel1badname} 
\be
h(x,t) \equiv H(0,t) - \frac{(x+ y_0(t))^2}{4 t} + \frac{y_0'(t)}{2} x + \frac{1}{4} \int_0^t ds \, y_0'(s)^2.
\label{eq:generallinearpotential}
\ee 
The height is thus a parabola centered at $x=x_m(t)$
\be
x_m(t) = t y_0'(t) - y_0(t) 
\ee 
plus droplet KPZ fluctuations. In the case of the KPZ equation \eqref{kpz2} with  $b(t)=b$ \eqref{eq:generallinearpotential} corresponds to the result discussed in the text
with $y_0(t)= b t^2$. Note that the droplet result (GUE-TW at large time) requires an initial condition
such that $Z(x,t=0)=e^{h(x,t=0)}$ decays sufficiently fast. Indeed, in \eqref{rel1badname} the field $H$ is probed at space point $x+ b t^2$ very far from the origin at large time. Let us consider an
initial condition $h(x,t=0)= - \phi(x)$ with $\phi(x)>0$. If the initial condition is e.g. a wedge, 
$\phi(y)= B |y|$, one expects that fluctuations of $h(x,t)$ for $x$ fixed (e.g. at $x=0$) at large $t$
will be given instead by the flat IC class, i.e. the GOE-TW distribution.

Let us write the solution $h_0(x,t)$ in presence of the linear potential, but in the absence of the noise.
From \eqref{eq:generallinearpotential} it gives a heuristic description of the ``mean profile'' in the presence of noise,
replacing $h(x,t)$ by its average. It reads 
\be \label{sup} 
e^{h_0(x,t)} = \int_{-\infty}^{+\infty} \frac{dz}{\sqrt{4 \pi t}} e^{ - \frac{(x+ b t^2 - z)^2}{4 t} + x b t - \phi(z) }
\ee 
Consider first the wedge, $\phi(z)=-|z|$. At large time one can use the saddle point method to estimate the
integral. The argument of the exponential is maximum at $z=z(t)= b t^2 - 2 t + x$: Indeed, $|z|$ can be replaced by $z$
and this argument can be approximated as $\simeq - (z-z(t))^2/(4 t) + (b t - 1)(x-t)$. The problem thus looks
like the standard KPZ problem with flat IC. Hence we expect the GOE-TW distribution. One sees that the
profile $h_0(x,t) \simeq b t (x-t)$ is linear, consistent with this expectation.

One can ask about the general form of the profile $h_0(x,t)$ for a larger class of IC.  At large time, we may approximate $h_0(x,t)$ by the maximum of the argument of the exponential in \eqref{sup}. The maximal argument is reached at $z=z_m(x)$ solution of $2 t \phi'(z_m(x)) = b t^2 + x - z_m(x)$. 
The profile $h_0(x,t)$ has a maximum at some $x=x_m$. Denoting $z_m=z_m(x_m)$  one finds that the maximum is reached at $x_m = b t^2 + z_m$ with $\phi'(z_m)= b t$. For the class $\phi(z) = \frac{1}{1+\delta} |z|^{1+\delta}$ with $\delta>0$, one thus finds
$x_m = b t^2 + (b t)^{1/\delta}$. Further, one may check that the curvature of the profile $h_0(x,t)$ (in the large time limit) is given by  $\partial_x^2 h_{0}(x,t) 
\simeq - \phi''(z_m)/(1+ 2 t \phi''(z_m))$. It is negative for $\delta>0$, which is a sign that we are in the droplet IC class. 
For $\delta \to 0^+$, this curvature vanishes, and one recovers the above result about the wedge: the profile is linear at large time and does not exhibit a maximum.

\bigskip

{\bf A solvable case of the KPZ equation in presence of a quadratic potential}. 
It is interesting to note that, although no exact solutions exist for the usual KPZ equation in presence of a time-independent
	quadratic potential $V(x,t)=a \frac{x^2}{2}$, it can be solved for certain time-dependent
	disorders $c(t)$. It is easy to solve for $a=a_c(t)$, using that $a_c(t) = - \frac{c(t)}{2} (1/c(t))''$.
	
	Consider $a<0$ i.e. a confining potential for the DP. The general solution is
	$c(t) = 1/(c_1 e^{\sqrt{2 |a|}  t}  + c_2 e^{- \sqrt{2 |a|}  t} )$. The simplest example is $V(x,t)=- \frac{x^2}{4 t_0^2}$ and
	$c(t) = e^{- t/t_0}$, the solution is then $h(x,t) = H( x e^{-t/t_0} , \frac{t_0}{2} (1-e^{-2 t/t_0} ) ) - \frac{x^2}{4 t_0} - \frac{t}{2 t_0}$
	with $H(y,\tau)$ solution of the standard KPZ equation with
	$H(x,0)=h(x,0)+ \frac{x^2}{4 t_0}$. At large $t$ is leads to standard finite time KPZ fluctuations.
	The asymptotic PDF of $h(x,t=+\infty)$ scaled by $(t_0/2)^{1/3}$ becomes the GUE-TW distribution 
	as $t_0 \gg 1$ (for droplet IC). 
	
	Interestingly, the opposite case, $c(t) = e^{t/t_0}$, leads to the solution
	$h(x,t) = H( x e^{t/t_0} , \frac{t_0}{2} (e^{2 t/t_0}-1 ) ) + \frac{x^2}{4 t_0} + \frac{t}{2 t_0}$
	with $H(y,\tau)$ solution of the standard KPZ equation with $H(x,0)=h(x,0)- \frac{x^2}{4 t_0}$.
	At large time it leads to TW type (and KPZ fixed point) type fluctuations for any value of $t_0$.

	Finally for $a<0$, the solvable cases are $c(t) = c_1/\cos( \sqrt{2 a} (t+t_0))$, 
	which, however, lead to diverging $c(t)$ for some periodic times.

\subsection{6) Directed polymer and its wandering exponent}

We recall that the partition sum $Z(x,t)$ of the $d=1+1$ continuum directed polymer (DP)
in a (time inhomogeneous) random potential $\sqrt{2 c(t)} \xi(x,t)$ with one fixed endpoint at $(x,t)$ and in presence
of an external potential $V(x,t)$ is solution of the stochastic heat equation (SHE), 
Equation \eqref{eq:SHEtdependent} in the text. The special solution, denoted $Z(x,t|x_0,0)$, 
with initial condition $Z(x,t=0|x_0,0)=\delta(x-x_0)$ is called the droplet IC solution
and corresponds to a DP with both endpoints fixed (point to point DP). 
More general IC conditions correspond 
to other DP geometries, for instance the flat IC $Z(x,t=0)=1$ corresponds to the point to line DP.

Similarly one defines the second DP problem, in the random potential
$\sqrt{2} \hat \xi(y,\tau)$ and in presence of a quadratic external potential
whose partition sum, $\hat Z(y,\tau)$, is solution of 
\be \label{she2}
\partial_\tau \hat Z = \partial_y^2 \hat Z + (- A(\tau(t)) \frac{y^2}{2}  + \sqrt{2} \hat \xi(y,\tau) ) \hat Z
\ee 
and one denotes the droplet solution, $\hat Z(y,\tau|y_0,0)$, with initial condition
$\hat Z(y,\tau=0|y_00)=\delta(y-y_0)$. 

When both $\xi$ and $\hat \xi$ are unit space-time white noises, the relation between the two partition sums is then,
from \eqref{transf1}, \eqref{transf2} in the text
\be \label{transfZ} 
Z(x,t) = \sqrt{c(t)} \hat Z(c(t) x,\tau(t)) e^{\frac{c'(t)}{4 c(t)} x^2} 
\quad , \quad 
Z(x,t | 0,0) = \sqrt{c(t)} \hat Z(c(t) x,\tau(t)|0,0) e^{\frac{c'(t)}{4 c(t)} x^2},
\ee 
(up to an immaterial multiplicative constant). The first identity leads to some correspondence between the DP geometries, as discussed in the text, see
\eqref{init},
and we see that the point to point DP are in correspondence in both problems. These relations
can also be checked directly from \eqref{she2}.

An important observable is the PDF of the endpoint of the DP. It is defined as the average
\be
P(x,t) = \overline{P_\xi(x,t)} = \overline{ \frac{Z(x,t)}{\int dx' Z(x',t)}} 
\ee
where $P_\xi(x,t)$ is the endpoint PDF in a given sample. From it one defines the moments
of the endpoint PDF $\langle x^n \rangle = \int dx x^n P(x,t)$, and 
the transverse wandering $\delta x(t) = \sqrt{\langle x^2 \rangle - \langle x \rangle^2}$.
Using the mapping \eqref{transfZ} we obtain
\be \label{PP} 
P(x,t) = \overline{ \frac{\hat Z(c(t) x,\tau(t)) e^{\frac{c'(t)x^2}{4 c(t)}} }{\int dx' 
\hat Z(c(t) x',\tau(t)) e^{\frac{c'(t)(x')^2}{4 c(t)}}}} \quad , \quad \hat Z(y,\tau) = e^{H(y,\tau)} 
\ee 

Let us restrict to the droplet solution, i.e. the point to point DP.  
Consider the case $c(t)=\left( \frac{t_0}{t+t_0} \right)^{\alpha}$, and $a(t)=a_c(t)$, as above.

Let us start with $\alpha<1/2$ and $t_0=O(1)$. The denominator in \eqref{PP} is $e^{h(x,t)}$, and in the 
limit $t \gg 1$ the statistics of $h(x,t)$ is described by Eq. \eqref{drop00}.
Given that the prefactor in e.g. \eqref{drop00} is large, the PDF in any given sample is concentrated around $\tilde x^*$ which realizes the maximum of the bracket in \eqref{drop00}. Let us define for $\omega>0$, ${\cal P}_\omega(\tilde x^*)$ the PDF of $\tilde x^* = {\rm argmax}_{z \in \mathbb{R}} \left[ {\cal A}_2(z) - \omega z^2 \right]$. It is a one-parameter universal distribution: for $\omega=1$ it is the known endpoint PDF for the standard DP \cite{greg2,quastelendpoint,baikgreg}, and it was calculated recently \cite{quastelinprep}
for other values of $\omega$.
Our conclusion is that for large $t$, and fixed $t_0=O(1)$, the endpoint distribution takes the form
\be \label{resP1} 
P(x,t) \simeq \frac{1}{x(t)} {\cal P}_{\frac{1-\alpha}{1-2 \alpha}}\left(\frac{x}{x(t)} \right) \quad , \quad \alpha < 1/2
\ee 
where $x(t) \sim t^{\zeta(\alpha)}$ is the correlation length scale given in Eq. \eqref{xt1}. For $\alpha<1/2$ the transverse wandering length of the DP is thus proportional to the correlation length scale, i.e. $\delta x(t) \propto x(t)$.

Let us study the limit of ${\cal P}_\omega$ for $\omega \to +\infty$, which is relevant for $\alpha \to 1/2^-$. 
In that limit the quadratic well is strong and the position of the maximum $\tilde x^*$ is close to zero. Hence one can rescale $z= \omega^{-2/3} u$ and one has 
\be
\max_{z \in \mathbb{R}} \left[ {\cal A}_2(z) - \omega z^2 \right] - {\cal A}_2(0) \simeq \omega^{-1/3}  \max_{u \in \mathbb{R}} \left[ \sqrt{2} B(u) - u^2 \right] 
\ee 
where $B(u)$ is the two sided Brownian motion, and we have used that the Airy process is locally Brownian, i.e. that
as $\epsilon \to 0$, ${\cal A}_2(\epsilon u)- {\cal A}_2(0) = \sqrt{2 \epsilon} B(u) + o(\epsilon)$ \cite{AiryBrownian}. 
We also use here and below the Brownian scaling, $B(a u) =\sqrt{a} B(u)$.
Hence one has $\tilde x^* = \omega^{-2/3} u^*$ where $u^*={\rm argmax}[\sqrt{2} B(u) - u^2 ]$. The PDF of $u^*$,
which we denote $P_0(u^*)$, is well known \cite{Groeneboom,PLDMonthus} 
\be
P_0(u) = g(u) g(-u) \quad , \quad g(u)= \int_{-\infty}^{+\infty} \frac{dw}{2 \pi} \frac{e^{- i w u}}{{\rm Ai}(i w)} 
\ee 
with $P_0(u) \sim |u| e^{-|u|^3/3}$ at large $|u|$. Hence one has, in the large $\omega \to +\infty$ limit
\be
\omega^{-2/3} {\cal P}_{\omega}(\omega^{-2/3} u) \to P_0(u)
\ee 

Let us consider now $\alpha=1/2$, the marginal case. We already surmise that as $\alpha \to 1/2^-$, since $\omega = \frac{1-\alpha}{1-2 \alpha} \to +\infty$, the distribution $P_0$ should arise. Let us show now how it works. Let us start from Eq. \eqref{drop01}. Anticipating the result, let us write $x = 2^{5/3} t_0^{1/6} t^{1/2} z$. One can rewrite the r.h.s. 
in \eqref{drop01} as $(2 t_0)^{1/3} [ \sqrt{2} B(z) - z^2]$. This means that in a given sample 
\be
P_\xi(x,t) \propto e^{(2 t_0)^{1/3} [ \sqrt{2} B(z) - z^2]} \quad , \quad z = \frac{x}{2^{5/3} t_0^{1/6} t^{1/2}}
\ee
up to a normalization constant. This identifies with the random Gibbs measure at "temperature" $(2 t_0)^{-1/3}$
of the so-called "toy model", much studied in the disordered system literature
\cite{MezDots,PLDMonthus,PLDMonthus2}.
If $t_0 \gg 1$, it is dominated by the maximum in the exponential and
the PDF of $x/x_1(t)$ is given by $P_0$ as $t/t_0 \gg 1$
\be
P(x,t) \simeq \frac{1}{x_1(t)} P_0( \frac{x}{x_1(t)}) \quad , \quad x_1(t) = 2^{5/3} t_0^{1/6} t^{1/2} 
\ee
Some results are also known for the ``finite temperature'' regime $t_0=O(1)$. Note that the
transverse wandering length of the DP is now set by the new length scale $\delta x(t) \propto x_1(t) \sim t^{1/2}$, i.e. leading to diffusion, 
different for the super-diffusive correlation length $x(t) \sim t^{1/2} [\log(t/t_0)]^{2/3}$.

For $\alpha>1/2$, and $t_0=O(1)$, the quadratic part $- \alpha x^2/(4 t)$ dominates the endpoint PDF, and the random part becomes negligible. Let us write $x= \hat x t^{1/2}$ with fixed $\hat x$ and large $t$.
Then we can neglect the dependence in $\hat x$ of the 
factor $\hat Z$ in \eqref{PP}, i.e. $\hat Z(c(t)x, \tau(t))$ is asymptotically constant as $x$ varies in the $t^{1/2}$ scale.
Conversely, this factor is expected to vary on scales $x=x(t) \sim t^{\alpha}\gg t^{1/2}$. More precisely, because $\alpha>1/2$, $c(t) t^{1/2}\to 0$ as $t$ goes to infinity, so that for fixed $\hat x$, the numerator of \eqref{PP} can be approximated by 
 \begin{equation} \label{aa} 
 \hat Z(c(t)x, \tau(t)) e^{\frac{c'(t)x^2}{4 c(t)}} \simeq \hat Z\left(0,\frac{t_0}{2 \alpha -1}\right)e^{-\frac{\alpha}{4} \tilde x^2}.
 \end{equation}  
Hence we obtain that at large $t$, $P(x,t) \simeq (\frac{\alpha}{4 \pi})^{1/2} e^{- \frac{\alpha}{4} \hat x^2}$.
Thus, as stated in the main text, the polymer wandering scale is $ \delta x(t) \sim t^{1/2}$, i.e. diffusive, and is typically different from the spatial correlation scale which is $x(t) \propto  t^{\alpha}$. Figure \ref{fig:decorrelation} provides a heuristic explanation for the relation between these spatial scales and the geometry of polymer paths.

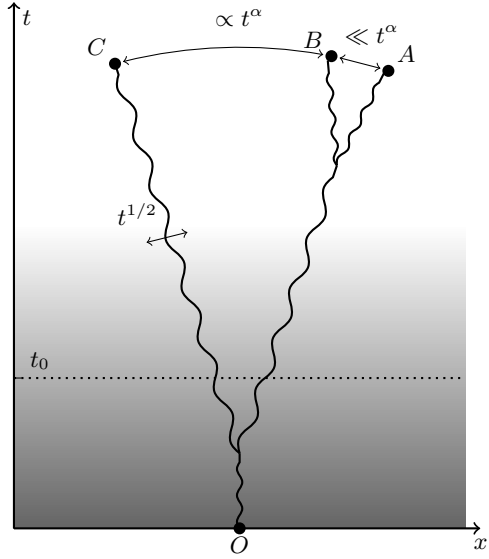
\begin{figure}[h]
	\begin{center} 
		\begin{tikzpicture}[scale=2]
		\usetikzlibrary{shapes,backgrounds, shadings, snakes}
		\tikzstyle{polymer} = [thick, snake=coil, segment aspect=0, segment length= 10pt,  segment amplitude=1pt]
		\tikzstyle{polymerwandering} = [thick, snake=coil, segment aspect=0, segment length= 20pt,  segment amplitude=2pt]
		\begin{scope}
		\clip (-1.5,0) rectangle (1.5,4);
		\shade[bottom color=black, top color=white, fill opacity=0.6] (-2,0) rectangle(2,2);
		\end{scope}
		\draw[thick, ->] (-1.5,0) --  (1.6,0);
		\draw[thick, ->] (-1.5,0) --  (-1.5,3.5);
		\draw (-1.5,3.4) node[right]{$t$};
		\draw (1.5,-0.1) node[right]{$x$};
		\draw (-1.5,1) -- (-1.45,1) node[above right] {$t_0$};
		\draw[thick, dotted] (-1.5,1) -- (1.5,1);
		\draw[polymer] (0,0) -- (90:0.5);
		\draw[polymerwandering] (90:0.5)--  (75:2.5);
		\draw[polymer]  (75:2.5) -- +(63:.7);
		\draw[polymer] (75:2.5) -- +(95:.7);
		\draw[polymerwandering]  (90:0.5) -- (105:3.2);
		
		\draw[<->] (108:2) -- (100:2);
		\draw (108:2.2) node{$t^{1/2}$};
		
		\draw[<->] (73:3.2) -- (78:3.2); 
		\draw (75:3.4) node{$ \ll t^{\alpha} $}; 
		
		\draw[<->] (80:3.2) arc (80:104:3.2);
		\draw (90:3.4) node{$ \propto t^{\alpha} $}; 
		
		\fill (0,0) circle(0.04) node[below] {$O$};
		\fill (72:3.2) circle(0.04) node[above right] {$A$};
		\fill (79:3.2) circle(0.04) node[above left] {$B$};
		\fill (105:3.2) circle(0.04) node[above left] {$C$};
		\end{tikzpicture}
	\end{center}
	\caption{The gray shading indicates the amplitude of the noise which decreases with time. The directed polymer partition function from $O$ to $A$ or $B$ is dominated by paths which will branch to $A$ or $B$ after exiting the area with large weights, thus the partition functions are highly correlated. However, the paths going to $B$ or $C$ branch much earlier and thus the partition functions are different. If the distance from $B$ to $C$ was much larger than $t^{\alpha}$, the partition function would fully decorrelate.}
	\label{fig:decorrelation}
\end{figure}

Finally let us discuss the case $t_0 \gg 1$. In the large $t$ limit with $s=t/t_0$ fixed we can now use Eq. \eqref{vardrop}, and we find $P(x,t) \simeq \frac{1}{x(t)} {\cal P}_{\omega(s)}(\frac{x}{x(t)} )$, where $x(t)=t_0^{2/3} \frac{2 \hat \tau(s)}{\hat c(s)}$ and $\omega(s)$ is defined in \eqref{vardrop}. For $s=O(1)$ this result is valid for any $\alpha$. For $\alpha<1/2$ the result is thus qualitatively similar to the one above in \eqref{resP1} for $t_0=O(1)$, the two results matching perfectly when $s=t/t_0 \to +\infty$, since $\omega(+\infty)=\frac{1-\alpha}{1-2 \alpha}$.
For $\alpha \geq 1/2$ and $s=O(1)$ the result is quite different from the result obtained above for $O(t_0)=1$.
They can still be matched as $s=t/t_0 \to \infty$, but the matching is more complicated since $\omega(s)$ diverges at large $s$. Let us focus on
$\alpha=1$ for simplicity, and consider the Eq. \eqref{var11} where we recall $\tilde x=\frac{x}{x(t)} = \frac{x}{2 t^{2/3} (1+ \frac{t}{t_0})^{1/3}}$. Let us rescale $\tilde x=(1+\frac{t}{t_0})^{-2/3} z$, and study the regime $t/t_0 \gg 1$.
Using that the Airy process is locally Brownian we have
\be
P_\xi(x,t) \propto e^{ t_0^{2/3} t^{-1/3} [ \sqrt{2} B(z) - z^2] } \quad , \quad z = \frac{x}{2 (t_0 t)^{1/3}} 
\ee
thus there is an intermediate regime $t_0 \ll t \ll t_0^2$ where the endpoint PDF behaves as
$P(x,t) \sim \frac{1}{x_2(t)} P_0(x/x_2(t))$ where $x_2(t) = 2 (t_0 t)^{1/3}$ is a new length scale,
intermediate between the correlation scale $x(t) \sim t$ and the diffusive scale $x_1(t) \sim t^{1/2}$. 
Finally for $t/t_0 \gg t_0 \gg 1$
one recovers the diffusive result obtained in \eqref{aa}.

\bigskip

\subsection{7) The case $a(t)=0$: absence of external potential} 

When $a(t)=0$, i.e. $V(x,t)=0$, the KPZ equation with a general time-dependent noise $c(t)$ maps to 
the KPZ equation with unit noise in presence of a quadratic potential $- \frac{A(\tau)}{2} y^2$, with
$A(\tau(t))=a_c(t)/c(t)^4 = - \frac{1}{2} c(t)^{-3} (1/c(t))''$. 

One can ask which $c(t)$ lead to $A(\tau)=A$ a positive constant. The general solution with $c(0)=1$ is the two parameter family (where one can choose $t_1 \geq t_0$) 
\be
c(t) = \frac{1}{\sqrt{ (1 + \frac{t}{t_0}) (1 + \frac{t}{t_1}) }}
 \quad , \quad 8 A = \left(\frac{1}{t_0}-\frac{1}{t_1}\right)^2 \quad , \quad 
\tau(t)= \frac{1}{\frac{1}{t_0} - \frac{1}{t_1}} \left(\log\Big(1+ \frac{t}{t_0}\Big) - \log\Big(1+ \frac{t}{t_1}\Big)\right)
\ee 
with $\tau(+\infty)= \frac{1}{\frac{1}{t_0} - \frac{1}{t_1}}  \log(t_1/t_0)$. 
The case $t_1=+\infty$, i.e. $c(t) = 1/\sqrt{1+t/t_0}$, with $A=1/(8 t_0^2)$, is the ``critical case'' studied in the main text. The case $t_1=t_0$, i.e. $c(t)=1/(1+ t/t_0)$, with $A=0$, is the case $\alpha=1$ studied above and in the main text. In the general case, and especially for $t_1 \gg t_0$, $c(t)$ exhibits a crossover between the two behaviors
$c(t) \sim 1/\sqrt{t}$ (on scale $t_0$) and $c(t) \sim 1/t$ (on scale $t_1$). 

Assume $t_1  \geq t_0 \gg 1$, hence the curvature of the quadratic well is small $A \ll 1$. Let us examine qualitatively
the problem of the point to line DP of length $\tau$ in a quadratic well $- \frac{A}{2} y^2$ with a unit white noise random potential (hence we restrict to the droplet IC). In the limit $A \ll 1$ the confinement due to the quadratic well acts only at large scale: by scaling one sees that it cuts the growth of the variance of the DP endpoint fluctuations, noted $\langle y^2 \rangle$, at a crossover time $\tau \simeq \tau_A= 1/\sqrt{8 A} \gg 1$.
This is obtained by considering \cite{footnote6} a segment of length $\tau$ of the DP, wandering over a distance $y$, and balancing the elastic energy, $\propto y^2/\tau$, with the potential energy, $\propto A y^2 \tau$. Hence one has
\be \label{y2} 
\langle y^2 \rangle \, \simeq \, \begin{cases} \, \, Y \tau^{4/3}  ~~~~~~~ \quad , \quad 1 \ll \tau \ll \tau_A \\
\, \, y_2 (8 A)^{-2/3} \quad , \quad \tau \gg \tau_A  \end{cases}
\ee 
where $Y$ and $y_2$ are numbers of order unity. Similarly, since segments of length $\tau_A$ become essentially uncorrelated, one expects that the free energy fluctuations of the point to point DP scale 
as 
\be
\delta H(0,\tau) \simeq \begin{cases}  \tau^{1/3} \chi_2 ~~~~ \quad , \quad 1 \ll \tau \ll \tau_A \\
\tau_A^{1/3} (\frac{\tau}{\tau_A})^{1/2}  \omega \quad , \quad \tau \gg \tau_A  \end{cases}
\ee 
where $\chi_2$ is GUE-TW distributed, and $\omega$ a unit Gaussian. 

Transporting these results to the original problem one finds for the free energy fluctuations, in the case where $t_1 \geq t_0 \gg 1$
\be \label{gen2} 
\delta h(0,t) \simeq \begin{cases}  \left( \frac{t_0 t_1}{t_1-t_0} \log(\frac{1+ \frac{t}{t_0}}{1+ \frac{t}{t_1}} ) \right)^{1/3} \chi_2 ~~~~ \quad , \quad \frac{t_1-t_0}{t_1 t_0} \ll \log(\frac{1+ \frac{t}{t_0}}{1+ \frac{t}{t_1}}) \ll 1 \\
\left( \frac{t_0 t_1}{t_1-t_0} \right)^{1/3}
\left(\log(\frac{1+ \frac{t}{t_0}}{1+ \frac{t}{t_1}} ) \right)^{1/2} 
 \omega \quad , \quad \log(\frac{1+ \frac{t}{t_0}}{1+ \frac{t}{t_1}}) \gg 1   \end{cases}
\ee 
For the case $t_1=+\infty$ one recovers the result given in the main text (Eq. 19)
\be
\delta h(0,t) \simeq \begin{cases}  t^{1/3} \chi_2 ~~~~ \quad , \quad 1 \ll t \ll t_0  \\
t_0^{1/3}
(\log \frac{t}{t_0})^{1/2} 
 \omega \quad , \quad t \gg t_0   \end{cases}
\ee 
In the case $t_1=t_0$, i.e. $\alpha=1$ and droplet IC, only the first regime exists (since $A=0$ and $\tau_A=+\infty$) and
\be
\delta h(0,t) \simeq  \left(\frac{t t_0}{t+t_0}\right)^{1/3} \chi_2 ~~~~ \quad , \quad 1 \ll t 
\ee
The result \eqref{gen2} interpolates between these cases, the interpolation parameter being $t_1/t_0$.
Note that for the second, purely gaussian, regime to exist one needs $\log(t_1/t_0) \gg 1$. If not
the PDF saturates at some non-universal value.\\

The study of the transverse wandering of the DP is a bit more delicate. Let us focus on
the model $t_1=+\infty$, i.e. $c(t)=1/\sqrt{1+ \frac{t}{t_0}}$, studied in the main text.
We study the case $t_0 \gg 1$. Going back to the height $H(y,\tau)$, i.e. the DP in quadratic well $-A y^2/2$, 
let us separate the height from its mean profile, using the result from the STS symmetry 
(applying \eqref{sts3} to the problem for $H(y,\tau)$). One defines
\be \label{sts10} 
H(y,\tau) = \tilde H(y,\tau) - \frac{y^2}{8 t_0} \coth\left(\frac{\tau}{2 t_0}\right) = \tilde H(y,\tau) 
-  \frac{x^2}{8 (t+t_0)} \left(1+ \frac{2 t_0}{t}\right) 
\ee
such that the one point PDF of $\tilde H(y,\tau)$ is independent of $y$. We recall that $\tau=t_0 \log(1+ \frac{t}{t_0})$
and that $y=x/\sqrt{1+ \frac{t}{t_0}}$. It implies for the original problem that
\be
h(x,t) = H(y,\tau) - \frac{x^2}{8(t+t_0)} = \tilde H(y,\tau) - \frac{x^2}{4 t} 
\ee 
The fact that the quadratic terms add up to $-x^2/(4 t)$ is a consequence of the STS symmetry, i.e. since $a(t)=0$,
the profile is the usual parabola independently of $c(t)$, see \eqref{sts0}. We know little about the process $\tilde H(y,\tau)$, but we can heuristically assume that
\bea
\tilde H(y,\tau) &\simeq& \tau^{1/3} {\cal A}_2(\frac{y}{2 \tau^{2/3}}) \quad , \quad 1 \ll \tau \ll t_0 \\
&\simeq& t_0^{1/3} \left(\frac{\tau}{t_0}\right)^{1/2} \omega +
t_0^{1/3} {\cal A}\left(\frac{y}{2 t_0^{2/3}}\right) \quad , \quad \tau \gg t_0
\eea 
where ${\cal A}(z)$ is a $O(1)$ unknown process, with a one-point PDF independent of $z$.
The picture is that for $\tau \gg t_0$ the DP is bounded by the quadratic well and only the last independent
segment of length $\propto t_0$ contributes to the spatial fluctuations. This is reasonable as it reproduces
the two limits in \eqref{y2}, and for $\tau \gg t_0$, using the large $\tau$ limit of \eqref{sts10},
it expresses the
PDF $P^A_\xi(y,\tau)$ of the endpoint $y$ for the DP in the quadratic well $-A y^2/2$ in a given sample as
\be \label{y22} 
P^A_\xi(y,\tau) \propto 
e^{t_0^{1/3} {\cal A}\left(\frac{y}{2 t_0^{2/3}}\right) - \frac{y^2}{8 t_0} } = e^{
t_0^{1/3} [{\cal A}(\tilde y) - \frac{1}{2} \tilde y^2] } \quad , \quad \tilde y=y/(2 t_0^{2/3})
\ee 
Since $t_0\gg1$, the mean endpoint PDF $P_A=\overline{P_\xi^A}$ identifies with the distribution of the arg-max of the term in the exponential, its variance being related to the prefactor $y_2$ in \eqref{y2}.

Going back to the original problem we see that: 
\begin{itemize}
	\item In the first regime $1 \ll t \ll t_0$, $\tau \simeq t$ and $c(t) \simeq 1$ and 
	$h(x,t) \simeq t^{1/3} [ {\cal A}_2(\hat x) - \hat x^2 ]$ with $\hat x=\frac{x}{2 t^{2/3}}$ leading to the
	endpoint distribution of the standard DP problem. 
	\item In the second regime $t \gg t_0 \gg 1$ we obtain
	\be
	h(x,t) \simeq t_0^{1/3} (\frac{t}{t_0})^{1/2} \omega +
	t_0^{1/3} [{\cal A}(\tilde x) - \tilde x^2]  \quad , \quad  \tilde x= \frac{x}{2 t_0^{1/6} t^{1/2}} 
	\ee 
	and the endpoint PDF can be written as $P(x,t) \propto e^{t_0^{1/3} [{\cal A}(\tilde x) - \tilde x^2] }$ leading for $t_0 \gg 1$ to a related, but slightly different, maximization problem from \eqref{y22}.
	This shows that the DP wandering length obeys $\delta x(t)^2 < c(t)^2 \langle y^2 \rangle$, but that both sides are of the same order, $\propto t_0^{4/3} t$, i.e. diffusive, as indicated in the main text. 
\end{itemize}

\section{II Inhomogeneous discrete model}

\subsection{1) Preliminaries on the gamma distribution} 

 Before analyzing the directed polymer model with inverse gamma weights discussed in the letter, we gather here some useful facts about the (inverse) gamma distribution. Let $w$ be a random variable with inverse gamma distribution of parameter $\gamma$, i.e. the random variable with  PDF 
 \begin{equation}
P(w) = \frac{1}{\Gamma(\gamma)} w^{-\gamma-1} e^{-1/w}.
 \end{equation}
 The moments of $w$ are given by $\mathbb E [w^n] = \Gamma(\gamma-n)/\Gamma(\gamma)$, hence 
the mean and variance of $G$ are given by
\begin{eqnarray}
\mathbb E [ w] &=& \frac{1}{\gamma-1} = \frac{1}{\gamma} + \frac{1}{\gamma^2}  +o(1/\gamma^2), \text{ for }\gamma>1, \\ \mathrm{Var}[ w] &=& \frac{1}{(\gamma-2) (\gamma-1)^2}  = \frac{1}{\gamma^3} + o(1/\gamma^3), \text{ for }\gamma>2,
\label{eq:meamvarianceinversegamma}
\end{eqnarray} 
where the approximations hold for large $\gamma$.

In the study of directed polymer models, one is often led to consider not only the distribution of Boltzmann weights but also on site energies. Let us define $E$ such that  $w=e^{E/\theta}$, so that here the PDF of the on site energy is thus $P(E) = \frac{1}{\theta \Gamma(\gamma)} 
e^{-\gamma E/\theta} e^{- e^{-E/\theta} }$. If we scale $\gamma = \theta \tilde \gamma$,  in the zero temperature  limit $\theta \to 0$, $P(E)$ converges to
$P(E)=\tilde \gamma e^{- \tilde \gamma E} \mathds{1}_{E>0}$, i.e. the PDF of an exponential random variable of parameter $\tilde\gamma$. We will adress below in which sense this limit corresponds to a zero temperature limit. For the moment, we simply remark that although $\gamma$ (or $\theta$) can be physically
interpreted as a temperature for $\gamma$ close to $0$, the relation between $\gamma$ and the physical temperature is more complicated in general. In particular, in Section II 6) we will explain that $\gamma$ should be interpreted as the square of the temperature $T^2$ as $\gamma\to\infty$. 
\bigskip 

We will also need the cumulants of on site-energies $E = \log w$ (where $w$ is still an inverse gamma random variable of parameter $\gamma$, and we have set $\theta=1$ for simplicity). A direct computation shows that 
\begin{equation}
\mathbb E\left[e^{u E}\right] = \mathbb E \left[ w^u \right] = \frac{\Gamma(\gamma-u)}{\Gamma(\gamma)}.
\end{equation} 
This implies that the cumulants of $E$ (in the sequel of the paper, we will use indifferently the notations $\kappa_n(X)$ or $\langle X^n\rangle_c$ to denote the $n$-th cumulant of a random variable $X$) are given by 
\begin{equation}
	\kappa_n(E) = \partial_{u}^n \log \mathbb E\left[ e^{u E} \right] \Big\vert_{u=0} = (-1)^n\psi^{(n-1)}(\gamma),
	\label{eq:loggammacumulants}
\end{equation}
where $\psi$ is the digamma function \eqref{eq:defdigamma}, and in particular, $\mathbb E \left[ E \right] = - \psi(\gamma) $ and $\mathrm{Var} \left[ E \right] = \psi'(\gamma)$.  Thus, for large $\gamma$, we have the approximations 
\begin{equation}
\mathbb E\left[ E \right] \simeq -\log \gamma, \;\;\; \mathrm{Var}  \left[ E \right] \simeq \frac{1}{\gamma}, \;\;\; \kappa_3(E) = \frac{1}{\gamma^2}, \;\;\; \kappa_4(E) \simeq \frac{2}{\gamma^3},
\end{equation} 
while for $\gamma\to 0$, we have 
\begin{equation}
\mathbb E\left[ E \right] \simeq \frac{1}{\gamma}  , \;\;\; \mathrm{Var}  \left[ E \right] \simeq \frac{1}{\gamma^2}, \;\;\; \kappa_3(E) = \frac{2}{\gamma^3}, \;\;\; \kappa_4(E) \simeq \frac{6}{\gamma^4}.
\end{equation}

\subsection{2) Fredholm determinant formula}

The three next sections are based on the following result from \cite[Corollary 1.8]{borodin2013log}. 
Fix $n\geqslant m\geqslant 1$ and real parameters $\alpha_i, \beta_j$ such that $\alpha_i+\beta_j>0$ for all $i,j$. For $u\in \mathbb C$ such that $\Re[u]>0$,  
\begin{equation}
\mathbb E[e^{-u Z(n,m)}]  = \det(I+K)_{\mathbb{L}^2(\mathcal{C})},
\label{eq:fredholmdetloggamma}
\end{equation}
where $\mathcal{C}$ is a positively oriented closed contour enclosing the set of $-\beta_j$ and no other singularity,  and 
\begin{equation}
K(v,v')  =  \int_{\delta-\I\infty}^{\delta+\I\infty} \frac{\mathrm{d}z}{2\I\pi}
\frac{\pi}{\sin(\pi(v-z))}\frac{1}{z-v'}\frac{H(z)}{H(v)}
\label{eq:kernelLogGamma}
\end{equation}
where 
\begin{equation}
H(z) =  u^z \frac{\prod_{i=1}^{n}\Gamma(\alpha_i-z)}{\prod_{j=1}^m\Gamma(z+\beta_j)}.
\end{equation} 
and the integration contour $\delta +i\mathbb R$ is such that  $\delta<\alpha_j$, the contour $\mathcal C$ lies to the left of $\delta +i\mathbb R$ and all the poles at $z= v+1, v+2, \dots$ lie to the right of $\delta +i\mathbb R$. In order to match the above result with \cite[Corollary 1.8]{borodin2013log}, we have simply set $\beta_j = -a_j$ ($\lbrace a_j \rbrace $ is a set of parameters used in \cite{borodin2013log}). Note that \cite{borodin2013log} assumes that $a_j\geqslant 0$ but this is unnecessary, the formula can be analytically continued to negative $a_j$ as long as $\alpha_i-a_j = \alpha_i+\beta_j>0$ for all $1\leqslant i \leqslant n$ and $1\leqslant j\leqslant m$.

We will focus on the case $n=m$ (though the asymptotics  when $n/m$ is an arbitrary constant are very similar). Asymptotic analysis of \eqref{eq:fredholmdetloggamma} in the homogeneous case, i.e $a=0$, have been performed in a number of works \cite{borodin2013log, borodin2015height, thiery2014log, krishnan2018tracy}, by Laplace's method. We adapt the same approach to the inhomogeneous case. It should be noted that the following asymptotic results do not constitute mathematical theorems. Mathematical proofs would require performing a more careful analysis of the function $H$ along the tails of the contours, and proving a number of estimates to justify the convergence. Such justifications can be found for instance in \cite{borodin2013log, borodin2015height,  krishnan2018tracy} in a similar context. Making all these justifications in the present context (inhomogeneous weights) constitutes a mathematical challenge, but from the physical point of view it does not seem necessary, and thus we will proceed via saddle point analysis without attempting to prove rigorously a theorem.  

\subsection{3) Asymptotic analysis}

\textbf{Analysis of the amplitude $\sigma_n$. }
Let us fix $\temp>0$ and set $\alpha_i=\beta_i=\temp i^a$. For that particular choice of parameters, we set $G(z) = \log H(z)$ in accordance with \eqref{eq:defG}. Recall the definition of the digamma function \begin{equation}
\psi(x)=\tfrac{d}{dx}\log(\Gamma(x))
\label{eq:defdigamma}
\end{equation}
Since $G''(0) = G^{(4)}(0) = 0$, we have by Taylor expansion, 
\begin{equation}
G(z)  = z\log(u) + z \frac{f_n}{\temp} + z^3 \frac{\sigma_n^3 }{3\temp^3} + o(z^4),
\label{eq:TaylorG}
\end{equation}
where
\begin{equation}
f_n= -2 \temp \sum_{i=1}^n \psi(\temp  i^a), \quad \sigma_n^3 = \temp^3 \sum_{i=1}^n -\psi''(\temp i^a). 
\end{equation}
We claim that the criterium to determine if we will observe Tracy-Widom fluctuations or a stabilization of the free energy (i.e. fluctuations of the free energy on the constant scale according to a non-universal distribution) is whether the quantity $\sigma_n$ 
diverges to infinity or not.  In principle, the true criterium should be that the term of order 3 in the Taylor expansion is dominant with respect to the remainder. This will happen if and only if $\sigma_n$ diverges. 
Assume for the moment that $\temp$ is fixed. The digamma function satisfies the asymptotics $ \psi(x) \simeq \frac{-1}{x^2}$ as  $x$ goes to $+\infty$. Thus, for $0\leqslant a < 1/2$, 
\begin{equation}
\sigma_n^3  \simeq \sum_{i=1}^n \frac{\temp}{i^{2a}} \simeq \frac{Tn^{1-2a}}{1-2a},\quad n\to\infty.
\end{equation} 
For $a>1/2$, $\sigma_n$ converges to a constant, and for $a=1/2$, $\sigma_n^3  \simeq \temp \log n$. 

\bigskip 

\textbf{Asymptotic analysis when $a>1/2$}
Assume  $a>1/2$ and $\temp>0$ is fixed. We will first show that  the kernel \eqref{eq:kernelLogGamma} converges to some kernel $\tilde K$. 
Let $\tilde u  = u e^{-2\sum_{i=1}^n  \psi(\temp i^a)} $. Then,
\begin{equation}
G(z) = z\log(\tilde u) +  \sum_{i=1}^n \log\Gamma(\temp i^a-z) - \log\Gamma(\temp i^a+z) + 2 z  \psi(\temp i^a).
\end{equation} 
Using the series expansions
\begin{equation}
\log \Gamma(z) = -\gamma z -\log z + \sum_{k=1}^{\infty} \frac{z}{k} -\log\left( 1+\frac{z}{k}\right)
\label{eq:seriesloggamma}
\end{equation}
and
\begin{equation}
\psi(z) = -\gamma -\frac 1 z +\sum_{k=1}^{\infty} \frac{1}{k} -\frac{1}{z+k}, 
\label{eq:seriespsi}
\end{equation}
we obtain that 
\begin{equation}
\lim_{n\to \infty}  G(z) -z\log(\tilde u)  =  \sum_{i=1}^{\infty} \sum_{k=0}^{\infty} \log\left(1+\frac{z}{\temp i^a+k} \right) - \log\left(1-\frac{z}{\temp i^a+k} \right) - \frac{2z}{\temp  i^a+k}=:G_{\infty}(z).
\label{eq:limitGn1}
\end{equation}	
Note that this double sum is well-defined because 
\begin{equation}
 \sum_{i=1}^{\infty} \sum_{k=0}^{\infty} \frac{1}{(\temp  i^a+k)^3} < \infty.
\end{equation}
Using the exponential growth of the sine function towards $\I\infty$, and the boundedness of $\Re [G_{\infty}(z)]$ on the contour $\I\mathbb R$, we deduce by dominated convergence that $K(v,v')$ converges as $n$ goes to infinity to $K_{\infty}(v,v')$ where
\begin{equation} 
 K_{\infty}(v,v')  =  \int_{-\I\infty}^{\I\infty} \frac{\mathrm{d}z}{2\I\pi}
\frac{\pi}{\sin(\pi(v-z))}\frac{1}{z-v'}\tilde u^{z-v}e^{G_{\infty}(z)-G_{\infty}(v)}.
\label{eq:nonuniversalkernel}
\end{equation}

The kernel $K$ (depending on $n$) is acting on $\mathbb L^2(\mathcal C)$, where the contour $\mathcal C$ is a positively oriented contour containing all points $-\temp i^a$ for $1\leqslant i\leqslant n$.  We may deform $\mathcal C$ to some infinite contour $\mathcal C_{\infty}$,  independent from $n$, which contains all  points $-\temp i^a$ for $i\geqslant 1$ and we have $\det(I+K)_{\mathbb L^2(\mathcal C)} =  \det(I+K)_{\mathbb L^2(\mathcal C_{\infty})}$. Applying dominated convergence to the Fredholm determinant expansion, we deduce the convergence of Fredholm determinants from the convergence of kernels, that is we arrive at 
 \begin{equation}
\det(I+K)_{\mathbb L^2(\mathcal C)} \xrightarrow[n\to\infty]{}\det(I+K_{\infty})_{\mathbb L^2(\mathcal C_{\infty})}.
 \end{equation}
 A rigorous mathematical justification would require some bounds on the kernel $K(v,v')$ valid  along the contour $\mathcal C_{\infty}$ towards infinity, we assume without justification that such bounds hold. 
 Finally, we conclude that the random variable $\tilde{\mathcal{Z}} = \mathcal Z(n,n) e^{2\sum_{i=1}^n  \psi(\temp i^a)}$ weakly converges to some probability distribution characterized by its Laplace transform $ \mathbb E[e^{-\tilde u \tilde {\mathcal Z}}] = \det(I+ K_{\infty})_{\mathbb L^2(\mathcal C_{\infty})}$. Note that in order to check that the limit is indeed a probability distribution, i.e. no mass has been  lost in the limit, it is enough to check that  $\det(I+K_{\infty})$ goes to $1$ as $u$ goes to $0$, which is readily verified.

\bigskip 

\textbf{Asymptotic analysis when $0\leqslant a\leqslant 1/2$ }
Assume now that $0\leqslant a\leqslant 1/2$ and $\temp>0$ is fixed. Let $ u  =  e^{2\sum_{i=1}^n  \psi(\temp i^a)- r \sigma_n/\temp}$. Then, since $\sigma_n$ goes to infinity, 
\begin{equation}
\mathbb E\left[e^{-u \mathcal Z}\right] \simeq \mathbb P\left( \frac{\log \mathcal Z(n,n) + 2 \sum_{i=1}^n  \psi(\temp i^a) }{\sigma_n/\temp} \leqslant r \right),
\end{equation}
as $n$ goes to infinity, provided the left-hand-side converges to some probability distribution function (see e.g. \cite[Lemma 4.1.39]{borodin2014macdonald}).  

We may analyze the Fredholm determinant $\det(I+K)_{\mathbb L^2(\mathcal C)}$ by Laplace's method.  Let us define  $\mathcal C_a^{\varphi}$ to be  an infinite contour in the complex plane going straight from $\infty e^{-\I\varphi}$ to $a$ and then to $\infty e^{\I\varphi}$.  Using \eqref{eq:TaylorG} and rescaling variables near $0$ by a factor $\sigma_n/\temp$, we see that $\det(I+K)_{\mathbb L^2(\mathcal C)}$ converges to $\det(I+K^{\rm GUE})_{\mathbb L^2(\mathcal C_0^{2\pi/3})}$  
where 
\begin{equation}
K^{\rm GUE}(v,v')  =  \int_{\mathcal C_1^{\pi/3}} \frac{\mathrm{d}z}{2\I\pi}
\frac{1}{v-z} \frac{1}{z-v'}\exp\left(\frac{z^3}{3} - r z -\frac{v^3}{3}+  r v \right).
\end{equation}
Note that the contours may be deformed to vertical lines as in the letter \eqref{eq:kernelGUEletter}. We recognize a well-known kernel such that $\det(I+K^{\rm GUE}) = F_{\rm GUE}(r)$, the CDF of the Tracy-Widom GUE distribution. Indeed, using $\frac{-1}{v-z} = \int_0^{\infty} \exp(\lambda (v-z))\mathrm d \lambda$ for $\Re[v-z]<0$, we may factorize the kernel $K^{\rm GUE}$ as $K^{\rm GUE} = - A B$ where 
\begin{equation}
A(v,\lambda) = \exp\left( \frac{-v^3}{3} +(r+\lambda)v \right), \quad B(\lambda' , v') =  \int_{\mathcal C_1^{\pi/3}} \frac{\mathrm{d}z}{2\I\pi}
 \frac{1}{z-v'}\exp\left(\frac{z^3}{3} - (r+\lambda' )z \right).
\end{equation}
Using the identity $\det(I-AB) = \det(I-BA)$, we find that $\det(I+K^{\rm GUE}) = \det(I-K_{\rm Ai})_{\mathbb L^2(r, \infty)}$, where $K_{\rm Ai}$ is the Airy kernel 
\begin{equation}
K_{\rm Ai}(\lambda, \lambda') = \int_{\mathcal C_0^{2\pi/3}} \frac{\mathrm{d}v}{2\I\pi}\int_{\mathcal C_1^{\pi/3}} \frac{\mathrm{d}z}{2\I\pi}  \frac{1}{z-v}\exp\left(\frac{z^3}{3} - \lambda  z -\frac{v^3}{3}+  \lambda' v \right).
\end{equation}

 Thus,
\begin{equation}
\lim_{n\to \infty} \mathbb P\left( \frac{\log \mathcal Z(n,n) + 2 \sum_{i=1}^n  \psi(\temp i^a) }{\sigma_n/\temp} \leqslant r \right) = \det(I+K^{\rm GUE}) = F_{\rm GUE}(r).
\end{equation}
Recall that  $\sigma_n^3 \simeq \frac{\temp}{ (1-2 a)} n^{1-2 a}$,
hence defining (minus) the free energy as $\mathcal F_n = \temp\log(\mathcal Z(n,n))$, the free energy fluctuations at large $n$ are \begin{equation}
\delta \mathcal F_n \simeq
\frac{\temp^{1/3}}{ (1-2 a)^{1/3}} n^{\frac{1-2 a}{3}} \chi_2.
\end{equation}

As we have already mentioned, we omit here the mathematical details to prove the convergence of the Fredholm determinant. Let us simply observe that it is reasonable to replace $G(z)$ by the first terms in its Taylor expansion: indeed  $G^{(5)}(0) = \mathcal O (n^{1-4a} +cst)$  which is negligible compared to $\sigma_n^3$. 

In the case $a=1/2$, $\sigma_n^3$ still diverges to $+\infty$ but slowly, in the scale $\log(n)$. The limit theorem still holds. 

\subsection{4) Zero-temperature limit}

In the log-gamma polymer, weights $w$ are distributed as inverse gamma random variables. Recall that  if we write Boltzmann weights as  $w = e^{E/\theta}$, where $w$ is an inverse gamma random variable of parameter $\gamma$, then scaling $\gamma = \theta\tilde \gamma$, the variable $E$ converges as $\theta$ goes to zero to an exponential random variable of parameter $\tilde \gamma$.
We now study this limit for the inhomogeneous model.

\bigskip 

Fix some $n\geqslant 1$. Consider the log-gamma polymer  model with weights with parameter $\gamma_{i,j} = \temp(a_i+b_j)$. When $\temp$ goes to zero, $\mathcal F_n = \temp\log(\mathcal Z(n,n))$ weakly converges to $L(n,n)$, the last passage time from $(1,1)$ to $(n,n)$ in a model with energies distributed as exponential random variables. More precisely, 
\begin{equation}
L(n,n) = \max_{\pi: (1,1)\to (n,n)} \sum_{(i,j)\in \pi} E_{i,j}
\label{eq:defLnn}
\end{equation} 
where $E_{i,j}$ are independent exponential random variables with parameter $a_i+b_j$. This is the model studied in \cite{johansson2008some}, which considers in particular the case $a_i=b_j=i^a$. 

Let us scale $u$ in \eqref{eq:kernelLogGamma} as $u=e^{-s\temp^{-1}}$. Then $\mathbb E[e^{-uZ}]$ converges to $\mathbb P(L(n,m) \leqslant s)$. It can be shown that the Fredholm determinant \eqref{eq:fredholmdetloggamma} converges as well so that 
\begin{equation}
\mathbb P(L(n,m) \leqslant s)  = \det(I+K^{\rm LPP})_{\mathbb L^2(\mathcal C)},
\end{equation}
where 
\begin{equation}
K^{\rm LPP}(v,v')  =  \int_{\delta-\I\infty}^{\delta+\I\infty} \frac{\mathrm{d}z}{2\I\pi} \frac{1}{v-z}\frac{1}{z-v'} e^{-s(z-v)}
\prod_{i=1}^n \frac{i^{a}-v}{i^a-z} \prod_{j=1}^m \frac{j^a+z}{j^a+v}.
\end{equation}
The contour $\mathcal C$ encloses all poles at $-j^a$ for all $j \geq 1$ and $\delta$ is chosen so that $\delta+\I\mathbb R$ passes to the right of the contour $\mathcal C$. Again, let us write $\frac{1}{v-z}   = -\int_0^{+\infty} e^{x(v-z)}dx$,
so that we can factorize the kernel as $K^{\rm LPP}=-AB$ with 
\begin{equation}
A(v,x) = e^{xv+sv } \frac{\prod_{i=1}^n i^a-v}{\prod_{j=1}^m j^a+v},\ \ \ B(x',v') = \int_{\delta-\I\infty}^{\delta+\I\infty} \frac{\mathrm{d}z}{2\I\pi} \frac{1}{z-v'} e^{-s z-x' z}
\frac{ \prod_{j=1}^m j^a+z}{\prod_{i=1}^n i^a-z}.
\end{equation} 
Using the identity  $\det(I-AB) = \det(I-BA)$ for Hilbert-Schmidt kernels, we may write $ \mathbb P(L(n,m) \leqslant s)  = \det(I-\tilde K^{\rm LPP})_{\mathbb L^2(s, \infty)}$ where
\begin{equation}
\tilde K^{\rm LPP}(x,y)  =  \int_{\mathcal C} \frac{\mathrm{d}v}{2\I\pi} \int_{\delta-\I\infty}^{\delta+\I\infty} \frac{\mathrm{d}z}{2\I\pi} \frac{1}{z-v} e^{-xz+yv}
\prod_{i=1}^n \frac{i^{a}-v}{i^a-z} \prod_{j=1}^m \frac{j^a+z}{j^a+v}.
\label{eq:defKtildeLPP} 
\end{equation}
Note that $\tilde K^{\rm LPP}$ is the same kernel as in \cite[Eq. 1.11]{johansson2008some}. We may deform the contours in \eqref{eq:defKtildeLPP} so that the contour for the variable $z$ becomes $\mathcal C_{1/2}^{\pi/4}$ and the contour for the variable $v$ becomes $\mathcal C_{-1/2}^{3\pi/4}$. The angles chosen do not matter much as long as the real part is increasing (resp. decreasing) along the tails of the $z$ contour (resp. $v$ contour). It was shown in \cite{johansson2008some} that the large time asymptotics of $L(n,n)$ depend on the value of $a$. Let 
$c_n=2 \sum_{i=i}^n i^{-a}$. For $a\in (1/3,1)$, fluctuations of $L(n,n)$ are of order $1$, and their distribution is characterized by the kernel 
\begin{equation}
\lim_{n\to \infty} \tilde K^{\rm LPP}(x+c_n, y+c_n) = \int_{\mathcal C_{-1/2}^{3\pi/4}} \frac{\mathrm{d}v}{2\I\pi} \int_{\mathcal C_{1/2}^{\pi/4}} \frac{\mathrm{d}z}{2\I\pi} \frac{e^{-xv+yz}}{v-z} \frac{e^{F_{\temp\to 0}(v)}}{e^{F_{\temp\to 0}(z)}},
\label{eq:limitzerotemperaturealarge}
\end{equation}
where 
\begin{equation}
F_{\temp\to 0}(z) =  \sum_{k=1}^{\infty} \log \left(1+\frac{z}{k^a} \right) - \log \left(1-\frac{z}{k^a} \right) -\frac{2z}{k^a} .
\label{eq:zerotemperaturefunction}
\end{equation}
We recover the cases a) and b) of \cite[Theorem 1.1]{johansson2008some}. 
Note that in that work it is proved that the spatial behavior is different according to $1/3 < a <1/2$ (non-trivial extended kernel) 
and $1/2 <a < 1$ (trivial extended kernel).

For $a \in [0,1/3]$, the kernel $\tilde K^{\rm LPP}$ converges to the Airy kernel is the sense that 
\begin{equation}
\lim_{n\to \infty} d_n \tilde K^{\rm LPP}(c_n+ d_n x, c_n+d_n y)  = \int_{\mathcal C_{-1}^{2\pi/3}} \frac{\mathrm{d}v}{2\I\pi}\int_{\mathcal C_1^{\pi/3}} \frac{\mathrm{d}z}{2\I\pi}
\frac{1}{z-v}\exp\left(\frac{z^3}{3} - x z -\frac{v^3}{3}+  y v \right),
\end{equation}
where 
$d_n = (2\log n)^{1/3}$ when $a=1/3$ and $d_n=\left(\frac{2n^{1-3a}}{1-3a} \right)^{1/3}$ otherwise. This means that for $a \in [0,1/3]$, 
$L(n,n) \simeq d_n \chi_2$ where $\chi_2$ follows the Tracy-Widom GUE distribution.

\subsection{5) Low temperature crossover}
\label{sec:zerotemperature}

In this Section, we study the case where the parameter $\temp$ goes to zero simultaneously as $n$ goes to infinity. Let us scale $u$ as $u=e^{-s/\temp}$. Then $\mathbb E[e^{-uZ}]$ can be approximated by  $\mathbb P(\temp \log(Z(n,n)) \leqslant s)$ with an error of order $\mathcal O(e^{-\temp^{-1}})$. It is convenient to rescale variables in the kernel \eqref{eq:kernelavecfonctionG} so that $ \mathbb E[e^{-uZ}] = \det(I+K),$ where 
\begin{equation}
K(v,v')  =  \int_{\delta-\I\infty}^{\delta+\I\infty} \frac{\mathrm{d}z}{2\I\pi}
\frac{\temp\pi}{\sin(\temp\pi(v-z))}\frac{1}{z-v'}e^{F_n(z)-F_n(v)}
\end{equation}
with
\begin{equation}
 F_n(z) = - z s + \sum_{i=1}^n \log\Gamma(\temp i^a-\temp z) - \log\Gamma(\temp i^a+\temp z).
\end{equation}
We have already seen that by Taylor approximation, 
\begin{equation}
 F_n(z) = -z s -z f_n  - \frac{ z^3}{3} \sigma_n^3 + \mathcal O(\temp^5z^5).
\end{equation}

Let $a\in(1/3,2/3)$. We know from the previous results that if $\temp$ goes to zero sufficiently fast, we should expect the free energy to behave as in the zero-temperature model, that is, converge to a non-universal distribution.  If, however,  the temperature goes to $0$ slowly enough, we expect the free energy to behave still as if $\temp$ was fixed and thus have fluctuations following the Tracy-Widom distribution. We will see that the threshold arises for $\temp$ of order  $\mathcal O(n^{-1+2a})$.

Let us scale $\temp$ as $\temp=A n^{-c}$. We need to determine for which range of $c$ the quantity $\sigma_n$ converges or diverges. For $c\geqslant a$, it is not difficult to show that $\sigma_n$ converges to a constant, using the asymptotics of the digamma function ($ \psi''(x) \simeq - 2/x^3 $ as $x$ goes to zero).
Consider now $c<a$. We decompose the series as a sum $\sigma_n^3=S_1+S_2$ as 
\begin{equation}
\sigma_n^3  = \underbrace{ \sum_{i=1}^{n^{c/a}}  - \temp^3  \psi''\left(A \left(\frac{i}{n^{c/a}}\right)^a\right)}_{(S_1)} + \underbrace{ \sum_{i=n^{c/a}+1}^n  - \temp^3 \psi''(A i^a n^{-c})}_{(S_2)}. 
\end{equation} 
Again, using approximation of the digamma function near $0$ and  Taylor-Maclaurin formula, one readily obtains that the first sum $S_1$ converges to  a constant. 
Since $\psi''(x)\simeq \frac{-1}{x^2}$ as $x$ goes to infinity, the sum $S_2$ is divergent only if 
\begin{equation}
 \sum_{i=n^{c/a}+1}^n   \frac{\temp^3}{( A i^a n^{-c})^2}  =   \sum_{i=n^{c/a}+1}^n  
\frac{ A n^{-c}}{ i^{2a}} \simeq  \frac{ A n^{-c}}{1-2a}n^{1-2a} 
\end{equation}
is divergent (recall that $a<1/2$).
Thus, the sum $S_2$, and consequently  $\sigma_n$ as well, is divergent only if $c<1-2a$, in which case 
\begin{equation}
\sigma_n^3 \simeq \frac{A}{1-2 a} n^{1-2 a -c}.
\end{equation} 
Otherwise, when $c\geqslant 1-2a$, $\sigma_n$ converges to a constant. We may now adapt the asymptotic analysis performed above in the finite temperature case. 
For $a\in (1/3, 1/2)$ and $\temp=A n^{-c}$, we have the following. 
\begin{itemize}
	\item If $c<1-2a$, then  
	\begin{equation}
	\lim_{n\to \infty} \mathbb P\left( \frac{\mathcal F_n -f_n}{\sigma_n } \leqslant s\right) =F_{\rm GUE}(s).
	\end{equation}
	Hence the free energy $\mathcal F_n$ fluctuates as
	\begin{equation}
	\delta \mathcal F_n  \simeq \frac{ A^{1/3}}{(1-2 a)^{1/3}} n^{\frac{1-2 a -c}{3}} \chi_2. 
	\end{equation} 
	\item If $c>1-2 a$, then 
	\begin{equation}
	 \lim_{n\to \infty} \mathbb P\left(\mathcal F_n -f_n \leqslant s \right)  = \det(I+K^{\temp\to 0}),
	\end{equation}
	where 
	\begin{equation}
	 K^{\temp\to 0}(v,v')  = \int_{\delta-\I\infty}^{\delta+\I\infty} \frac{\mathrm{d}z}{2\I\pi} \frac{1}{v-z}\frac{1}{z-v'}
	e^{-s(z-v)} e^{F_{\temp\to 0}(z)-F_{\temp\to 0}(v)} 
	\end{equation}
	with 
	\begin{align}
	F_{\temp\to 0}(z) &=  \lim_{n\to \infty}  \sum_{i=1}^n \log\Gamma(\temp i^a-\temp z) - \log\Gamma( \temp i^a+ \temp z) + 2 z \temp  \psi(\temp i^a),\\
	&= \sum_{i=1}^{\infty} \log\left(1+\frac{z}{i^a}\right) - \log\left(1-\frac{z}{i^a}\right) -\frac{2z}{i^a}.
	\label{eq:Finfinity}
	\end{align}
	Indeed, using the series representations \eqref{eq:seriesloggamma} and \eqref{eq:seriespsi}, 
	\begin{equation}
	\sum_{i=1}^n \log\Gamma(\temp i^a-\temp z) - \log\Gamma(\temp i^a+\temp z) + 2 z \temp  \psi(\temp i^a) =
	\sum_{i=1}^n \sum_{k=0}^\infty \log\left( 1 + \frac{\temp z}{\temp  i^a +k} \right) + \log\left( 1 - \frac{\temp z}{\temp i^a +k} \right) -\frac{2 \temp z}{\temp  i^a +k},
	\end{equation}	
	and only the terms corresponding to $k=0$ remain in the limit. 
	We recover exactly \eqref{eq:limitzerotemperaturealarge} \eqref{eq:zerotemperaturefunction} which shows that the free energy fluctuations have the same distribution as in the zero temperature case. 
	
	\item If $c=1-2a$, we set $\temp =A n^{-1+2a}$. Again we have that 
	\begin{equation}
	 \lim_{n\to \infty} \mathbb P\left( \mathcal F_n - f_n \leqslant s \right)  = \det(I+K^{\rm cross})_{\mathbb{L}^2(\mathcal C_{-1/2}^{3\pi/4})},
	\end{equation}
	where 
	\begin{equation}
	 K^{\rm cross}(v,v')  = \int_{\delta-\I\infty}^{\delta+\I\infty} \frac{\mathrm{d}z}{2\I\pi} \frac{1}{v-z}\frac{1}{z-v'}
	e^{-s(z-v)} e^{F_{\rm cross}(z)-F_{\rm cross}(v)} 
	\end{equation}
	with 
	\begin{equation}
	F_{\rm cross}(z) =  \lim_{n\to \infty}  \sum_{i=1}^n \log\Gamma( \temp i^a- \temp z) - \log\Gamma( \temp i^a+ \temp z) + 2 z  \temp   \psi( \temp i^a).
	\end{equation}
	The function $F_{\rm cross}$ interpolates between the zero temperature case \eqref{eq:Finfinity} and  a cubic behaviour as in the Airy kernel. It depends on $A$ as
	\begin{equation}
	 F_{\rm cross}(z)  = \frac{ A z^3}{3(1-2a)} + F_{\temp\to 0}(z).
	\end{equation} 
	
	Indeed,
	\begin{multline}
	\sum_{i=1}^n \log\Gamma(\temp i^a-\temp z) - \log\Gamma(\temp i^a+\temp z) + 2 z \temp   \psi(\temp i^a) =\\
	\sum_{i=1}^n \sum_{k=0}^\infty \log\left( 1 + \frac{ A n^{-1+2a}z}{ A n^{-1+2a} i^a +k} \right) + \log\left( 1 - \frac{ A n^{-1+2a}z}{ A n^{-1+2a} i^a +k} \right) -\frac{2 A n^{-1+2a}z}{ A n^{-1+2a} i^a +k}.
	\label{eq:crossoverfunctionsum}
	\end{multline}	
	In order to determine the limit, consider separately the case $k=0$, for which the main contribution is given by terms corresponding to small $i$, and the terms corresponding to $k\geqslant 0$ for which the main contribution comes from large $i$. The term $k=0$ simplifies and yields $F_{\temp \to 0}(w)$. In the terms corresponding to $k\geqslant 1$, we may use the Taylor expansion of the logarithm and series expansion of the polygamma function so that 
	\begin{equation}
	 \eqref{eq:crossoverfunctionsum} = - 2\sum_{i=1}^n \sum_{j=1}^\infty \frac{z^{2j+1} ( A n^{-1+2a})^{2j+1}}{(2j+1)!} \psi^{(2j)}( A n^{-1+2a}i^a+1),
	\end{equation}   
	where we used the identity $\sum_{k=1}^{+\infty} \frac{1}{(k+x)^{2 j+1}} = -\frac{1}{(2 j)!} \psi^{(2j)}(1+x)$.
	Using the large $x$ asymptotics $\psi^{(2j)}(x) \simeq - (2j-1)!/x^{2j}$, it is equivalent to 
	\begin{equation}
	 \sum_{j=1}^{\infty} \sum_{i=1}^n A n^{-1+2a} \frac{z^{2j+1}}{j(2j+1)} \frac{1}{i^{2aj}}.
	\end{equation}
	We see that only the term corresponding to $j=1$ will contribute to the limit and this yields 
	\begin{equation}
	 \sum_{i=1}^n A n^{-1+2a} \frac{z^{2j+1}}{3}\frac{1}{i^{2a}} \xrightarrow[n\to \infty]{}   \frac{ A z^3}{3(1-2a)}.
	\end{equation}
\end{itemize}
	\textbf{Remark. } As in Section II 2), the Fredholm determinant  $\det(I+K^{\rm cross})$ can be written, using $\det(I+AB) = \det(I+BA)$ as  a Fredholm determinant $\det(I+\tilde K^{\rm cross})$ where the kernel $\tilde K^{\rm cross}$ acts on $\mathbb L^2(\mathbb R)$. The kernel  $\tilde K^{\rm cross}$ is a limit of the Schur process correlation kernel, which usually occurs in zero temperature models. More specifically, $\tilde K^{\rm cross}$ corresponds to a limit of the so called Airy kernel with two sets of parameters introduced in \cite[Remark 2]{borodin2008airy}.

\subsection{6) Discrete model with arbitrary weight distribution and KPZ scaling theory}

For a generic interface model in the KPZ univerality class, the interface height $h(x,t)$, starting from an initial condition in the droplet class, is expected under mild hypotheses to obey a limit theorem of the form \cite{spohn2012kpz}
\begin{equation}
 h(vt,t) \simeq t \phi(v) + \left(\tfrac 1 2 \lambda A^2 t \right)^{1/3} \chi_2,
 \label{eq:KPZlimittheorem}
\end{equation}
for large times, where the function $\phi(v)$ and the coefficients $\lambda, A$ are model-dependent and we provide their definition below. A necessary condition for this limit to hold is that the limit profile $\phi$ is curved at $v$.  In certain cases, these coefficients  can be computed explicitly.  We refer the reader to \cite{spohn2012kpz} and \cite{krug1992amplitude} for details about KPZ scaling theory. The aim of this section is to explain how the KPZ scaling theory needs to be modified in the time dependent inhomogeneous case. We start by recalling KPZ scaling theory for directed polymers in the homogeneous case. 

\bigskip 
\textbf{Homogeneous case. } Directed polymer models fit in the KPZ scaling theory framework. Consider the partition function of a polymer model $\mathcal Z(n,m)$ as defined in the main text of the letter \eqref{eq:defZ}. As in the text of the letter, it will be more convenient to work with space-time coordinates $\uptau = n+m$ and $\varkappa = n-m$ and we define $Z_d(\varkappa, \uptau) = \mathcal Z(n,m)$. 

In this context, the analogue of the interface height is the free energy, so that in this section, 
\begin{equation}
h(\varkappa, \uptau)  = \log Z_d(\varkappa, \uptau).
\label{eq:defh}
\end{equation}
We define the slope field associated to $h$ as  $u(\varkappa, \uptau) = \frac 1 2 \left(h(\varkappa+1, \uptau) - h(\varkappa-1, \uptau) \right)$.
Let us assume that translation invariant and stationary distributions of the slope field are known and parametrized by the density $\rho = \mathbb E [u(0,\uptau)]$, and let us denote the corresponding measure by $\mu_{\rho}$. For the log-gamma polymer with parameter $\gamma$, these stationary measures  \cite{seppalainen2012scaling} are parametrized by a real number $\vartheta \in (0,\gamma)$, such that under $\mu_{\rho}$, the slope field $u(\varkappa)$ is i.i.d. as $\varkappa$ varies, and distributed as
\begin{equation}
 \frac{1}{2} \left(  \log G(\gamma-\vartheta) -\log G(\vartheta) \right),
\end{equation}
where $G(\gamma-\vartheta)$ and $G(\vartheta)$ are independent gamma distributed random variables with parameters respectively $\gamma-\vartheta$ and $\vartheta$. Hence the density $\rho$ is related to the parameter $\vartheta$ via 
\begin{equation}
 \rho = \frac{1}{2} \left(\psi(\gamma-\vartheta) - \psi(\vartheta)  \right). 
\end{equation}
More precisely, the stationary measures introduced in \cite{seppalainen2012scaling} are such that increments $h(\uptau,\varkappa)-h(\uptau-1,\varkappa+1)$ are distributed as $-\log G(\gamma-\vartheta)$, and increments $h(\uptau, \varkappa)-h(\uptau, \varkappa-1)$ are distributed as $-\log G(\vartheta)$. For a fixed time $\uptau$, all these increments are independent as $\varkappa$ varies.

One also defines the instantaneous current $j(\rho)$, which equals the increment of $h(\varkappa, \tau)$ per unit of time under the stationary slope field $\mu_{\rho}$. For the log-gamma polymer with parameter $\gamma$, we have 
\begin{equation}
j(\rho(\vartheta)) = \frac{-1}{2} \mathbb E[\log \tilde G(\gamma-\vartheta)  +\log \tilde G(\vartheta)] = \frac{-1}{2}\left(\psi(\gamma-\vartheta) + \psi(\vartheta)  \right).
\label{eq:jloggamma}
\end{equation}
(Note that increments $-\log \tilde G(\gamma-\vartheta), -\log \tilde G(\vartheta)$ along the time direction are not independent, but this does not matter for the computations.)

In general, the function $\phi$ appearing in \eqref{eq:KPZlimittheorem} is the Legendre transform of the function $j(\rho)$, that is  \cite[Eq. (3.13)]{spohn2012kpz}
\begin{equation}
 \phi(v) = \inf_{\rho \in \mathbb R} \left\lbrace v \rho -j(\rho)\right\rbrace. 
\end{equation}
For the log-gamma polymer model, it can be computed explicitly, and we find that $\phi(0) = - \psi(\gamma/2)$. At velocity $v=0$, the density $\rho$ that optimizes the variational problem above is $\rho=0$ and the corresponding value of $\vartheta$ is $\vartheta=\gamma/2$. More generally, the relation between $v$ and $\rho$ is determined by $v=\partial j/\partial \rho$.

\bigskip 
Now we may explain the coefficients $\lambda$ and $A$ which appear in the magnitude of fluctuations in \eqref{eq:KPZlimittheorem}. We define the curvature of the limit shape $\lambda$ by  $\lambda \equiv \lambda(\rho) = j''(\rho)$. Implicitly, $\lambda$ depends on the velocity $v$ through the local density $\rho$ around the location $\varkappa=\uptau v$. The coefficient $A$ is the  integrated covariance of the slope field, 
\begin{equation}
A \equiv A(\rho) = 2 \left(\sum_{j\in \mathbb Z} \mathbb E_{\mu_{\rho}} \left[  u(0)u(j)\right]  - \rho^2 \right).
\label{eq:defA}
\end{equation}
For the log-gamma polymer model, the stationary slope field is i.i.d. in space so that 
\begin{equation}
 A(\rho) = 2\mathrm{Var} \left[ \frac{1}{2} \left( \log G(\gamma-\vartheta) -\log G(\vartheta) \right) \right]  = \frac{1}{2} \left( \psi'(\gamma-\vartheta) + \psi'(\vartheta) \right).
 \label{eq:Aloggamma}
\end{equation}
Note that our definition of $A$ in \eqref{eq:defA} differs from the definition given in Eq. (2.8) of \cite{spohn2012kpz} by a factor $2$. This is due to the fact that we work on the square lattice and strictly speaking, our height field $h(\varkappa, \uptau)$ is defined only when $\varkappa$ and $\uptau $ have the same parity. In any case, the coefficient $A$ should measure the size of lateral increments under the stationary measure. More precisely, if the density field is distributed under $\mu_{\rho}$, then the variance of the height difference $ h(\varkappa)-h(\varkappa+d)$ between two points at distance $d$ should scale as $A(\rho) d$ as $d$ grows. It is then easy to check that \eqref{eq:defA} is the correct definition for discrete directed polymers.

Using \eqref{eq:Aloggamma} and \eqref{eq:jloggamma}, one obtains that for the log-gamma polymer model, 
\begin{equation}
 \frac{1}{2}\lambda A^2 = \frac{A^2}{2} \frac{1}{\partial_{\vartheta} \rho} \frac{\partial}{\partial \vartheta} \left( \frac{1}{\partial_{\vartheta} \rho} \frac{\partial j}{\partial \vartheta} \right) = \frac{-1}{2} \frac{\psi'(\gamma-\vartheta)\psi''(\vartheta) + \psi'(\vartheta)\psi''(\gamma-\vartheta)}{\psi'(\gamma-\vartheta)+ \psi'(\vartheta)} .
\end{equation}
At $v=0$, the corresponding $\vartheta$ is $\vartheta=\gamma/2$ and one finds $\frac{1}{2}\lambda A^2 = \frac{-1}{2} \psi''(\gamma/2)$, so that 
\begin{equation}
h(0,t) \simeq - t \psi(\gamma/2)  +  \left(\frac{-1}{2}\psi''(\gamma/2)  t \right)^{1/3} \chi_2,
\label{eq:limittheoremloggamma}
\end{equation}
as $t$ goes to infinity, where we recall that $h$ was defined in \eqref{eq:defh}. The asymptotics \eqref{eq:limittheoremloggamma} was first proved in \cite{borodin2013log} for $v=0$ and obtained for arbitrary $v$ in \cite{thiery2014log, footnote8}. 

\bigskip 

Let us consider a general polymer model with weights $w=e^{\Enorm/T}$, where we assume that the distribution of energies $\Enorm$ has variance $1$. We will keep henceforth the notation $\Enorm$ for on-site energies with are assumed to be normalized to have variance $1$, while we use the letter $E$ to denote on-site energies which may depend on some parameter $\gamma$, the location, etc. To put the log-gamma polymer in this framework, one has to assume that $P(\Enorm)$ depends on $T$ (indeed, $\Enorm$ is distributed as $T$ times the log of an inverse gamma random variable with parameter $\gamma$). In any case, we define the temperature $T$ as
\begin{equation}
T=\frac{1}{\sqrt{\mathrm{Var} \log w}}.
\label{eq:deftemperature}
\end{equation} 
Since the variance of the log of an inverse gamma variable of parameter $\gamma$ is given by $\psi'(\gamma)$ (see \eqref{eq:loggammacumulants}), the parameter $\gamma$ is related to the temperature $T$ via $\psi'(\gamma)=1/T^2$, so that 
$ \gamma\propto T  $ for $\gamma \to 0$ and $\gamma\propto T^2$ for $\gamma\to\infty$. 

For an arbitrary distribution on weights $w$ with $5$ finite moments, one expects the limit theorem \eqref{eq:KPZlimittheorem} to still hold. Stationary distributions $\mu_{\rho}$ should exist under mild assumptions but they are in general not known explicitly so that  we cannot compute exactly the coefficients $\lambda$ and $\rho$.

However, in the large temperature regime, one expects, that as $T$ goes to infinity
\begin{equation}
\frac{1}{2} \lambda A^2 \simeq \frac{2}{T^4},
\label{eq:largetemperatureestimate}
\end{equation} 
provided the distribution of weights $w$ has a sufficient number of finite moments \cite{footnote7}. This estimate is based on the universality of convergence of directed polymer free energy at high temperature to the KPZ equation \cite{calabrese2010free, alberts2014intermediate} (see also \cite{krishnan2018tracy}). In the case of the log-gamma polymer, one can check that indeed as $\gamma$ goes to infinity, 
\begin{equation}
  \frac{1}{2} \lambda A^2 = \frac{-1}{2} \psi''(\gamma/2) \simeq \frac{2}{\gamma^2} \simeq \frac{2}{T^4}.
\end{equation}

\bigskip 

\textbf{Inhomogeneous case. }  Let us consider now a polymer model with weights $w=e^{\Enorm/T}$, where we assume that $T= T(\uptau)$ may depend onthe location of the site through the time $\uptau$. Again, for the log-gamma polymer to fit in this framework, one has to assume that the distribution of $E$ also depends on $\uptau$. The parameter $\gamma$ of the log-gamma polymer now depends on $\uptau$, and it is related to the temperature via 
\begin{equation}
\psi'(\gamma(\uptau))=\frac{1}{T(\uptau)^2}.
\end{equation}
In any case, $1/T(\tau) =\sqrt{\mathrm{Var} \log w}$ where the distribution of $w$ is now $\uptau$ dependent. 

Let us focus on the fluctuations of $h(\varkappa=0, \uptau)$. In the time dependent case, we expect that the scalings in \eqref{eq:KPZlimittheorem} will be modified, and the Tracy-Widom GUE limit distribution will occur only when the size of fluctuations of $h(0,\uptau)$ grows to infinity as $\uptau$ goes to infinity. Otherwise, the fluctuations of $h$ would be determined by a finite number of weights and we expect a non-universal distribution. 

The simplest functional of the fluctuations that is linear in time is the third cumulant, denoted $\langle h^3 \rangle_c$. In the homogeneous case,  $\langle h^3 \rangle_c \propto \lambda A^2 \uptau$. We expect that in the time dependent case, 
\begin{equation}
\langle h^3 \rangle_c \propto  \sum_{t=1}^{\uptau} \frac 1 2 \lambda(t) A(t)^2,
\label{eq:thirdcumulanth}
\end{equation} 
at least when the functions $\lambda(t)$ and $A(t)$ vary slowly enough, and the divergence of this quantity as $\uptau$ goes to infinity is a necessary criteria for $h(0,\uptau)$ to have Tracy-Widom distributed fluctuations. In particular, for the log-gamma polymer model, 
\begin{equation}
 \sum_{t=1}^{\uptau} \frac 1 2 \lambda(t) A(t)^2 = \frac 1 2 \sum_{t=1}^{\uptau} -\psi''(\gamma(t)/2). 
\end{equation}
Modulo some constant (due to the fact that we use the parameter $\uptau$ instead of $n$), this sum is asymptotically equivalent to the quantity $\sigma_n$ defined in \eqref{eq:sigman}, and we have seen that for the solvable inhomogeneous log-gamma polymer model studied in the previous sections, Tracy-Widom fluctuations occur if and only if $\sigma_n$ diverges. 

\bigskip 

However, as we have already mentioned, explicit expressions of the quantities $\lambda(t)$ and $A(t)$ are, in general, not available. If inhomogeneities are chosen so that $T(\uptau)$ goes to infinity as $\uptau$ goes to infinity, we may use the estimate \eqref{eq:largetemperatureestimate}, and we find that 
\begin{equation}
 \sum_{t=1}^{\uptau} \frac 1 2 \lambda(t) A(t)^2 \simeq  \sum_{t=1}^{\uptau} \frac{2}{T(t)^4},
\end{equation}
whenever the series are divergent (when the series are convergent the series may converge to different values). Recall that $1/T^4 = (\mathrm{Var} \log w)^2$. Hence, a general criteria to predict the occurrence of Tracy-Widom fluctuations is whether the sum of the squares of logarithms of weights converges or diverges along the polymer path.  

\medskip 
\textbf{Example 1: } For a model with weights $w=e^{\Enorm/T}$ where $\Enorm$ has a fixed distribution with variance $1$ and $T(t) = t^{a'}$, the series $\sum \frac{1}{T(t)^4}$ is divergent for $a'\leqslant 1/4$ and convergent for $a'>1/4$. Hence we expect that the free energy has Tracy-Widom fluctuations when $a'\leqslant 1/4$ and non-universal fluctuations determined by weights close to the origin when $a'>1/4$. 

\medskip 
\textbf{Example 2:} For the log-gamma polymer model with $\gamma(t)=t^a$. We have that $(\mathrm{Var} \log w )^2 = \psi'(\gamma)^2$. Since $\psi'(\gamma) \sim 1/\gamma$ as $\gamma\to\infty$, the series $\sum \psi'(\gamma)^2$ is divergent for $a\leqslant 1/2$ and convergent for $a>1$. Hence we expect that the free energy has Tracy-Widom fluctuations when $a\leqslant 1/2$ and non-universal fluctuations determined by weights close to the origin when $a>1/2$. This is exactly what we have proved for the model with inhomogeneities given as $\gamma_{i,j} = i^a+j^a$. 

\medskip 
\textbf{Example 3:} 
For the (homogeneous) KPZ equation itself (i.e. \eqref{kpz3} with all coefficients time independent) the KPZ scaling holds with $A=D/(2 \nu)$ and here $D=2 c$. In the units used 
here (i.e. in \eqref{kpz} of the Letter) we thus have $\nu=1$, $\lambda=2$
and $A=c$, hence $\frac{1}{2} A^2 \lambda=c^2$. The analogue of \eqref{eq:thirdcumulanth} 
for the time inhomogeneous KPZ equation $c \to c(t)$ thus yields 
$\langle h^3\rangle_c \propto \int_0^t c(u)^2 du$. We have seen in some cases that the divergence of this quantity is the exact criterion for TW type fluctuations at large times (and believed to hold more generally).

\section{III Numerical results}

We consider in this section two models:
\begin{itemize}
	\item A zero temperature model, with on site energies distributed as exponential variables of parameter $\gamma_{i,j} = (i+j)^a$. This corresponds to the zero temperature limit of the log-gamma polymer model discussed in the letter.  We denote by $L(n,n)$ (see \eqref{eq:defLnn}) the optimal energy (last passage time). 
	\item A positive temperature model, with Boltmann weights $w=e^{E/T}$ with $T=1$ where on site energies $E$ are distributed as exponential variables of parameter $\gamma_{i,j} = (i+j)^{a'}$. We denote by $\mathcal Z(n,m)$ or $Z_d(\varkappa, \uptau)$ its partition function, as in the letter. 
\end{itemize}
The positive temperature model converges (in the limit $T\to 0$) to the zero temperature model, i.e. $a=a'$. 
\subsection{1) Zero-temperature model}

In Fig. \ref{fig:DPTW} in the letter, we have shown the difference between the empirical CDF of the optimal energy $L(n,n)$  (for the zero temperature model with $\gamma_{i,j} = (i+j)^a$) and the CDF of the Tracy-Widom GUE distribution, for various polymer lengths $n$, and $a= 0.3, 0.4$. The empirical CDF have been centered and scaled to compare to the Tracy-Widom CDF. We have reproduced the results along with the additional case $a=0.2$ in Fig. \ref{fig:DPTWsuppmat}. 
\begin{figure}[h]
	\centering
	\includegraphics[width=5.5cm]{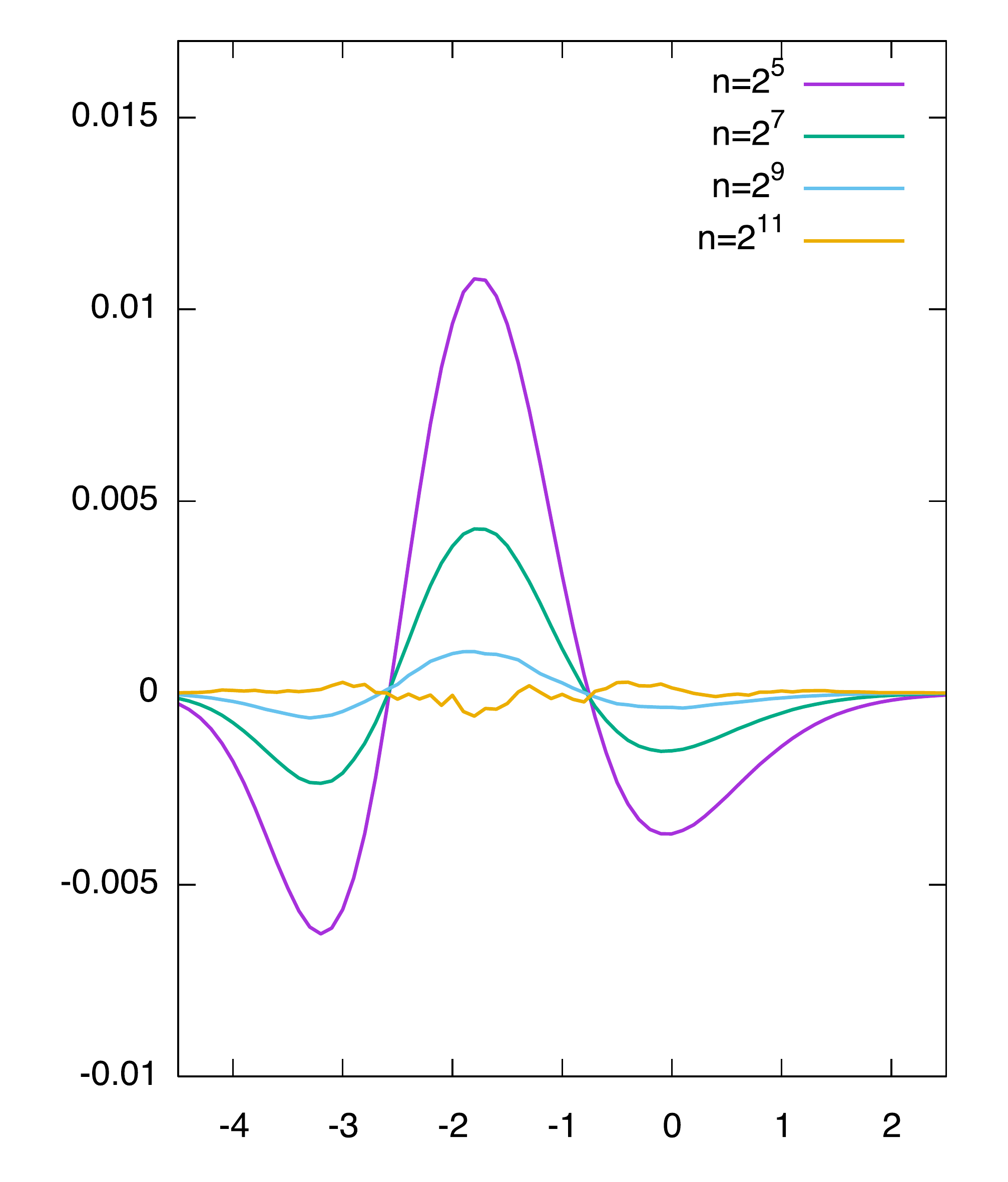}
	\includegraphics[width=5.5cm]{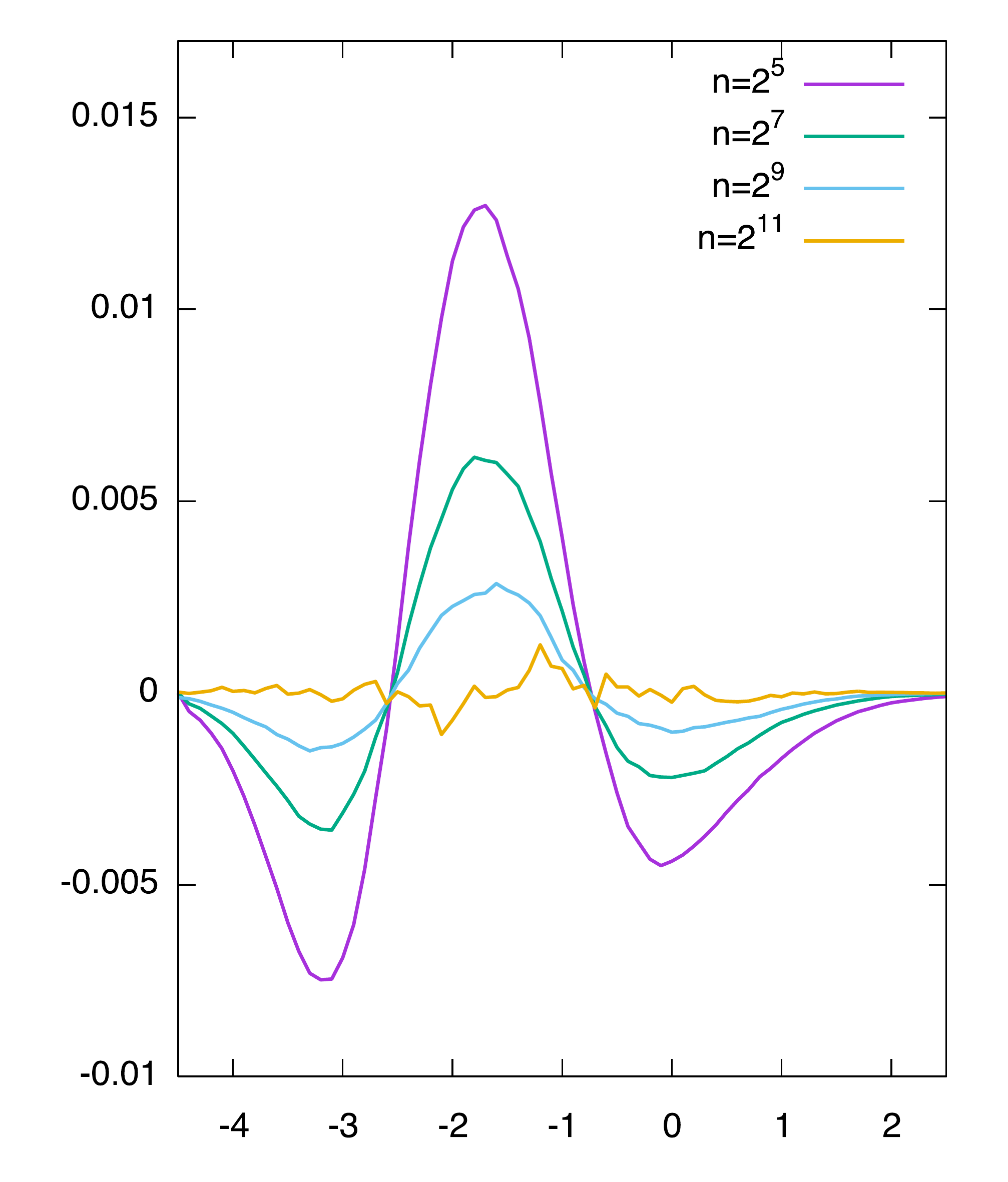}
	\includegraphics[width=5.5cm]{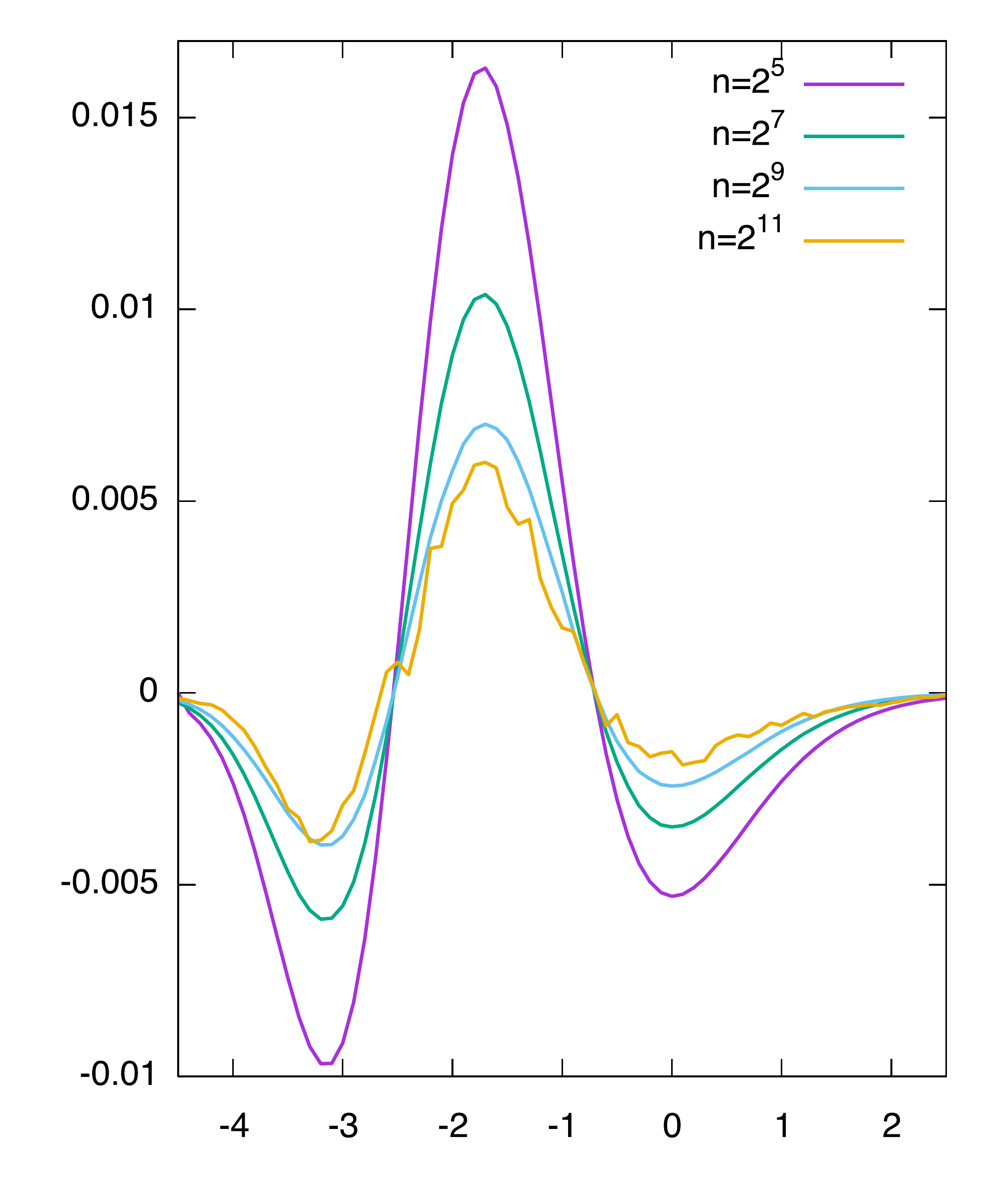}
	
	(a) \hspace{5cm} (b) \hspace{5cm} (c)
	\caption{Difference between the empirical CDF of the optimal energy in the zero temperature model (with exponential energies with parameter $\gamma_{i,j} = (i+j)^a$) and the CDF of the GUE TW distribution. The optimal energy is centered and scaled to have the same mean and
		variance as the GUE TW distribution. (a): case $a=0.2$, for various polymer lengths $n$. (b): case $a=0.3$ and the same polymer lengths. (c): case  $a=0.4$.}
	\label{fig:DPTWsuppmat}
\end{figure}
One can clearly see that regarding the bulk of the distribution, the convergence to the Tracy-Widom distribution seems to hold in the cases $a=0.2$ and $a=0.3<1/3$ but not in the case $a=0.4>1/3$ (where the difference of CDF seems to converge to a non-zero limit).

We have also investigated the tail of the distributions in Fig. \ref{fig:DPtailTW}. We find that for $a=0.2$ the tails of the empirical PDF of $L(n,n)$ (centered and scaled) seem to match the tails of the Tracy-Widom GUE distribution, while they are slightly different in the case $a=0.4$. 
\begin{figure}[h]
	\includegraphics[width=8cm]{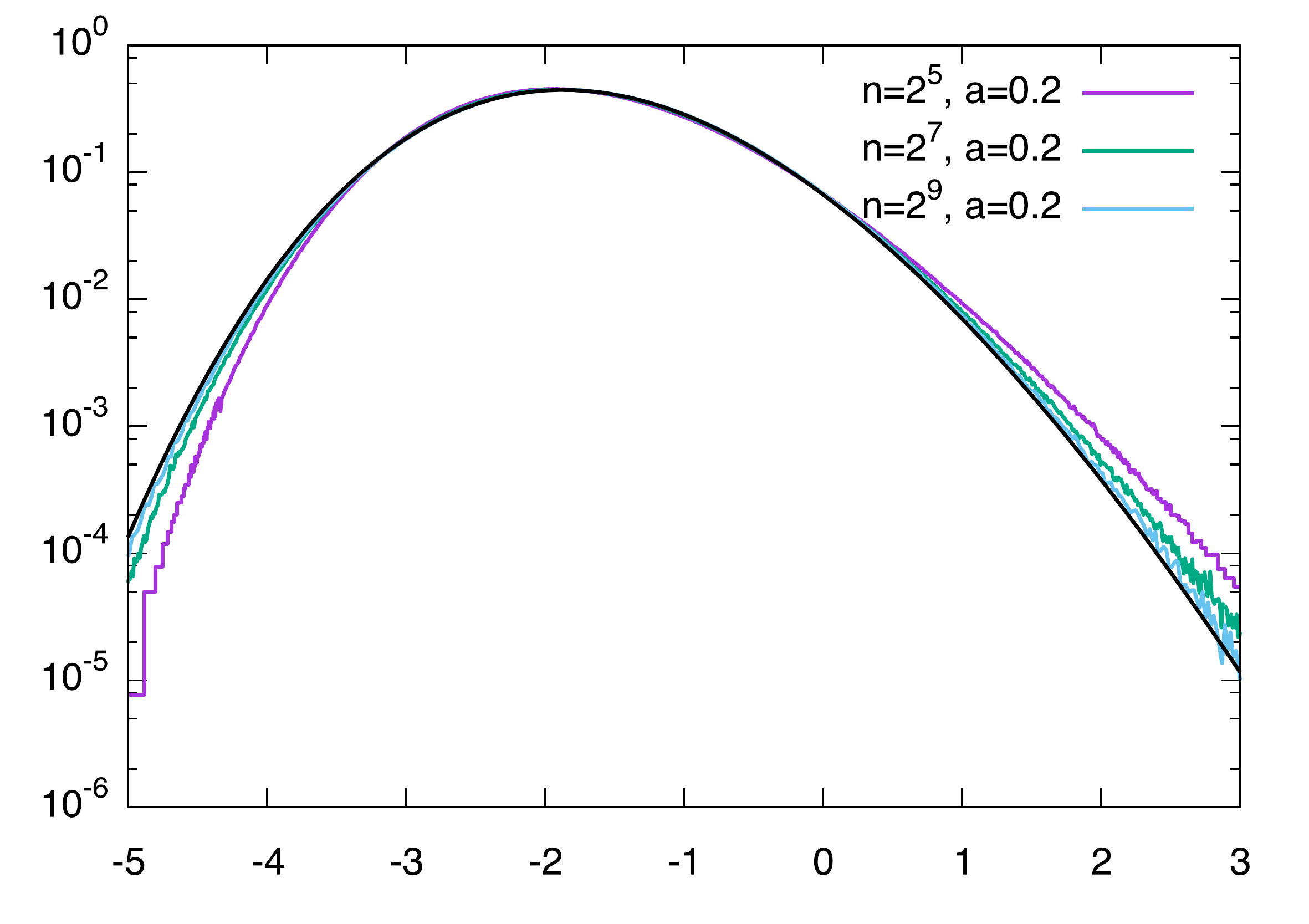}
	\includegraphics[width=8cm]{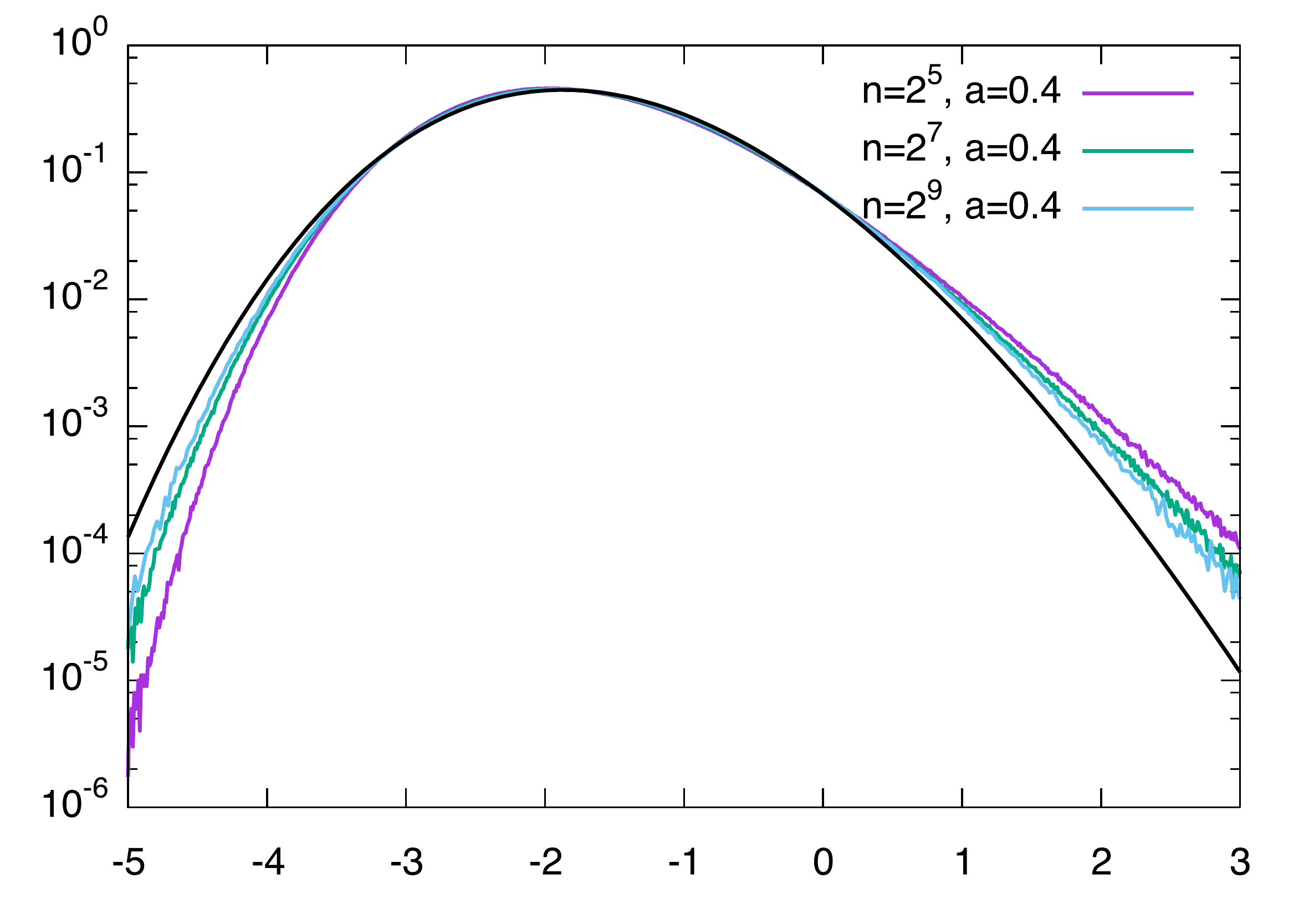}
	
	(a) \hspace{7cm} (b) 
	\caption{Tails of the empirical PDF of the optimal energy in the zero temperature model (same as in Fig. \ref{fig:DPTWsuppmat}), centered and scaled to the same mean and variance as the Tracy-Widom distribution (whose PDF is drawn in black for comparison). (a): case  $a=0.2$, for various polymer lengths $n$. (b): case  $a=0.4$. 
	}
	\label{fig:DPtailTW}
\end{figure}

\subsection{2) Positive temperature model}

Now we consider the positive temperature model. Let us emphasize that 
instead of simulating the log-gamma polymer model (which demands a lot of computational resources in order to reach large polymer sizes), we have performed simulations of  the polymer model with Boltzmann weights $w=e^{E}$ where on-site energies $E$ are distributed as exponential random variables with parameter $\gamma_{i,j} = (i+j)^{a'}$. According to the discussion made in Section II 6), we may view  this parameter $\gamma_{i,j}$ as a (local) temperature, so that the model corresponds to example 1 of section II 6). 

The results are shown in in Fig. \ref{fig:HTsuppmat}. There is a strong evidence that the fluctuations are TW distributed for $a'=0.2 <1/4$ and converge to another limit when $a'=0.3>1/4$ and $a'=0.4$. 
\begin{figure}[h]
	\centering
	\includegraphics[width=5.5cm]{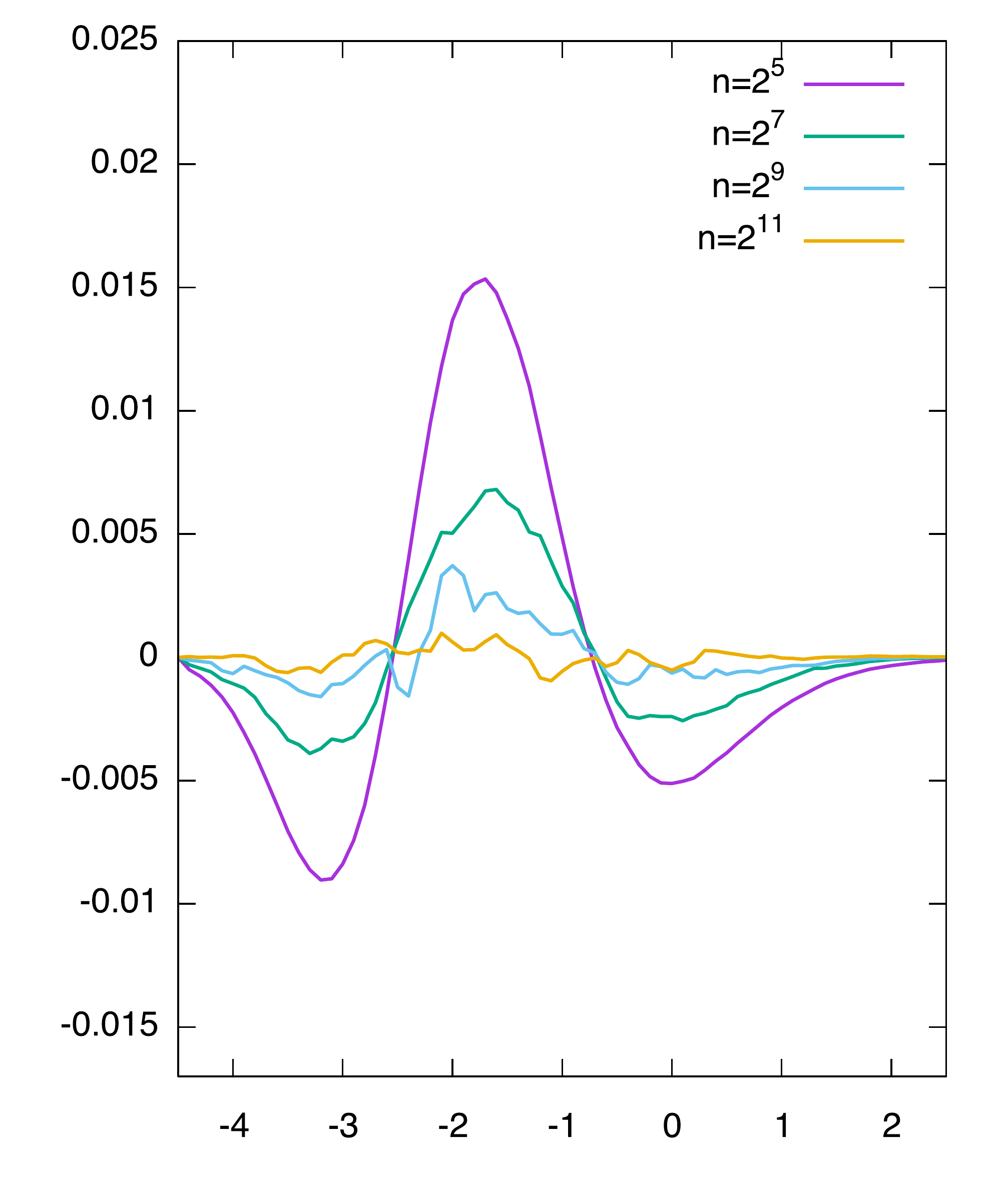}
	\includegraphics[width=5.5cm]{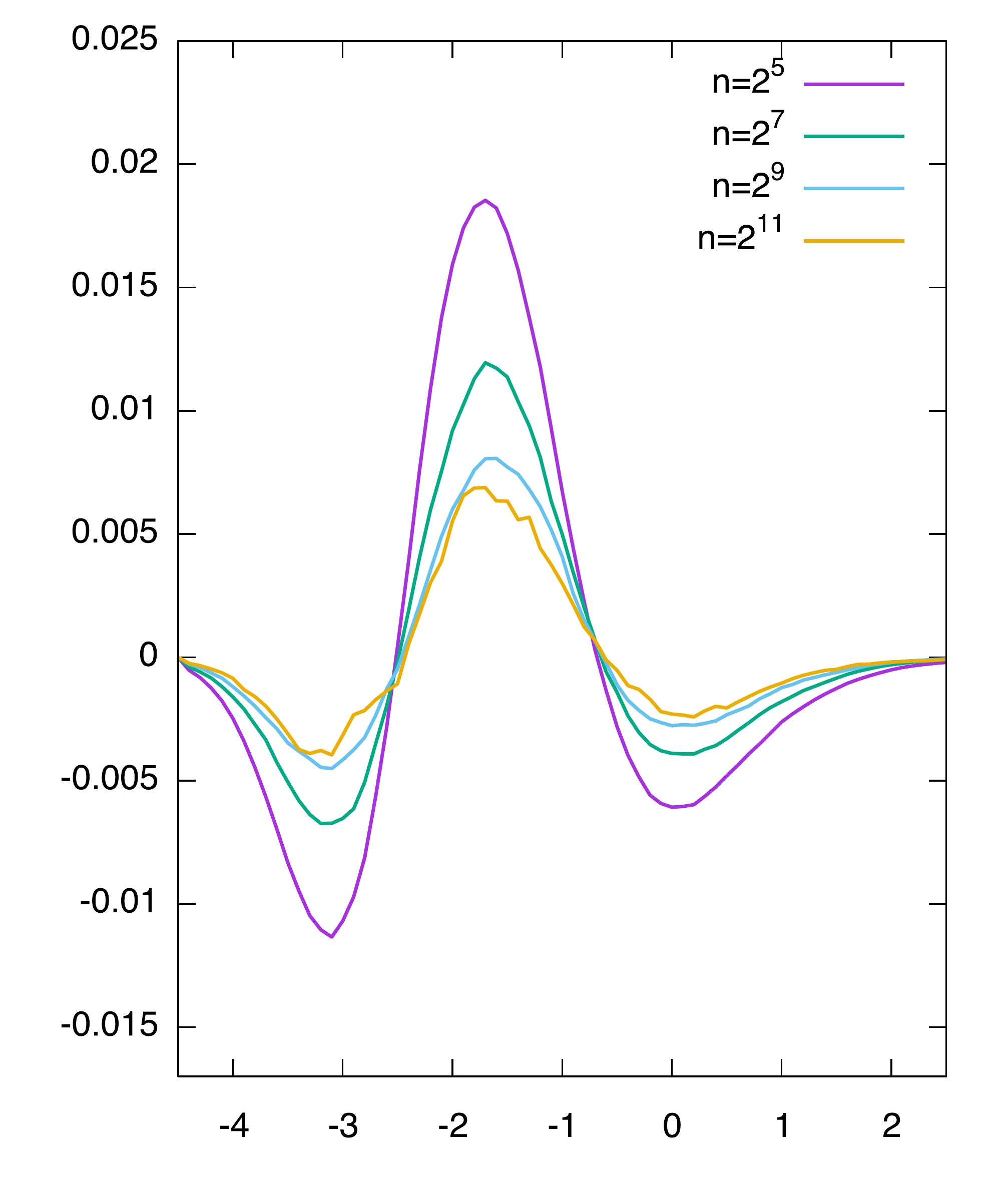}
	\includegraphics[width=5.5cm]{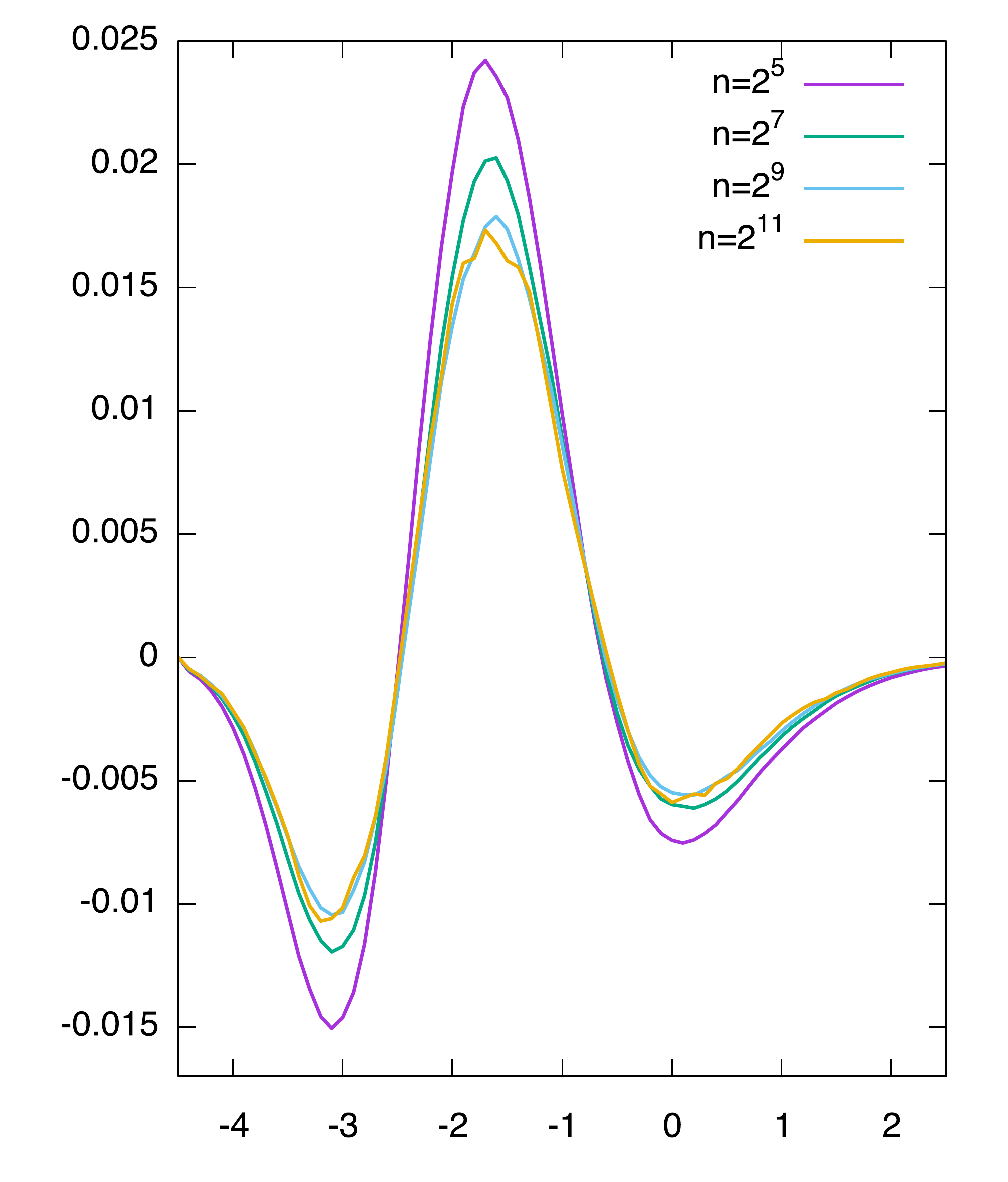}
	
		(a) \hspace{5cm} (b) \hspace{5cm} (c)
	\caption{Difference between the empirical CDF of the positive temperature free energy and the CDF of the GUE TW distribution (centered and scaled to the same mean and
		variance). (a): case $a'=0.2$, for various polymer lengths $n$. (b): case $a'=0.3$ and the same polymer lengths. (c): case  $a'=0.4$.}
	\label{fig:HTsuppmat}
\end{figure}

\subsection{3) Profile of the third-cumulant as the temperature varies}
 
Previous sections indicate that the transition between Tracy-Widom fluctuations or non-universal ones occurs between $a=0.3$ and $a=0.4$ (we expect $a_c=1/3$) for the zero temperature model considered in Section III 1), and between $a'=0.2$ and $a'=0.3$ (we expect $a'_c=1/4$) for the positive temperature model considered in Section III 2).  

Let us consider first the positive temperature model (with exponential on-site energies of parameter $\gamma_{i,j} = (i+j)^{a'}$).  In order to confirm the critical value $a'_c=1/4$, we use the criterium that Tracy-Widom fluctuations should occur if and only if $\langle \log Z_d(\varkappa=0, \uptau) ^3 \rangle_c$ diverges as $\uptau$ goes to infinity. Recall that for the homogeneous model with $\gamma_{i,j} \equiv \gamma$, we know that 
\begin{equation}
  \left\langle \log Z_d(\varkappa=0, \uptau) ^3 \right\rangle_c\simeq \uptau \, B(\gamma)  \, \langle \chi_2^3\rangle_c.
\end{equation}
For the positive temperature model with inhomogeneous parameters $\gamma_{i,j} = (i+j)^{a'}$, we expect (as in Section II 6)) that 
\begin{equation}
\left\langle \log Z_d(\varkappa=0, \uptau) ^3 \right\rangle_c\propto \langle \chi_2^3\rangle_c \sum_{t=1}^{\uptau} B(t^{a'}).
\label{eq:sumtoestimate}
\end{equation}  
To determine when the above series diverges, we need to estimate $B( \gamma)$. 
Note that from Section II 6),  $B(\gamma) = \frac 1 2 \lambda A^2 \left\langle \chi_2^3\right\rangle_c$ but we cannot compute explicitly  the coefficients $\lambda, A$ for the positive temperature model with exponential on-site energies. However, we expect that
\begin{equation}
B(\gamma) \simeq \frac{2}{\gamma^4}\left\langle \chi_2^3\right\rangle_c, \text{ as }\gamma \to\infty, \;\;\;\; B(\gamma) \simeq \frac{8 }{\gamma^3}\left\langle \chi_2^3\right\rangle_c, \text{ as }\gamma \to 0. 
\label{eq:thirdcumulantestimates}
\end{equation} 
Indeed, for the positive temperature model with homogeneous weights of parameter $\gamma$, we have that  at large temperature $\frac 1 2 \lambda A^2\simeq \frac{2}{T^4}$ (see \eqref{eq:largetemperatureestimate}). By our definition of the temperature \eqref{eq:deftemperature},  for this model the temperature is related to $\gamma $ via $T=\gamma$, and we recover that $B(\gamma) \simeq \frac{2}{\gamma^4}\left\langle \chi_2^3\right\rangle_c$ as in \eqref{eq:thirdcumulantestimates}. 
Thus, we deduce that for the inhomogeneous positive temperature model with parameter $\gamma_{i,j} = (i+j)^{a'}$, the series \eqref{eq:sumtoestimate} diverges if and only if $a'\leqslant 1/4$, hence the transition occurs at $a'_c=1/4$. 

\bigskip 
Now we turn to the zero temperature model. We denote by $L(\varkappa, \uptau)$ the optimal energy. Recall that for the homogeneous polymer model with $\gamma_{i,j}\equiv \gamma$, we have $\gamma \log  Z_d(\varkappa, \uptau) \simeq \mathcal L (\varkappa, \uptau)$ for small $\gamma$, where $\mathcal L$ is the optimal energy in the homogeneous zero temperature model with exponential energies of parameter $1$. For the latter model, it is known \cite{johansson2000shape} that $\mathcal L (\varkappa=0, \uptau) \simeq 2 \uptau + 2 \uptau^{1/3} \chi_2$ at large times $\uptau$, hence for small $\gamma$, 
we have \begin{equation}
 B(\gamma) \simeq \frac{8 }{\gamma^3}\left\langle \chi_2^3\right\rangle_c,
\end{equation}
as in \eqref{eq:thirdcumulantestimates}.  
Going back to the inhomogeneous model with $\gamma_{i,j} = (i+j)^a$, we expect that 
\begin{equation}
\left\langle L(\varkappa=0, \uptau)^3\right\rangle_c \propto \langle \chi_2^3\rangle_c  \sum_{t=1}^{\uptau} \frac{8}{t^{3a}},
\end{equation}
which diverges if and only if $a\leqslant 1/3$. Thus, 
for the zero temperature model, the transition occurs at $a_c=1/3$.

We have checked that the estimates \eqref{eq:thirdcumulantestimates} are consistent with third cumulants  obtained via numerical simulations in Fig. \ref{fig:thirdcumulant}. Our theoretical predictions seem correct, although values of $B(\gamma)$ obtained by numerical simulation are slightly above the predicted ones -- we believe that this is due to the small size of polymers that we have used (512) in order to produce Fig. \ref{fig:thirdcumulant}.
\begin{figure}
	\includegraphics[width=11cm]{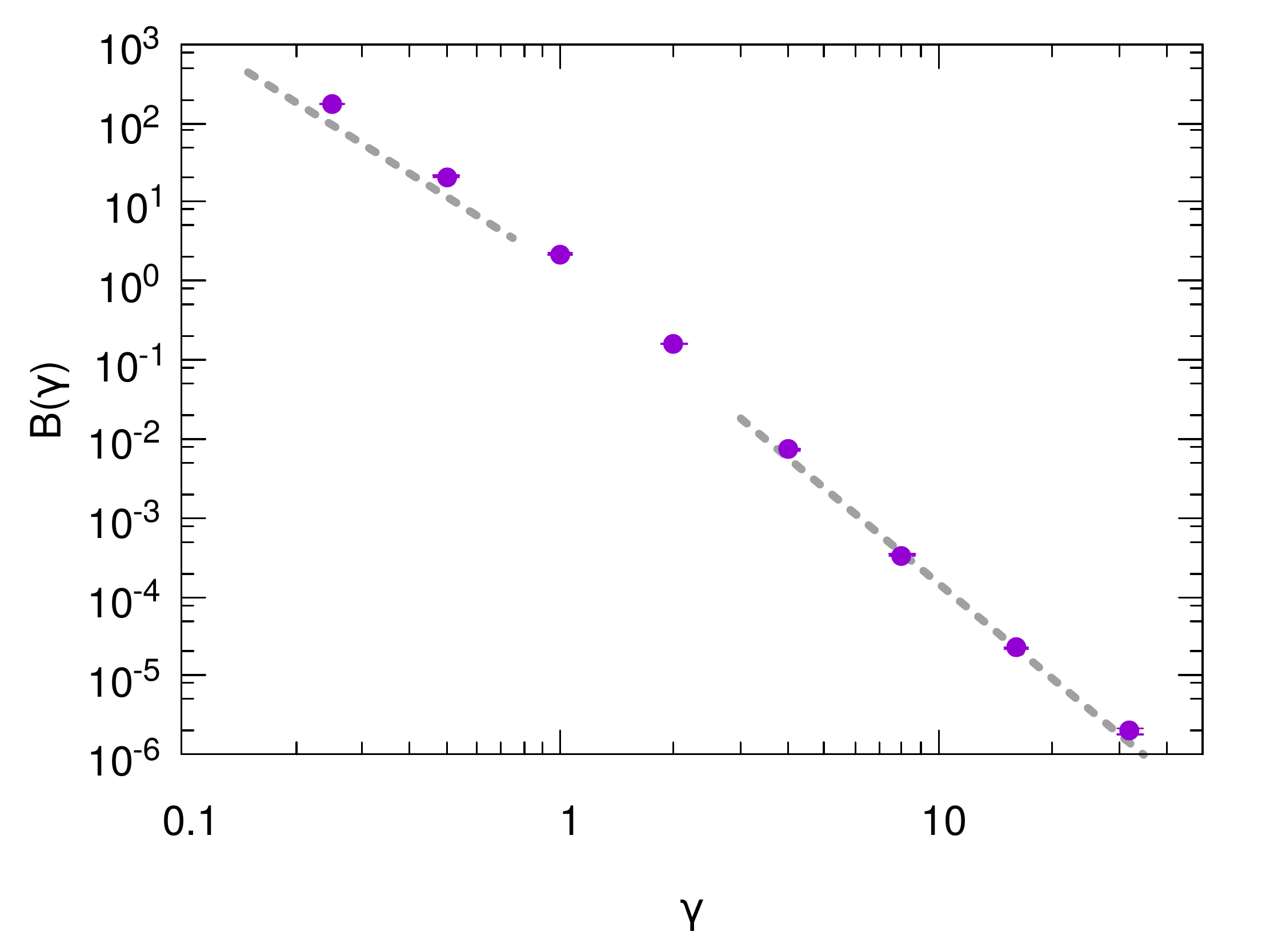}
	\caption{Log-log plot of $B(\gamma)$ for various values of $\gamma$ and directed polymers of length $512$. The dotted lines correspond to the curves $\frac{2}{\gamma^4}\left\langle \chi_2^3\right\rangle_c$ for large $\gamma$ and $\frac{8}{\gamma^3}\left\langle \chi_2^3\right\rangle_c$ for small $\gamma$. 
	}
	\label{fig:thirdcumulant}
\end{figure}

\section{IV Convergence to the KPZ equation}

In this section, we provide some details regarding the convergence of the discrete recurrence \eqref{eq:discreterecurrence} to the SHE. 

\subsection{1) Case $\gamma=\sqrt{n}/ c\big(\frac{i+j}{2n}\big)$}

Let us consider the scalings from \eqref{eq:scalingsloggamma} and assume that  $\gamma_{\varkappa, \tau}=\sqrt{n}/ c(t)$. Recall that we have set $\eta_{\varkappa, \tau}= \frac{2 w_{\varkappa, \tau}}{C_{\tau}}-1 $ where $C_{\tau} = 2 \mathbb E[w_{\varkappa, \tau}] = \frac{2}{\sqrt{n}/c(\tau/2n)-1}$. Using \eqref{eq:meamvarianceinversegamma}, we obtain that 
\begin{equation}
\mathbb E[\eta_{\varkappa, \tau}] = 0, \quad \mathrm{Var} [n\eta_{\varkappa, \tau}] \simeq n^{3/2}c(t).
\end{equation}
 The family of independent variables $n\eta_{\varkappa, \tau}$ converges (in the sense of distributions) to a white noise $\xi(x,t)$ of variance $c(t)/2$. Recalling the rescaled partition function $Z_r$ defined in 
 the main text as $Z_r(\varkappa, \uptau) = Z_d(\varkappa, \uptau) \left(\prod_{s=1}^{\uptau}C_{s}\right)^{-1}$,
 and multiplying \eqref{eq:discreteSHE} by $n$, we obtain 
 \begin{equation}
n \nabla_{\tau} Z_r(\varkappa, \tau) = \frac{1+\eta_{\varkappa, \tau}}{2}\,  n\Delta_{\varkappa} Z_r(\varkappa, \tau-1) + n\eta_{\varkappa, \tau} \, Z_r(\varkappa, \tau-1). 
\label{eq:discreteSHE2}
 \end{equation}
Let us define $Z(x,t)  = \lim_{n\to \infty} Z_r(2nt,x \sqrt{n}).$ Assuming the limit exists and converges in a suitably strong sense, the continuum limit of \eqref{eq:discreteSHE2} yields 
\begin{equation}
 \partial_{t} Z(x,t) = \partial_x^2 Z(x,t) + \sqrt{2c(t)} \xi(x,t)Z(x,t),  
\label{eq:SHEsuppmat}
\end{equation}
so that $h(x,t) = \log Z(x,t)$ solves 
\begin{equation}
\partial_t h = \partial_x^2 h + (\partial_x h)^2  + \sqrt{2 c(t)} ~ \xi(x,t).
\end{equation}

Our derivation of the continuum stochastic PDE \eqref{eq:SHEsuppmat} is somewhat formal. In order to make it more rigorous from the mathematical point of view, one should adapt the arguments of \cite{alberts2014intermediate} dealing with the homogeneous case:   write a a Feynman Kac-type representation for $Z_r(\tau, \varkappa)$, expand it as a chaos series, and  justify that this series converges to the chaos series solution associated to \eqref{eq:SHEsuppmat}.

\subsection{2) Case $\gamma=\frac{\sqrt{n}}{ 2c\left(\frac{i}{n}\right)}+\frac{\sqrt{n}}{ 2c\left(\frac{j}{n}\right)}$}

Previously, we had set up the renormalization factor $C_\tau$ so that the coefficient in front of the Laplacian -- that is $\frac{1+\eta_{\varkappa, \tau}}{2}$ --  has exactly mean $1/2$. However now, the expectation of $w_{\varkappa, \tau}$ depends on both $\varkappa$ and $\tau$, so that we cannot set $C_{\tau} = 2 \mathbb E[w_{\varkappa, \tau}]$. We will choose $C_{\tau}$ as $C_{\tau}=\frac{2}{\sqrt{n}/c\big(\frac{i+j}{2n}\big)-1}$ so that it matches with $2 \mathbb  E[w_{\varkappa, \tau}]$ at the first order in $n$. Recall that $i=(\tau+\varkappa)/2 = (2nt+\sqrt{n}x)/2$ and $j=(\tau-\varkappa)/2 = (2nt-\sqrt{n}x)/2$. Using \eqref{eq:meamvarianceinversegamma} and Taylor expansion of the function $c(t)$, we find 
\begin{equation}
\mathbb E[\eta_{\varkappa, \tau}] =  x^2 \frac{c(t)c''(t)-2(c'(t))^2}{8c(t)^2n} + o(1/n), \quad \mathrm{Var} [n\eta_{\varkappa, \tau}] \simeq n^{3/2}c(t).
\end{equation}
In this case, the family or random variables $n\eta_{\varkappa, \tau}$ converges to 
\begin{equation}
\sqrt{c(t)/2}\xi(x,t) + \frac{c(t)c''(t)-2(c'(t))^2}{8c(t)^2}. 
\end{equation}
so that in the continuum limit, \eqref{eq:discreteSHE} becomes 
\begin{equation}
\partial_{t} Z(x,t) = \partial_x^2 Z(x,t) + \left(  \sqrt{2c(t)}\xi(x,t) +  x^2\frac{c(t)c''(t)-2(c'(t))^2}{4c(t)^2}\right)Z(x,t). 
\end{equation}
so that $h(x,t) = \log Z(x,t)$ solves 
\begin{equation}
\partial_t h = \partial_x^2 h + (\partial_x h)^2  + a_c(t)\frac{x^2}{2} + \sqrt{2c(t)} ~ \xi(x,t),
\end{equation}
where $a_c(t)$ was defined in \eqref{eq:defac}.

\subsection{3) Comparison of the two approaches: change of variable and  discretization}

In this section, we explain how the results obtained from asymptotic analysis of exact formulas for the log-gamma polymer in Section II are consistent with the results obtained from the change of variables on the KPZ equation. Recall that we defined in the letter two inhomogeneous DP models with log-gamma weights whose free energy converges to the inhomogeneous KPZ equation, Model I and Model II defined in \eqref{eq:model1}. 

In the case $a=1/2$, the inhomogeneity parameters of Model II become 
\begin{equation}
 \gamma_{i,j} = \frac 1 2  \left( (i+t_0n)^a + (j+t_0n)^a \right)
\end{equation}
On one hand, as $t_0$ goes to zero, we recover the model studied above, i.e. $\gamma= i^a+j^a$ (the factor $1/2$ is inconsequential) and we have seen that the free-energy fluctuations are  Tracy-Widom distributed for large $n$ as long as $a\leqslant 1/2$ (see \eqref{eq:cvtoTW}). 

One the other hand, we have also seen in the previous section that the free energy of Model II converges to the solution of the inhomogeneous KPZ equation \eqref{kpz2} with $c(t) = \frac{1}{(t+t_0)^{a}}$ and  $V(x,t)=a_c(t) \frac{x^2}{2}$.  By the change of variables \eqref{transf1}, \eqref{transf2}, the solution $h(x,t) = \log Z(x,t)$ where $Z$ solves \eqref{eq:SHEtdependent} with $V(x,t)=a_c(t) \frac{x^2}{2}$,	can be mapped to the droplet solution of the  homogeneous KPZ equation \eqref{kpz} at time $\tau(t) = \log(1+t/t_0)$. As $t_0 \to 0$, $\tau(t)$ goes to $+ \infty$ and we recover  Tracy-Widom GUE  fluctuations as well.

\section{V Bethe ansatz}

Most of the results of this section, which we give here for completeness, were obtained by T. Thiery. 

\subsection{1) Time inhomogeneous evolution}

Consider $Z(x,t)$ satisfying the SHE equation \eqref{eq:SHEtdependent}
in the text with $V(x,t)=0$ and a white noise variance $c(t)$. The moments
\be
{\cal Z}(x_1,\dots,x_n;t) = \overline{Z(x_1,t) \dots Z(x_n,t)} 
\ee 
satisfy the imaginary time quantum mechanical evolution equation 
$\partial_t {\cal Z} = - H_n(t) {\cal Z}$ where $H_n(t)$ is a time dependent version of the attractive Lieb-Liniger
Hamiltonian 
\be
H_n(t) = -  \sum_{i=1}^n \partial_{x_i}^2 - 2 c(t) \sum_{i<j} \delta(x_i-x_j) 
\ee 
We consider the infinite line.
Let us denote $\Psi_\mu(c) \equiv \Psi_\mu(\vec x, c)$, with $\vec x = \{ x_1,\dots,x_n \}$, the 
eigenstates for a fixed value of $c(t)=c$. They are known from the Bethe ansatz to be the string states, 
parameterized by (i) the (integer) number of strings $1 \leq n_s \leq n$ (ii) the (integer) 
sizes of each string $m_j \geq 1$, $j=1,\dots,n_s$ with $\sum_{j=1}^{n_s} m_j=n$,
(iii) the real momenta of each strings $k_j$. Their corresponding eigenenergy is $E_\mu=\sum_{j=1}^{n_s} (m_j k_j^2- \frac{c^2}{12} m_j(m_j^2-1))$.
We denote these eigenstate labels collectively as $\mu$. 
We denote $\hat{\Psi}_\mu(c)=\Psi_\mu(c)/||\Psi_\mu(c)||$ the normalized eigenstates .

For the time dependent problem we are interested in the solution
$\Psi(t) \equiv \Psi(\vec x,t)$ of
\bea \label{eqt} 
\partial_t \Psi(\vec x,t) = - H_n(t) \Psi(\vec x,t)
\eea 
Since the $\Psi_\mu(c)$ form a basis for any $c$ we can always decompose, for each $t$
\bea
\Psi(t) = \sum_\mu a_\mu(t) \hat{\Psi}_\mu(c(t)) \quad , \quad  a_\mu(t) = \langle \hat{\Psi}_\mu(c(t)) | \Psi(t) \rangle 
\eea 
Inserting into \eqref{eqt} one finds
\bea
 \sum_\mu ( \partial_t a_\mu(t) \hat{\Psi}_\mu(c(t)) 
+ a_\mu(t) \dot c(t) \partial_c \hat{\Psi}_\mu(c(t)) ) = - \sum_\mu a_\mu(t) E_\mu(c(t)) \hat{\Psi}_\mu(c(t))
\eea
We can decompose
\bea
\partial_c \hat{\Psi}_\mu(c) = \sum_{\mu'} A_{\mu',\mu}(c) \hat{\Psi}_{\mu'}(c) 
\quad , \quad A_{\mu',\mu}(c) = \langle \hat{\Psi}_{\mu'}(c) | \partial_c \hat{\Psi}_\mu(c) \rangle
\eea 
Note that since the $|\hat{\Psi}_\mu \rangle$ form an orthonormal basis, 
one has $\partial_c \langle \hat{\Psi}_{\mu'} | \hat{\Psi}_\mu \rangle=0$ and 
the matrix $A_{\mu',\mu}(c)$ is anti-hermitian, i.e. $A_{\mu',\mu}(c)= - A^*_{\mu,\mu'}(c)$.

Projected on the basis we obtain the evolution equation for the $a_\mu(t)$ as
\bea
\partial_t a_\mu(t) = - E_\mu(c(t)) a_\mu(t) 
-  \dot c(t) \sum_{\mu'} A_{\mu,\mu'}(c(t))  a_{\mu'}(t) 
\eea 
which is an exact equation. 

In time-dependent problems the adiabatic limit is often discussed. In that limit $\dot c(t) \to 0$ and one
may want to approximate
$a_\mu(t) \simeq e^{- \int_0^t ds E_\mu(c(s))} a_\mu(0)$, leading to the adiabatic approximation
\be \label{a0} 
\Psi(t) \simeq \sum_\mu  e^{- \int_0^t ds E_\mu(c(s))}  \hat{\Psi}_\mu(c(t))  \langle \hat{\Psi}_\mu(c(0)) | \Psi(0) \rangle
\ee 
Here we note that the factor 
\be \label{a1doublon} 
e^{- \int_0^t ds E_\mu(c(s))} = e^{- \tau(t) \sum_{j=1}^{n_s} (m_j^3-m_j) - t \sum_{j=1}^{n_s} m_j k^2_j}  \quad , \quad \tau(t)=\int_0^t ds \, c(s)^2
\ee
depends both on the original time, and the "new" time $\tau(t)$ defined in the main text.
Usually the adiabatic limit can be controlled when there is a gap in the spectrum 
\cite{ComparatAdiabatic}. Here however because of the momenta $k_j$ the spectrum has 
a continuous part, so the general validity of the approximation is unclear. We point out further references \cite{IntegrableTimeDependent,ErmakovBethe} 
on related questions.

\subsection{2) One-string states}

The calculation of $A_{\mu,\mu'}(c)$ is usually quite non-trivial. 
Consider now the {\it 1 string states}, i.e. $n_s=1$, where all
$n$ bosons are in a single bound state of momentum $k$. They are parameterized
by a single momentum $\Psi_\mu \equiv \psi_k$. We now show that for these
states 
\be
A_{k,k'}(c) = 0 
\ee
Indeed, the un-normalized $1$string states are given by
\bea
\psi_k(\vec x) = n! e^{- \frac{c}{2} \sum_{i<j} |x_i - x_j |  } e^{i k \sum_i x_i} 
\eea
with energy $E = n k^2 - \frac{c^2}{12}n(n^2-1)$.
They are orthogonal for $k \neq k'$, and their inverse square norm is $|| \psi_k ||^{-2} = \frac{c^{n-1}}{n! n^2}$. 
It is easy to see that one has the explicit form
\bea
\partial_c \psi_k(\vec x) = - \frac{1}{2} \sum_{i<j} |x_i - x_j | \,  \psi_k(\vec x)
\eea 
Hence one has
\bea
\partial_c \hat \psi_k(\vec x) = f(\vec x) \hat \psi_k(\vec x)
\eea 
where the function $f(\vec x)= - \frac{1}{2} \sum_{i<j} |x_i - x_j |  + || \psi_{k}(c) || \partial_c || \psi_{k}(c) ||^{-1}
= - \frac{1}{2} \sum_{i<j} |x_i - x_j |  + \frac{1}{2} (n-1) c$, 
is real and independent of $k$. From this it follows that the matrix
$A_{k' ,k }(c) = \langle \hat{\psi}_{k'}(c)  | \partial_{c} \hat{\psi}_k(c) \rangle
= \int d\vec x \hat{\psi}^*_{k'}(\vec x,c) f(\vec x) \hat{\psi}_{k}(\vec x,c)
= A_{k ,k' }^*(c)$. Since it must also be anti-hermitian, it follows
that $A_{k' ,k }(c)=0$. \\

{\bf Tail of the one-point PDF}. It has been found, in a number of situations in the time-independent case
(see discussion in \cite{DeNardisPLD2timelong,DeLucaPLDNpaths,PLDLateTimes})
that if one restricts the sum over the eigenstates $\mu$
to the subspace of the single string states $n_s=1$, one obtains, in the large time limit, the exact right tail of the PDF of $h(0,t)$ for large positive values. It is thus tempting to surmise, heuristically, that a similar feature holds 
in the time-dependent case. Since $A_{k,k'}(c) = 0$, if one projects the evolution onto this subspace, the evolution
is then identical to the adiabatic one \eqref{a0}-\eqref{a1doublon}, where the sum over states $\sum_\mu$ 
becomes an integral $\int \frac{dk}{2 \pi}$ since the states are labeled by $k$. It would then lead to the conjecture
that the leading (stretched exponential) behavior of the tail is the same as the one for the time independent case, 
up to the replacement of $c^2 t$ by $\int_0^t du c(u)^2$, that is 
replacement of $t$ by $\tau(t)$. Indeed the integration over the momentum $k$ in \eqref{a1doublon} only leads to 
simple prefactors depending on the initial condition \cite{PLDLateTimes}. 
Note that, since there is a gap between the (low lying) 1-string states 
and the other eigenstates (with two or more strings), some further control on the full problem 
may even be possible
in the adiabatic limit. This is the case notably for the flat initial condition, when there is a clean gap between the
ground state (a single string with $m_1=n$ bosons and zero momentum $k=0$), and the
first excited states (a two string state) \cite{PCPLDFlat}.

{}

%
%
%
%

\end{widetext} 

\begin{thebibliography}{1}
\bibitem{footnote2} $\delta h:=  h(0,t) - \mathbb E[h(0,t)]$. We use the convention that $\propto$ means ``is proportional to'' and $\simeq$ means ``is equivalent to''.
\bibitem{footnote1}
	Our $c$ here is usually denoted $\bar c$ in the delta Bose gas, and is minus the conventional $c$ there.
\bibitem{footnote3} We emphasize that the character $\uptau$ denotes the time in the discrete model, not to be confounded with the continuous time $\tau(t)$.
\end{thebibliography}
\begin{thebibliography}{1}
\bibitem{footnote4} Everywhere we use the Cole-Hopf solution to the KPZ equation,
	i.e. from the solution of the SHE with Ito convention.
\bibitem{footnote5} Note that the deterministic term $- \frac{(1-\alpha) x^2}{4 t}$ can be read from the solution
	of the noiseless KPZ in the external potential $a(t) \frac{x^2}{2}$, given by \eqref{noiseless}. This is
	a general property. It holds also for the standard KPZ equation. Same remark for the deterministic term
	in \eqref{vardrop}.
\bibitem{footnote6} 
	The reason why this estimate works is due to the so-called statistical tilt symmetry (which forbids the renormalization of the parameter $A$, and of the elastic coefficient).
\bibitem{footnote8}
	Note that in \cite{thiery2014log} the discrete space variable $\varkappa$ is defined as $(m-n)/2$ hence $\varphi$ there is equivalent to $v/2$ there, and $\vartheta$ here identifies with $\gamma/2+k_{\varphi}$ there.
\end{thebibliography}
\end{document}